\documentclass{article}\usepackage[paper=a4paper,top=20mm, bottom=30mm, inner=20mm, outer=20mm]{geometry}

\usepackage{natbib}

\usepackage{mathtools,amssymb,mathrsfs,bm,centernot,siunitx}
\usepackage[scr=boondox]{mathalpha}	
\usepackage{graphicx}
\usepackage{caption,subcaption,float}
\usepackage{tabularx,multirow}
\usepackage[inline]{enumitem}

\usepackage{xcolor}

\usepackage{calc,etoolbox}

\usepackage[inline]{enumitem}

\makeatletter

\@ifclassloaded{article}{
	\numberwithin{equation}{section}
	\numberwithin{figure}{section}
	\numberwithin{table}{section}
	
}{}

\@ifpackageloaded{hyperref}{}{\usepackage[hidelinks]{hyperref}}
\@ifpackageloaded{cleveref}{}{
	\usepackage[compress]{cleveref}

	\crefformat{section}{\S#2#1#3}
	\crefformat{subsection}{\S#2#1#3}
	\crefformat{subsubsection}{\S#2#1#3}
	\crefrangeformat{section}{\S\S#3#1#4 to~#5#2#6}
	\crefmultiformat{section}{\S#2#1#3}{ and~\S#2#1#3}{, \S#2#1#3}{ and~\S#2#1#3}
	
	\@ifclassloaded{jfm}{
		\crefname{figure}{figure}{figures}
		\Crefname{figure}{Figure}{Figures}
	}{
		\crefname{figure}{fig.}{figs.}
		\Crefname{figure}{Fig.}{Figs.}
	}
	
	\crefname{equation}{}{}
	\crefformat{equation}{(#2#1#3)}
	\crefmultiformat{equation}{(#2#1#3)}{ and (#2#1#3)}{, (#2#1#3)}{ and (#2#1#3)}
}

\@ifclassloaded{WileyNJD-v2}{}{
	\usepackage[title]{appendix}
	\newenvironment{myappendices}
	{
		\begin{appendices}
		\crefalias{section}{appsec}
	}{
		\end{appendices}
	}
	
	\crefname{appsec}{Appendix}{Appendices}
	\Crefname{appsec}{Appendix}{Appendices}
}

\usepackage{xspace}

\DeclareRobustCommand\onedot{\futurelet\@let@token\@onedot}
\def\@onedot{\ifx\@let@token.\else.\null\fi\xspace}

\def\eg{e.g\onedot} 
\def\ie{i.e\onedot} 
\def\cf{\textit{cf}\onedot} 
\def\etc{etc\onedot}

\def\sf{s.f\onedot}

\@ifclassloaded{svjour3}{}{
	
}

\newcommand{\floattag}[1]{%
  \@namedef{the\@captype}{#1}%
  \@namedef{theH\@captype}{#1}%
  \addtocounter{\@captype}{-1}}

\makeatother
\input{Templates/StandardMathsMacros}
\graphicspath{{Figures/}}

\usepackage{comment}

\newcommand{\reavg}[1]{
	{\:\!\overline{\vphantom{\reflc{\phi}}#1}\:\!}
}
\newcommand{\reflc}[1]{
	\vphantom{\phi}#1'
}
\newcommand{\spavg}[3]{
	\mathrlap{\left\langle #1 \right\rangle}
	\mathrlap{\left. \phantom{\left\langle #1 \right.} \! \right._{#2}^{#3}}
	\mathmakebox[\maxof
		{\widthof{$\displaystyle \left\langle #1 \right\rangle$}}
		{\widthof{$\displaystyle \left. \phantom{\left\langle #1 \right.} \! \right._{#2}^{#3}$}}
	]{}
}
\newcommand{\sravg}[3]{
	\spavg{\reavg{#1}}{#2}{#3}
}

\newcommand{\ElTu}[1]{\breve{#1}}
\newcommand{\equl}{\text{eq}}

\newlength{\multlinedwidth}
\colorlet{darkred}{red!50!black}
\newlist{planlist}{enumerate}{1}
\setlist[planlist]{label=(\arabic*),leftmargin=10mm,rightmargin=5mm,labelsep=1mm,itemindent=0mm,before=\color{magenta},resume}
\usepackage{printlen}

\begin{document}

\title{Self-stratifying turbidity currents}

\author{Edward W.G. Skevington and Robert M. Dorrell}

\date{March 3, 2023}

\maketitle

\begin{abstract}
	Turbidity currents, seafloor flows driven by the excess density of suspended particles, are key conveyors of sediment, nutrient, and pollutant from the continental margins to deep ocean, and pose critical submarine geohazard risks. Due to their vast scale and extreme aspect ratio, extant models are constrained to highly simplified depth-averaged theory and fail to capture observed behaviour. We propose a novel depth-averaged model capturing the internal energy balance and the vertical profiles of velocity, depth, and turbulent kinetic energy. The vertical profiles change as the current evolves: it self stratifies. This enables the critical new insight that turbidity current propagation is enabled by bidirectional cascades between mean-flow kinetic, turbulent, and gravitational potential energies. 

	The model is generalised for fully confined ‘canyon’ flow (no lateral overspill), and partially confined ‘channel’ flow (lateral overspill over bounding levees). ‘Quasi-equilibrium’ solutions for self-stratifying turbidity currents are constructed. These solutions are weekly unstable and connected to a slowly evolving manifold, wherein environmental currents are likely found. Equilibrium solutions, found for channel flow, are not stable either. Levee overspill removes dilute, low momentum fluid, rejuvenating the flow, which can cause a positive feedback loop where the fluid becomes increasingly concentrated. We test the new theory by modelling flow in the Congo canyon-channel system, for the first time simulating a supercritical turbidity current that travels 100s \unit{km} to the distal reaches of a real-world system. It is shown that self-stratification enhances material and momentum fluxes, determining the environmental impacts and risks from such flows. 
\end{abstract}

\section{Introduction} \label{sec:intro}


In submarine environments, a suspended particle load can cause an excess of density over the surrounding ambient water, generating a current travelling down a slope under gravity. These \emph{turbidity currents} play a dominant role in oceanic sediment transport, transporting nutrients and pollutants from the continental margin to the deep ocean, preserving a record of paleo-environments, and posing a hazard to submarine infrastructure such as cables and pipes \citep{ar_Carter_2015,ar_Hsu_2008}. The run-out of these currents is impressive, some currents traversing thousands of kilometres \citep{ar_Lewis_1994,ar_Savoye_2009}, and the cumulative deposits can be enormous, up to $\qty{e7}{km^3}$ \citep{ar_Curray_2002}. Consequently, the dynamics of these currents is of practical interest.

The primary feature of turbidity currents is known as \emph{auto-suspension}: the particles are able to remain in suspension over a long period of time. 
That is, the time scale of settling is substantially longer than it would be in a quiescent fluid. 
Many turbidity currents go substantially beyond this threshold, being actively erosional in some location.
This is explained, at least in part, through energetic considerations: the energy gained though downhill motion exceeds the energy required to uplift the particles \citep{ar_Bagnold_1962}.
The behaviour of a full auto-suspending current is captured in a simplified model by \citet{ar_Parker_1986}. 
They modelled two dimensional currents using a depth averaged framework that captured the volume of fluid, suspended sediment load, momentum, and turbulent kinetic energy (TKE).
This facilitated a more detailed understanding of the motion of turbidity currents, known as \emph{self-acceleration}: the component of gravity acting downslope balances the turbulent drag against the bed and accelerates the quiescent ambient entrained by the upper shear layer, the shear local to the bed erodes particles (or at least prevents them from settling) to maintain the excess density, and the shear in the upper and lower layers generates the TKE required to maintain the particle load in suspension. 

While this model captures many aspects of turbidity currents, it does omit several important features of real currents. It assumes that the current takes on a \emph{top-hat} profile, a choice of \emph{transverse structure} with uniform concentration, velocity, and TKE within the current sharply dropping to zero outside, neglecting the complex and evolving transverse structure of the current. Additionally, it does not capture the effect of levee overspill, while real world currents often exceed the depth of their confining levees. Critically, it only works on steep slopes, predicting a stalling current on shallower slopes due to inadequate generation of mean-flow kinetic energy (MKE) from the downhill motion (see \cref{sec:example}).

These shortcomings manifest themselves when examining real world flows. Typically, after the current has travelled down a comparatively steep canyon in the continental shelf, it reaches a very flat region of the abyssal plane. For example, the Congo system reduces from a slope of around $1\%$ to $0.1\%$ over $\qty{1000}{km}$ \citep{ar_Savoye_2009}, the Hikurangi channel experiencing a similar decrease in slope over its $\qty{1500}{km}$ length \citep{ar_Lewis_1994}. Despite this, the currents can be very rapid and energetic even on shallow slopes \citep{ar_Vangriesheim_2009,ar_Simmons_2020}. Of particular interest are the flushing currents \citep{ar_Canals_2006,ar_AzpirozZabala_2017} which travel the entire length of the channel, eroding previously deposited sediment and transporting it to the distal fan. The self-accelerating mechanism cannot, alone, explain the run-out of currents in such flat systems, and this is conjectured to be a fundamental shortcoming of of the top-hat approximation \citep{ar_Skevington_F006_TCFlowPower}.


To amend this situation, some authors have augmented Parker's model, with the vision to explain the behaviour of real world currents in more challenging environments. \citet{ar_Pittaluga_2018} introduced levee overspill to the three equation model of \citet{ar_Parker_1986} (omitting the dynamics of TKE). Similar to its predecessor, this model uses a top-hat assumption for the transverse structure of the current in the channel. It imposes critical flow at the crest of the levees, consistent with considerations from open channel hydraulics. This results in a model which is capable of reproducing both experimental data from a flume, and field data from the Monterey Canyon over distances of \qty{100}{km}. Despite these practical successes, there are some issues with the consistency of their model. The momentum equation includes an entrainment term from the ambient, so that the entrained fluid has a velocity equal to that of the current. This artificial boost to the momentum will have significantly altered the predicted dynamics.

\citet{ar_Traer_2018_p1} proposed a four equation model capturing both mass lost due to lateral overspill by pressure and that lost due to the sinuosity of the current. This model is then investigated in \citet{ar_Traer_2018_p2}, demonstrating a range of interesting behaviours comparable to those seen in real physical flows. However, the levee overspill is derived from a consideration of viscous flow dynamics, not inviscid, which results in the wrong dependence on the excess depth of the current (compare eq. (6) from \citet{ar_Traer_2018_p1} with eq. (18) from \citet{ar_Pittaluga_2018}). Additionally, as with \citet{ar_Pittaluga_2018}, there is an inconsistency where the transverse structure of the concentration is included in the overspill but not in the pressure.


In this work, we develop the fundamental theory necessary to build an internally consistent turbidity current model. \Citet{ar_Pittaluga_2018} and \citet{ar_Traer_2018_p1} showed that the transverse structure of the concentration is significant for the levee overspill, and should be included in such models. \Citet{ar_Skevington_F006_TCFlowPower} have conjectured that the transverse structure is crucial for explaining auto-suspension in long run-out systems. These observations are consistent with results from the related filed of submarine salinity currents \citep{ar_Dorrell_2014}. Here, we incorporate the transverse structure of all fields in a consistent way for both the bulk flow and levee overspill. We derive our four equation model in \cref{app:deriv}, taking care to account for the possibility that the transverse structure may evolve as a function of time and space. We also carefully derive the levee overspill in \cref{app:deriv_levee}, formally justifying the imposition of criticality, and how this is affected by the transverse structure. The newly derived, general purpose model is presented in \cref{sec:model}, and can be supplemented with any desired transverse structure (and entrainment, drag, and erosion), opening up new avenues of investigation.

The model is derived in such a way as to not require any assumptions about the turbulent production, or buoyancy production, present in the current. However, it is possible to retrospectively deduce expressions for both of these quantities from the model, and see how they depend on both the structure of the current and the levee overspill (\cref{sec:production}). We find that, upon the full inclusion of a (similarity form) transverse structure, the production is significantly different to that in a top-hat model. In particular, \cref{eqn:egytransfer_shape} (for flows without levee overspill) includes additional terms, indicating non-negligible additional sources of TKE \citep[\cf][]{ar_Skevington_F006_TCFlowPower}. The characteristic structure of the system is analysed in \cref{sec:time}, demonstrating the effect of the transverse structure on the characteristic speeds. 

From here on we build a full model in the general framework developed. Closures are presented in \cref{sec:turbclose}, taking care to capture the effect of drag appropriately (some previous models have used drag closures which rely on an equilibrium between TKE and mean-flow kinetic energy (MKE)). A closure for the turbulent dissipation is also given, generalising the approach of \citet{ar_Parker_1986}. We then present a simple model for the transverse structure, focusing on the uplift of particles by turbulent diffusion (\cref{sec:vert}). Consequently, we incorporate sophisticated new dynamics into into the energetics of the system. For example, in a steady flow (only variation in space) as the TKE increases, particles are uplifted and gravitational potential energy (GPE) is stored, the uplift creating an adverse pressure gradient slowing the flow. When the TKE decreases, this GPE is released into the MKE accelerating the flow. This provides a new understanding of how energy is stored to be used further down the system.

The steady, spatially varying dynamics of a fully confined current is then investigated (\cref{sec:phase}), where we establish the conditions for the flow to be in pseudo-equilibrium (\ie only the depth of the current increases downstream). However, this is shown to not be the only type of flow that persists over long time; a slow manifold is present in the system which enables the slow evolution of the current over distances many orders of magnitude greater than the depth. This demonstrates that flows need not be in pseudo-equilibrium to travel great distances relatively unchanged, indeed the pseudo-equilibrium state is actually unstable. We then move onto the full equilibrium possible for currents undergoing levee overspill (\cref{sec:equil}). However, the most interesting case is that of a current in a narrowing channel, characteristic of natural systems, where we see that the enhancement of concentration and momentum afforded by the levee overspill dramatically alters the flow state, and makes possible flows which increase in concentration while being depositional. We finish our investigation by modelling flows along the Congo canyon (\cref{sec:example}). This demonstrates, for the first time, supercritical and depositional flow to the distal end of the channel, in agreement with real world observations \citep{ar_AzpirozZabala_2017}. We show that such a flow is possible only with the inclusion of transverse structure, pure top-hat models become subcritical and stall prior to commencing overspill.

This paper includes appendices. In \cref{app:ETshape} we compare our method of quantifying the transverse structure to that developed by \citet{ar_Ellison_1959} and \citet{ar_Parker_1986}, in \cref{app:deriv} we derive the model for the flow in the bulk, and in \cref{app:deriv_levee} we justify the criticality criterion for the levee overspill.
\section{General model} \label{sec:model}

\begin{figure}
	\centering
	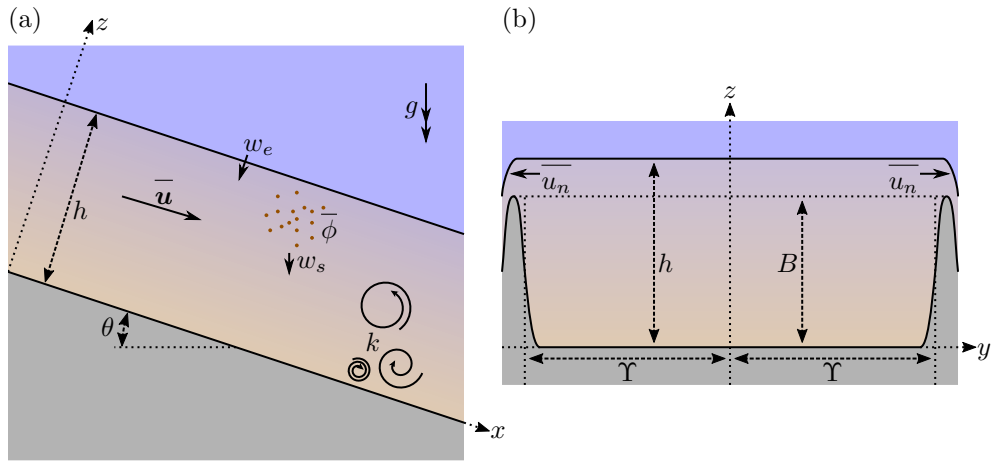
	\caption{The configuration for the turbidity current, with the bed in grey, ambient in blue, and current in brown fading toward blue in the less concentrated upper regions. (a) is a side view, showing the plane $y=0$ at the centre of the channel, while (b) shows a plane at constant $x$ for a current undergoing overspill.}
	\label{fig:configuration}
\end{figure}

We consider a dilute particle-driven gravity current flowing along an idealised rectangular channel, which may either partially confine or totally confine the current, as depicted in \cref{fig:configuration}. A general model for a current with arbitrary varying internal structure is derived in \cref{app:deriv} and presented here.

We define a coordinate system $(\vecb{x},t)=(x,y,z,t)$, where $x$ is the longitudinal direction and the axis lies along the thalweg of the channel; $y$ is the lateral direction, the levees existing at $y = \pm \Upsilon(x)$; $z$ is the transverse (bed normal) direction, the levees extending up to elevation $z = B(x)$; and $t$ is time. The channel, and coordinate system, are at an angle $\theta$ to the horizontal ($\theta>0$ meaning that the positive $x$ direction is downhill) which is slowly varying (terms proportional to $\dv*{\theta}{x}$ are assumed negligible). We denote the Reynolds averaged velocity of the fluid by $\reavg{\vecb{u}} = (\reavg{u},\reavg{v},\reavg{w})$, the Reynolds averaged normal velocity of the fluid over the levees by $\reavg{u_n}$, the Reynolds averaged volumetric concentration of particles by $\reavg{\phi}$, and the turbulent kinetic energy per unit mass (TKE) by $k$. We assume that the Reynolds averaged flow is symmetric across the channel (under $y \mapsto -y$). The elevation $z=h(x,t)$ is called the depth of the current, and is defined as the location above which the longitudinal and lateral velocities ($\reavg{u}$ and $\reavg{v}$) become negligible. Previous authors have used an integral scale based on $\reavg{u} \vphantom{a}^2$ to set the flow depth \citep{ar_Ellison_1959,ar_Parker_1986,ar_Islam_2010}; we establish the necessity of our approach in \cref{app:deriv_3D}. We define the normalised lateral and transverse coordinates $\mathscr{y} \eqdef y/\Upsilon$ and $\mathscr{z} \eqdef z/h$, and we express the quantities describing the flow as
\begin{subequations}
\allowdisplaybreaks[1]
\begin{align}
	\reavg{u}(x,y,z,t) 		&= \xi_u(x,\mathscr{y},\mathscr{z},t) \cdot U(x,t),
\\
	\reavg{v}(x,y,z,t) 		&= \xi_v(x,\mathscr{y},\mathscr{z},t) \cdot V(x,t),
\\
	\reavg{\phi}(x,y,z,t) 	&= \xi_\phi(x,\mathscr{y},\mathscr{z},t) \cdot \Phi(x,t),
\\
	k(x,y,z,t) 				&= \xi_k(x,\mathscr{y},\mathscr{z},t) \cdot K(x,t),
\\
	\reavg{u_n}(x,z,t) 		&= \xi_n(x,\mathscr{z},t) \cdot U_n(x,t),
\end{align}
where $\xi_u$, $\xi_v$, $\xi_\phi$, and $\xi_k$ satisfy
\begin{align} \label{eqn:profile_normalisation}
	\int_0^{\mathrlap{1}} \int_0^1 \xi_{\omitdummy}(x,\mathscr{y},\mathscr{z},t) \dd{\mathscr{y}} \dd{\mathscr{z}} &= 1,
	&&\text{and}&
	\int_0^1 \xi_n(x,\mathscr{z},t) \dd{\mathscr{z}} &= 1.
\end{align}
\end{subequations}
These expressions define $\xi_u$, $\xi_v$, $\xi_\phi$, $\xi_k$, and $\xi_n$, which capture the lateral and transverse structure of the current. In particular, the expressions defined in \cref{tab:shape_factors} are the features of the structure that influence the averaged properties of the current. Conversely, $U$, $\Phi$, $K$, and $U_n$ capture the longitudinal and temporal evolution of the channel averaged current. Note that our choice of normalisation \cref{eqn:profile_normalisation} and shape factors (\cref{tab:shape_factors}) are different to those used by some authors due to the different flow depth, see \cref{app:ETshape}. 

\begin{table}
	\centering
	$
	\renewcommand*{\arraystretch}{1.3}
	\begin{array}{ l | r@{}c@{}l | c | r@{}l | r@{}l }
								&	\multicolumn{3}{c|}{\text{Definition}}																						& 	\text{Top-hat}		&	\multicolumn{2}{c|}{\text{PGF\&Y}} & \multicolumn{2}{c}{\text{I\&I}}	\\ 
	\hline
		\sigma_{z\phi}			&	2 	\int_0^{\mathrlap{1}} \int_0^\infty					& \mathscr{z} \xi_\phi 			&	\dd{\mathscr{z}} \dd*{\mathscr{y}}			&	1					& 0&.67	& 0&.61	\\
		\sigma_{uu}				& 		\int_0^{\mathrlap{1}} \int_0^\infty					& \xi_u^2 						&	\dd{\mathscr{z}} \dd*{\mathscr{y}}			&	1 					& 1&.50	& 1&.50 	\\
		\sigma_{uuu}			& 		\int_0^{\mathrlap{1}} \int_0^\infty					& \xi_u^3 						&	\dd{\mathscr{z}} \dd*{\mathscr{y}}			&	1 					& 2&.48	& 2&.52 	\\
		\sigma_{u\phi}			& 		\int_0^{\mathrlap{1}} \int_0^\infty					& \xi_u \xi_{\phi}				&	\dd{\mathscr{z}} \dd*{\mathscr{y}}			&	1 					& 1&.52	& 1&.34 	\\
		\sigma_{uk}				& 		\int_0^{\mathrlap{1}} \int_0^\infty					& \xi_u \xi_k					&	\dd{\mathscr{z}} \dd*{\mathscr{y}}			&	1 					& ?&	& 1&.15 	\\
		\sigma_{uz\phi}			& 	2	\int_0^{\mathrlap{1}} \int_0^\infty					& \xi_u \mathscr{z} \xi_{\phi}	&	\dd{\mathscr{z}} \dd*{\mathscr{y}}			&	1 					& 0&.63	& 0&.70	\\
		\tilde{\sigma}_{uz\phi}		& 	2 \int_0^{\mathrlap{1}} \int_0^{\mathrlap{\infty}} \, \int_0^{\mathscr{z}_1}	& 
			\multicolumn{2}{@{}l|}{\eval*{\xi_u}_{\mathscr{z}_2} \cdot \eval*{\xi_{\phi}}_{\mathscr{z}_1} \dd{\mathscr{z}_2} \dd{\mathscr{z}_1} \dd*{\mathscr{y}}} & 1					& 1&.07	& 1&.03	\\
		\varsigma_{uz\phi}		&  
			&\multicolumn{2}{@{}l|}{\textfrac{1}{2} \ppar*{ \sigma_{uz\phi} + \tilde{\sigma}_{uz\phi} }} & 1	& 0&.85	& 0&.86	\\
	\hline
		\varsigma_{\phi}		& 		\int_0^1 								& \xi_{\phi} 			& \dd{\mathscr{y}} \big|_{\mathscr{z} = 0} 				&	\varsigma_\phi		& 3&.03	& 2&.72	\\
	\hline
		\varsigma_{u'}			&		\int_0^1 								& \pdv*{\xi_{u}}{\mathscr{z}} 	& \dd{\mathscr{y}} \big|_{\mathscr{z} = 0}		& \text{divergent} \\
	\hline
		\sigma_{v_y z\phi}' 	& \int_0^{\mathrlap{1}} \int_0^{\mathrlap{\infty}} \, \int_0^{\mathscr{z}_1} 		& 
			\multicolumn{2}{@{}l|}{ \eval*{ \pdv*{\xi_v}{\mathscr{y}} }_{\mathscr{z}_2} \eval*{\xi_{\phi}}_{\mathscr{z}_1} \dd{\mathscr{z}_2} \dd{\mathscr{z}_1} \dd{\mathscr{y}} }	&	1	&	\\
	\hline
		\varsigma_{z\phi}		&	2	\int_{\mathscr{z}_B}^\infty				&
			\multicolumn{2}{@{}l|}{ (\mathscr{z}-\mathscr{z}_B) \xi_{\phi} \dd{\mathscr{z}} \big|_{\mathscr{y} = 1} }	&	(1-\mathscr{z}_B)^2 \\ 
		\sigma_{nn}				&		\int_{\mathscr{z}_B}^\infty				& \xi_n^2						&	\dd{\mathscr{z}} \big|_{\mathscr{y} = 1}	&	(1-\mathscr{z}_B)^{-1}		\\
		\sigma_{nu}				& 		\int_{\mathscr{z}_B}^\infty				& \xi_n \xi_u					&	\dd{\mathscr{z}} \big|_{\mathscr{y} = 1} 	&	1					\\ 
		\sigma_{nuu}			& 		\int_{\mathscr{z}_B}^\infty				& \xi_n \xi_u^2					&	\dd{\mathscr{z}} \big|_{\mathscr{y} = 1} 	&	1					\\
		\sigma_{n\phi}			& 		\int_{\mathscr{z}_B}^\infty				& \xi_n \xi_{\phi}				&	\dd{\mathscr{z}} \big|_{\mathscr{y} = 1} 	&	1					\\
		\sigma_{nk}				& 		\int_{\mathscr{z}_B}^\infty				& \xi_n \xi_k					&	\dd{\mathscr{z}} \big|_{\mathscr{y} = 1} 	&	1					\\
		\sigma_{nz\phi}			& 	2	\int_{\mathscr{z}_B}^\infty				& \xi_n \mathscr{z} \xi_{\phi}	&	\dd{\mathscr{z}} \big|_{\mathscr{y} = 1} 	&	1+\mathscr{z}_B		\\
		\tilde{\sigma}_{nz\phi}		& 	2 \int_{\mathscr{z}_B}^{\mathrlap{\infty}} \int_{\mathscr{z}_B}^{\mathscr{z}_1}	&
			\multicolumn{2}{@{}l|}{\eval*{\xi_n}_{\mathscr{z}_2} \cdot \eval*{\xi_{\phi}}_{\mathscr{z}_1}	\dd{\mathscr{z}_2} \dd*{\mathscr{z}_1} \big|_{\mathscr{y} = 1} }	&	1-\mathscr{z}_B	\\
		\varsigma_{nz\phi}		&  
			&\multicolumn{2}{@{}l|}{\textfrac{1}{2} \ppar*{ \sigma_{nz\phi} + \tilde{\sigma}_{nz\phi} }} &	1	\\
	\end{array}$
	\caption{Definitions of the shape factors. Note that the shape factors involving $\xi_n$ have no meaning when $\mathscr{z}_B \geq 1$ (the current is not flowing over the levee so $U_n = 0$), and for the top-hat case we simplify manipulations by using the same expressions as when $\mathscr{z}_B < 1$. The final two columns includes values from \cite{ar_Parker_1987} and \cite{ar_Islam_2010}, see \cref{app:ETshape}.}
	\label{tab:shape_factors}
\end{table}

There are two special cases of flows we will consider, along with the general case. Firstly, when the structure of the current is in similarity form, the functions $\xi_\phi$, $\xi_u$, and $\xi_k$ do not depend on $x$ or $t$, meaning the shape factors in the first three sections of \cref{tab:shape_factors} ($\sigma_{z\phi}$ to $\varsigma_{u'}$) are constants, these shape factors describing properties of the along-channel flow, and not the lateral/overbank flow. Empirical values for these shape factors are included in \cref{tab:shape_factors}. However, there are inconsistencies in \cite{ar_Parker_1987} and \cite{ar_Islam_2010} where formally equivalent expressions yield different results: we have endeavoured to correct these for the table, see \cref{app:ETshape}. Secondly, for \emph{top-hat} flow, the structure takes the form
\begin{subequations}
\allowdisplaybreaks[1]
\begin{align}
	\xi_u = \xi_{k} &=
	\begin{cases}
		1,								&	 -1 < \mathscr{y} < 1 \text{ and } 0 < \mathscr{z} < 1,		\\
		0, 								& 	\mathscr{z} > 1,
	\end{cases}
	\\
	\xi_{\phi} &=
	\begin{cases}
		1,								&	-1 < \mathscr{y} < 1 \text{ and } 0 < \mathscr{z} < 1,		\\
		\varsigma_\phi					& 	-1 < \mathscr{y} < 1 \text{ and } \mathscr{z} = 0,			\\
		0, 								& 	\mathscr{z} > 1,
	\end{cases}
	\\
	\xi_v &=
	\begin{cases}
		2\mathscr{y},					&	 -1 < \mathscr{y} < 1 \text{ and } 0 < \mathscr{z} < 1,		\\
		0, 								& 	\mathscr{z} > 1,
	\end{cases}
	\\
	\xi_n &= 
	\begin{cases}
		\ppar{1-\mathscr{z}_B}^{-1},	&	\mathscr{z}_B < \mathscr{z} < 1,		\\
		0, 								& 	\mathscr{z} > 1,
	\end{cases}
\end{align}\end{subequations}
where $\mathscr{z}_B \eqdef B/h$, which yields the special values in \cref{tab:shape_factors}.

We now present the system of four equations derived in \cref{app:deriv} using the notation defined above. 
Firstly, conservation of fluid volume is
\begin{subequations}\label{eqn:1Dsys_vol}
\begin{gather}	\label{eqn:1Dsys_vol_DE}
	\pdv{h}{t} + \pdv{}{x} 
	\Big(
		\underbrace{  h U   \vphantom{\big)}}_{\mathclap{\text{volume flux}}}
	\Big) 
	= S_h,
\end{gather}
\vspace{-4mm}
\begin{flalign}\label{eqn:1Dsys_vol_S}
\text{where}&&
	S_h &=
	- \underbrace{  \frac{h}{\Upsilon} \ppar[\bigg]{ \dv{\Upsilon}{x} U + U_n }	}_{\mathclap{\text{lat. flow of volume}\quad}}   
	+ \underbrace{  w_e 										\vphantom{\bigg)}	}_{\mathclap{\qquad\text{entrainment velocity}}}.
&
\end{flalign}
\end{subequations}
Conservation of particle volume is
\begin{subequations}\label{eqn:1Dsys_part}
\begin{gather}	\label{eqn:1Dsys_part_DE}
	\pdv{}{t}\ppar*{h \Phi} + \pdv{}{x} 
	\Big(
		\underbrace{  \sigma_{u\phi} h U \Phi	\vphantom{\big)}}_{\mathclap{\text{particle flux}}}
	\Big)
	= S_{\Phi},
\end{gather}
\vspace{-4mm}
\begin{flalign}\label{eqn:1Dsys_part_S}
\text{where}&&
	S_{\Phi} &= 
	- \underbrace{	\frac{h}{\Upsilon} \ppar[\bigg]{ \sigma_{u\phi} \dv{\Upsilon}{x} U	+ \sigma_{n\phi} U_n } \Phi		}_{\mathclap{\text{lat. flow of particles}}}
	- \underbrace{	\varsigma_{\phi} w_s \Phi \cos\theta											\vphantom{\bigg)}	}_{\mathclap{\text{deposition}}}
	+ \underbrace{	E_s																				\vphantom{\bigg)}	}_{\mathclap{\text{erosion}}}.
&
\end{flalign}
\end{subequations}
Conservation of momentum (per unit mass) is
\begin{subequations}\label{eqn:1Dsys_mom}
\begin{gather}\label{eqn:1Dsys_mom_DE}
	\pdv{}{t} \ppar*{h U}
	+ \pdv{}{x} 
	\Big(
		  \underbrace{ \sigma_{uu} h U^2 										\vphantom{\big)}}_{\mathclap{\text{momentum flux}}}
		+ \underbrace{ \textfrac{1}{2} \sigma_{z\phi} R g h^2 \Phi \cos\theta	\vphantom{\big)}}_{\mathclap{\text{pressure}}}
	\Big)
	= S_U,
\end{gather}
\vspace{-4mm}
\begin{flalign}\label{eqn:1Dsys_mom_S}
\text{where}&&
	S_U &=
	- \underbrace{	\frac{h}{\Upsilon} \ppar[\bigg]{ \sigma_{uu} \dv{\Upsilon}{x} U + \sigma_{nu} U_n } U	}_{\mathclap{\text{lat. flow of momentum}}}
	- \underbrace{	u_\star^2															\vphantom{\bigg)}	}_{\mathclap{\text{basal drag}}}
	+ \underbrace{	R g h \Phi \sin\theta.												\vphantom{\bigg)}	}_{\mathclap{\qquad\text{downslope gravity}}}
&
\end{flalign}
\end{subequations}
The total energy per unit mass is defined to be the sum of the mean-flow kinetic energy (MKE), the turbulent kinetic energy (TKE), and the gravitational potential energy (GPE) that would be released if the suspended particles were moved in the $z$ direction (constant $x$,$y$) to the bed. This total energy satisfies the conservation equation
\begin{subequations}\label{eqn:1Dsys_egy}
\begin{multline}\label{eqn:1Dsys_egy_DE}
		\pdv{}{t} \Big( h \Big[
			  \underbrace{\textfrac{1}{2} \sigma_{uu} U^2						\vphantom{\big)}}_{\mathclap{\text{MKE}}}
			+ \underbrace{K														\vphantom{\big)}}_{\mathclap{\text{TKE}}}
			+ \underbrace{\textfrac{1}{2} \sigma_{z\phi} R g h \Phi \cos\theta	\vphantom{\big)}}_{\mathclap{\text{GPE}}}
		\Big]\Big)
		+ \pdv{}{x} \Big(
			  \underbrace{\textfrac{1}{2} \sigma_{uuu} h U^3 					\vphantom{\big)}}_{\mathclap{\text{MKE flux}}}
			+ \underbrace{\sigma_{uk} h U K 									\vphantom{\big)}}_{\mathclap{\text{TKE flux}\quad}}
			+ \underbrace{\varsigma_{uz\phi} U R g h^2 \Phi \cos\theta			\vphantom{\big)}}_{\mathclap{\qquad \text{GPE flux and pressure work}}}
		\Big)
		= S_T,
\end{multline}
\vspace{-4mm}
\begin{multline}\label{eqn:1Dsys_egy_S}
	\text{where}\qquad
	S_T =
	- \underbrace{	\frac{h}{\Upsilon} \ppar[\bigg]{ \sigma_{uuu} \dv{\Upsilon}{x} U + \sigma_{nuu} U_n } \textfrac{1}{2} U^2			}_{\mathclap{\text{lat. flow of MKE}}}	\\
	- \underbrace{	\frac{h}{\Upsilon} \ppar*{ \sigma_{uk} \dv{\Upsilon}{x} U + \sigma_{nk} U_n } K										}_{\mathclap{\text{lat. flow of TKE}}}
	- \underbrace{	\frac{h}{\Upsilon} \ppar*{ \varsigma_{uz\phi} \dv{\Upsilon}{x} U + \varsigma_{nz\phi} U_n } R g h \Phi \cos\theta	}_{\mathclap{\text{lat. flow of GPE}}}	\\
	+ \underbrace{	\sigma_{u\phi} U R g h \Phi \sin\theta															\vphantom{\bigg)}	}_{\mathclap{\text{work by downslope gravity}}}	\qquad\quad
	- \underbrace{	h \epsilon_T 																					\vphantom{\bigg)}	}_{\mathclap{\text{turbulent dissipation}}} \qquad
	- \underbrace{	w_s R g h \Phi \ppar*{\cos\theta}^2.															\vphantom{\bigg)}	}_{\mathclap{\text{GPE loss to settling}}}
\end{multline}\end{subequations}

Each pair of equations \cref{eqn:1Dsys_vol,eqn:1Dsys_part,eqn:1Dsys_mom,eqn:1Dsys_egy} includes a PDE describing the rate of change of the local amount of the conserved quantity, this change driven by advection by fluxes, pressure in the case of momentum and energy, and a source term. Note that the advection is altered by the presence of a shape factor, representing the correlation between the different fields, \eg in \cref{eqn:1Dsys_part_DE} a positive correlation between velocity and concentration means that the bulk particle load is advected faster than the bulk volume of fluid. The source terms contain many physical processes, including: the lateral flow (lat. flow) as driven by the changing width of the channel $\dv*{\Upsilon}{x}$ and the levee overspill at depth-average speed $U_n$; entrainment of ambient fluid at speed $w_e$; erosion of the bed at a rate $E_s$; drag from the bed with shear velocity $u_\star$; and turbulent dissipation both in the body of the fluid and through the bed at a depth-average rate $\epsilon_T$.

\subsection{Comparison to Parker's model}


The approach used here to model the energy is necessarily different to that of \cite{ar_Parker_1986} where top-hat profiles were used to simplify the fluxes. The important differences are in how we handle the TKE production, the turbulent buoyancy production, and the mean flow buoyancy production. Here, channel averaged quantities are respectively defined as
\begin{subequations}\begin{align}
	\mathcal{P} 	&\eqdef - \frac{1}{h \Upsilon} \int_0^{\mathrlap{\Upsilon}} \int_0^\infty \reavg{\reflc{u_i}\reflc{u_j}} \pdv{\reavg{u_i}}{x_j} \dd{z} \dd{y},	
\\
	\mathcal{B}_K 	&\eqdef   \frac{R g \cos{\theta}}{h \Upsilon} \int_0^{\mathrlap{\Upsilon}} \int_0^\infty \reavg{\reflc{w}\reflc{\phi}} \dd{z} \dd{y},	
\\
	\mathcal{B}_E 	&\eqdef   \frac{R g \cos{\theta}}{h \Upsilon} \int_0^{\mathrlap{\Upsilon}} \int_0^\infty \reavg{w}\reavg{\phi} \dd{z} \dd{y}.
\end{align}\end{subequations}
We understand $\mathcal{P}$ as the rate of conversion of MKE to TKE, while $\mathcal{B}_K$ is the rate of conversion of TKE to GPE, and $\mathcal{B}_E$ is the rate of conversion of MKE to GPE. In Parker's model, all three are calculated explicitly for the case of top-hat flow profiles from equations governing conservation of volume, particles, and MKE, and then substituted into the equation for conservation of TKE. Conversely, here we eliminate the three quantities between the three equations governing conservation of each type of energy, yielding a general model without assumptions on the transverse structure. Parker's model can be derived from our generalised model as follows. First, we substitute the top-hat shape factors from \cref{tab:shape_factors}, and eliminate the MKE and GPE terms from \cref{eqn:1Dsys_egy} using \cref{eqn:1Dsys_vol,eqn:1Dsys_part,eqn:1Dsys_mom} and, neglecting terms of order $\dv*{\theta}{x}$, obtain
\begin{subequations}
\allowdisplaybreaks[1]
\begin{gather}
	\pdv{h}{t} + \pdv{}{x}\ppar*{h U} = -\frac{h}{\Upsilon} \ppar*{ \dv{\Upsilon}{x} U + U_n } + w_e,
\\
	\pdv{}{t}\ppar*{h \Phi} + \pdv{}{x} \ppar*{h U \Phi}
	=- \frac{h}{\Upsilon} \ppar*{ \dv{\Upsilon}{x} U	+ U_n } \Phi
	- \varsigma_{\phi} \Phi w_s \cos\theta
	+ E_s,
\\
	\begin{multlined}[b][\multlinedwidth]
		\pdv{}{t} \ppar*{h U}
		+ \pdv{}{x} \ppar*{h U^2 + \textfrac{1}{2} R g h^2 \Phi \cos\theta  }
		= -\frac{h}{\Upsilon} \ppar*{ \dv{\Upsilon}{x} U + U_n } U
		- u_\star^2
		+ g R h \Phi \sin\theta,
	\end{multlined}
\\
	\begin{multlined}[b][\multlinedwidth]
		\pdv{}{t} \ppar*{ h K }
		+ \pdv{}{x} \ppar*{ h U K }
		= 
		- \frac{h}{\Upsilon} \ppar*{ \dv{\Upsilon}{x} U + U_n } K
		+ \frac{B}{\Upsilon} U_n R g h \Phi \cos\theta
	\\
		- h \epsilon_T
		+ U u_\star^2
		+ \textfrac{1}{2} \ppar*{ U^2 - R g h \Phi \cos\theta } w_e
	\\
		- \textfrac{1}{2} E_s R g h \cos\theta
		- \ppar*{ 1 - \textfrac{1}{2} \varsigma_{\phi} } w_s R g h \Phi \ppar*{\cos\theta}^2.
	\end{multlined}
\end{gather}\end{subequations}
Then we remove the effect of the channel by setting $\dv*{\Upsilon}{x} = U_n = 0$, and approximate $\cos\theta \simeq 1$, $\sin\theta \simeq \theta$ for shallow slopes, yielding equations (3), (4), (5), and (8) from \citet[][]{ar_Parker_1986}.

\subsection{Dimensionless parameters and critical flow} \label{sec:model_param}


We discus the dimensionless parameters to be used throughout this article. There are four groups which govern the internal dynamics of the flow (neglecting the effect of the levees): the slope angle $\theta$ along with
\begin{subequations} \label{eqn:dimless_param}
\begin{align}
	\Fro &\eqdef \ppar*{ \frac{\sigma_{uu}}{\sigma_{z\phi}} }^{1/2} \frac{U}{\ppar*{R g h \Phi \cos\theta}^{1/2}},
\\
	\beta &\eqdef \textfrac{1}{2} \varsigma_\phi \sigma_{z\phi} \frac{ w_s R g \Phi \ppar*{\cos\theta}^2 }{ \mathcal{B}_K },
\\
	\gamma &\eqdef \ppar*{\textfrac{1}{2} \sigma_{uu}}^{1/2} \frac{U}{\sqrt{K}}.
\end{align}\end{subequations}
The parameter $\beta$ acts like a Rouse number, and is computed as the ratio of the rate of the GPE loss (per unit mass) to settling to that gained by turbulent work. The change in GPE (per unit mass) is
\begin{gather*}
	\pdv{}{t} \text{GPE} 
	= \pdv{}{t} \textfrac{1}{2} \sigma_{z\phi} R g h \Phi \cos\theta
	= \frac{1}{h \Upsilon} \int_0^{\mathrlap{\Upsilon}} \int_0^\infty R g z \pdv{\reavg{\phi}}{t} \cos\theta \dd{z} \dd{y}.
\intertext{In similarity flow, so that $\xi_\phi$ is time invariant, the evolution of concentration driven by settling alone is}
	\eval*{h \pdv{\reavg{\phi}}{t}}_{\text{settling}} 
	= \eval*{\xi_\phi \cdot \pdv{}{t} \ppar*{h \Phi}}_{\text{settling}} 
	= - \xi_\phi \cdot \varsigma_\phi w_s \Phi \cos\theta = - \varsigma_\phi w_s \reavg{\phi} \cos\theta,
\shortintertext{thus}
	\eval*{\dv{}{t} \text{GPE}}_{\text{settling}}
	= \frac{-\varsigma_\phi}{h^2 \Upsilon} \int_0^{\mathrlap{\Upsilon}} \int_0^\infty w_s R g z  (\cos\theta)^2\reavg{\phi} \dd{z} \dd{y}
	= - \textfrac{1}{2} \varsigma_\phi \sigma_{z\phi} w_s R g \Phi (\cos\theta)^2,
\end{gather*}
and we arrive at the expression for $\beta$ given, which is discussed further in \cref{sec:vert}. 

The construction of the other parameters can be elucidated through substitution of the shape factors from \cref{tab:shape_factors} and rearranging, which yields
\begin{align} 
	\label{eqn:dimless_param_integral}
	\Fro^2 &= \frac{ \int_0^{\mathrlap{\Upsilon}} \int_0^\infty \textfrac{1}{2} \reavg{u}^2 \dd{z} \dd{y} }{ \int_0^{\mathrlap{\Upsilon}} \int_0^\infty R g z \reavg{\phi} \cos\theta \dd{z} \dd{y} },
&
	\gamma^2 &= \frac{ \int_0^{\mathrlap{\Upsilon}} \int_0^\infty \textfrac{1}{2} \reavg{u}^2 \dd{z} \dd{y} }{ \int_0^{\mathrlap{\Upsilon}} \int_0^\infty k \dd{z} \dd{y} }.
\end{align}
We identify $\gamma^2$ as the ratio of MKE to TKE. For the Froude number $\Fro$, shape factors have been included in \cref{eqn:dimless_param} so that $\Fro^2$ is the ratio of MKE to GPE, and for this reason we term it the \emph{energetic Froude number}. For similarity flow, this Froude number simplifies the classification of criticality, the critical flow depth $h_c$ defined to be that which minimises the channel averaged flow energy $E$ (MKE+TKE+GPE+pressure) for a given volume flux $q = Uh$, $K$, and $\Phi$. Explicitly
\begin{gather}
	E = \frac{\sigma_{uu} q^2}{2 h^2} + K + \sigma_{z\phi} R g h \Phi \cos\theta,
\quad\text{thus}\quad
	0 = \pdv{E}{h} = - \frac{\sigma_{uu} q^2}{h^3} + \sigma_{z\phi} R g \Phi \cos\theta
\notag\\
\text{precisely when} \qquad
	h = h_{\text{crit}} \equiv \ppar*{\frac{\sigma_{uu} q^2}{\sigma_{z\phi} R g \Phi \cos\theta}}^{1/3}
\quad\text{and}\quad
	\Fro^2 = 1.
\end{gather}
(This is equivalent to the result in \cref{app:deriv_levee_quant} for lateral flow.) By the discussion in \cref{sec:time}, we also have that a characteristic of the system are stationary when $\Fro=\pm 1$ for similarity flow. This shows that the \emph{energetic Froude number} correctly captures the defining property of the Froude number: the flow speed is equal to the celerity of long waves at $\Fro=\pm1$ \citep{ar_Baines_2003}. This has been shown not to be the case for other choices of `Froude number', such as that defined in terms of Ellison-Turner variables (see \cref{app:ETshape}) \citep{ar_Sumner_2013}, or a modification of that definition \citep{ar_Dorrell_2016}, therefore these cannot strictly be considered Froude numbers. We propose the \emph{energetic Froude number} as a superior replacement.


\subsection{Levee overspill}

We now present the levee overspill rate, which is derived in detail in \cref{app:deriv_levee}. If the fluid is not deep enough to overspill ($h \leq B$) then we can simply set $U_n=0$. When there is overspill ($h>B$) it can be shown that under some reasonable assumptions the flow at the levee crest must be critical (calculating the energetic Froude number with purely the lateral component of velocity). Rewriting \cref{eqn:levee_downslope_momy_crit_v} using the notation of this section results in 
\begin{align} \label{eqn:overspill_rate}
	U_n = \ppar*{ \frac{\varsigma_{z\phi}}{\sigma_{nn}} R g h \Phi \cos\theta }^{1/2}.
\end{align}
For a top-hat flow we obtain the outflow flux
\begin{align}
	h U_n = \ppar*{h-B}^{3/2} \ppar*{ R g \Phi \cos\theta }^{1/2},
\end{align}
which is of the form used by \citet{ar_Pittaluga_2018}. For the general case, the shape factors of \cref{tab:shape_factors} ($\varsigma_{z\phi}$ to $\varsigma_{nz\phi}$) need to be evaluated, with the integrals being performed at the location of the levee crest.

\section{Confined pseudo-equilibrium flow} \label{sec:quasiequ}


We begin our analysis of \cref{eqn:1Dsys_vol,eqn:1Dsys_part,eqn:1Dsys_mom,eqn:1Dsys_egy} by considering the steady flow of fluid down a deep channel of constant width and slope ($\pdv*{}{t} = \dv*{\theta}{x} = \dv*{\Upsilon}{x} = 0$, $B\to\infty$). Under such conditions we may expect the fluid to reach a \emph{pseudo-equilibrium} configuration after flowing a sufficient distance, where the transverse structure enters similarity form and the dimensionless parameters \cref{eqn:dimless_param} become independent of $x$,
\begin{align} \label{eqn:quasiequ_defn}
	\dv{\Fro}{x} = \dv{\beta}{x} = \dv{\gamma}{x} = 0.
\end{align}
The depth, $h$, may still vary with $x$.
Using the definitions \cref{eqn:dimless_param} and the buoyancy production \cref{eqn:production_GPE_from_TKE}, the conditions
\begin{align} \label{eqn:quasiequ_defn_simp}
	\dv{}{x} \ppar*{h \Phi} = \dv{U}{x} = \dv{K}{x} = 0
\end{align}
imply \cref{eqn:quasiequ_defn}, as used by \citet{ar_Parker_1986}. (These deductions are consistent with the expressions for $\mathcal{B}_K$ from either \cref{eqn:production_GPE_from_TKE} or \cref{eqn:simplified_beta}, or otherwise assuming that after substitution of \cref{eqn:quasiequ_defn_simp} $\mathcal{B}_K/\Phi$ is a function of $h \Phi$, $U$, $K$ and $\dv*{h}{x}$.)
Simplifying the governing system \cref{eqn:1Dsys_vol_DE,eqn:1Dsys_part_DE,eqn:1Dsys_mom_DE,eqn:1Dsys_egy_DE} using the pseudo-equilibrium conditions \cref{eqn:quasiequ_defn_simp}
\begin{subequations} \label{eqn:quasiequ_condn_genS}
\begin{align}
	U \dv{h}{x} &= S_h,
\\
	0 &= S_{\Phi},
\\
	\ppar*{ \sigma_{uu} U^2 + \textfrac{1}{2} \sigma_{z\phi} R g h \Phi \cos\theta } S_h &= U S_U,
\\
	\ppar*{ \textfrac{1}{2} \sigma_{uuu} U^2 + \sigma_{uk} K + \varsigma_{uz\phi} R g h \Phi \cos\theta } S_h &= S_T,
\end{align}
\end{subequations}
thus
\begin{subequations}
\allowdisplaybreaks[1]
\begin{gather}
	\label{eqn:quasiequ_vol}
	\dv{h}{x} = \frac{w_e}{U},
\\
	\label{eqn:quasiequ_part}
	0
	=
	\varsigma_{\phi} w_s \Phi \cos\theta
	- E_s,
\\			
	\label{eqn:quasiequ_mom}
	\ppar*{ \sigma_{uu} U^2 + \textfrac{1}{2} \sigma_{z\phi} R g h \Phi \cos\theta } w_e
	= U R g h \Phi \sin\theta
	- U u_\star^2,
\\			
	\label{eqn:quasiequ_egy}
	\begin{multlined}[b][\multlinedwidth]
		\ppar*{ \textfrac{1}{2} \sigma_{uuu} U^2 + \sigma_{uk} K + \varsigma_{uz\phi} R g h \Phi \cos\theta } w_e	\\
		=
		\sigma_{u\phi} U R g h \Phi \sin\theta - h \epsilon_T - w_s R g h \Phi (\cos\theta)^2.
	\end{multlined}
\end{gather}
\end{subequations}
This set of flow conditions will be used as the canonical example of turbidity current dynamics,
\section{TKE and buoyancy production} \label{sec:production}


In our derivation of the system \cref{eqn:1Dsys_vol,eqn:1Dsys_part,eqn:1Dsys_mom,eqn:1Dsys_egy} in \cref{app:deriv}, several additional equations were produced. In particular, equations for conservation of MKE,
\begin{subequations}	\label{eqn:1Dsys_MKE}
\begin{gather}	\label{eqn:1Dsys_MKE_DE}
	\pdv{}{t} \ppar*{ \textfrac{1}{2} \sigma_{uu} h U^2 } 
	+ \pdv{}{x} \ppar*{ \textfrac{1}{2} \sigma_{uuu} h U^3 + \textfrac{1}{2} \tilde{\sigma}_{uz\phi} U R g h^2 \Phi \cos\theta }
	= S_E,
\shortintertext{where}\label{eqn:1Dsys_MKE_S}
	\begin{aligned}
		S_E &= \tilde{S}_E - h \mathcal{P} - h \mathcal{B}_E,
	\\
		\tilde{S}_E &= 
		- \frac{h}{\Upsilon} \ppar*{ \sigma_{uuu} \dv{\Upsilon}{x} U + \sigma_{nuu} U_n } \textfrac{1}{2} U^2	\\&\quad
		- \frac{h}{\Upsilon} \ppar*{ \tilde{\sigma}_{uz\phi} \dv{\Upsilon}{x} U + \tilde{\sigma}_{nz\phi} U_n } \textfrac{1}{2} R g h \Phi \cos\theta
		+ \sigma_{u\phi} U R g h \Phi \sin{\theta};
	\end{aligned}
\end{gather}\end{subequations}
conservation of TKE,
\begin{subequations}	\label{eqn:1Dsys_TKE}
\begin{gather}	\label{eqn:1Dsys_TKE_DE}
	\pdv{}{t} \ppar*{ h K }
	+ \pdv{}{x} \ppar*{ \sigma_{uk} h U K }
	= S_K,
\shortintertext{where}	\label{eqn:1Dsys_TKE_S}
	\begin{aligned}
		S_K &= \tilde{S}_K
		+ h \mathcal{P} - h \mathcal{B}_K,
	&
		\tilde{S}_K &=
		- \frac{h}{\Upsilon} \ppar*{ \dv{\Upsilon}{x} \sigma_{uk} U + \sigma_{nk} U_n } K - h \epsilon_T;
	\end{aligned}
\end{gather}\end{subequations}
and conservation of GPE,
\begin{subequations}	\label{eqn:1Dsys_GPE}
\begin{gather}	\label{eqn:1Dsys_GPE_DE}
	\pdv{}{t}\ppar*{\textfrac{1}{2} \sigma_{z\phi} R g h^2 \Phi \cos\theta}
	+ \pdv{}{x} \ppar*{\textfrac{1}{2} \sigma_{uz\phi} U R g h^2 \Phi \cos\theta}
	= S_G,
\shortintertext{where}	\label{eqn:1Dsys_GPE_S}
	\begin{aligned}
		S_G &= 
		\tilde{S}_G
		+ h \mathcal{B}_E + h \mathcal{B}_K,
	\\
		\tilde{S}_G &= 
		- \frac{h}{\Upsilon} \ppar*{ \sigma_{uz\phi} \dv{\Upsilon}{x} U + \sigma_{nz\phi} U_n } \textfrac{1}{2} R g h \Phi \cos\theta 
		- w_s R g h \Phi \ppar*{ \cos\theta }^2.
	\end{aligned}
\end{gather}\end{subequations}
Summing these three equations yields the earlier equation for the conservation of total energy \cref{eqn:1Dsys_egy}, in particular $S_T = S_E + S_K + S_G$. 

The system \cref{eqn:1Dsys_vol,eqn:1Dsys_part,eqn:1Dsys_mom,eqn:1Dsys_egy} may be used to eliminate the time derivatives of the dependent variables $h$, $\Phi$, $U$, and $K$ from \cref{eqn:1Dsys_MKE,eqn:1Dsys_TKE,eqn:1Dsys_GPE}, and we may then deduce expressions for $\mathcal{P}$, $\mathcal{B}_K$, and $\mathcal{B}_E$ in terms of the dependent variables, their spatial derivatives, the boundary effects, and the values and derivatives of the shape factors. These expressions are very large, and will not be presented here. However, for the case of a similarity current in which many of the shape factors are constant, the manipulations become sufficiently compact and we report them below.

Using the governing system, \cref{eqn:1Dsys_vol_DE,eqn:1Dsys_part_DE,eqn:1Dsys_mom_DE,eqn:1Dsys_egy_DE}, we eliminate the time derivatives from \cref{eqn:1Dsys_MKE_DE,eqn:1Dsys_GPE_DE} to yield the balances between the source terms
\begin{subequations}
\allowdisplaybreaks[1]
\begin{gather}
	\begin{multlined}[b][\multlinedwidth] \label{eqn:source_MKE_from_GE}
		S_E =
		- \textfrac{1}{2} \sigma_{uu} U^2 S_h
		+ \sigma_{uu} U S_U
			+ \textfrac{1}{2} \sigma_{uu} U^2 \pdv{hU}{x}
		- \sigma_{uu}^2 U \pdv{hU^2}{x}
	\\
		+ \textfrac{1}{2} \sigma_{uuu} \pdv{hU^3}{x}
		- \textfrac{1}{2} \sigma_{uu} \sigma_{z\phi} U \pdv{R g h^2 \Phi \cos\theta}{x}
		+ \textfrac{1}{2} \tilde{\sigma}_{uz\phi} \pdv{U R g h^2 \Phi \cos\theta}{x},
	\end{multlined}
\\
	\begin{multlined}[b][\multlinedwidth] \label{eqn:source_GPE_from_GE}
		S_G
		= \textfrac{1}{2} \sigma_{z\phi} R g h \cos\theta \ppar*{ \Phi S_h + S_{\Phi} }
		- \textfrac{1}{2} \sigma_{u\phi} \sigma_{z\phi} R g h \cos\theta \pdv{h U \Phi}{x}
	\\
		- \textfrac{1}{2} \sigma_{z\phi} R g h \Phi \cos\theta \pdv{hU}{x}
		+ \textfrac{1}{2} \sigma_{uz\phi} \pdv{U R g h^2 \Phi \cos\theta}{x}.
	\end{multlined}
\end{gather}
\end{subequations}
and $S_K = S_T - S_E - S_G$. Note these expressions do not depend on the form of the source terms, and thus are applicable to models which include other physical effects.

Due to the structure of the source terms $S_E$, $S_K$, and $S_G$, it is not possible to rearrange them into expressions for $\mathcal{P}$, $\mathcal{B}_K$, and $\mathcal{B}_E$. Instead, two of these can be expressed in terms of the remaining one. We choose to express $\mathcal{P}$ and $\mathcal{B}_K$ in terms of $\mathcal{B}_E$, which yields 
\begin{align} \label{eqn:production_rearranged}
	h \mathcal{P} &= - h \mathcal{B}_E - \ppar*{ S_E - \tilde{S}_E },
&
	h \mathcal{B}_K &= - h \mathcal{B}_E + \ppar*{ S_G - \tilde{S}_G },
\end{align}
with \cref{eqn:1Dsys_TKE_S} then stating simply $S_E + S_K + S_G = \tilde{S}_E + \tilde{S}_K + \tilde{S}_G$. The reason for this choice is that $\mathcal{B}_E$ is a property of the mean flow, and can therefore be deduced from the equations describing the mean flow, whereas the others are properties of the turbulence requiring closure. To deduce $\mathcal{B}_E$, we employ the continuity equation
\begin{gather}
	\pdv{\reavg{u}}{x} + \pdv{\reavg{v}}{y} + \pdv{\reavg{w}}{z} = 0,
\qquad \text{thus} \qquad
	\eval*{\reavg{w}}_{z=z_1} = - \int_0^{z_1} \eval*{ \pdv{\reavg{u}}{x} + \pdv{\reavg{v}}{y} }_{z_2} \dd{z_2},
\shortintertext{and}		\label{eqn:production_GPE_from_MKE}
	\begin{aligned}
		\mathcal{B}_E 
		&= - \textfrac{1}{2} R g \Phi \cos{\theta}
		\pgrp[\bigg]{ \tilde{\sigma}_{uz\phi} h \pdv{U}{x} + \pbrk*{\tilde{\sigma}_{uz\phi} - \sigma_{uz\phi}} U \pdv{h}{x} + 2 \sigma_{v_y z\phi}' \frac{h V}{\Upsilon} }.
	\end{aligned}
\end{gather}

Returning to \cref{eqn:production_rearranged}, expressions for $S_E$ and $S_G$ have been deduced from the governing equations in \cref{eqn:source_MKE_from_GE,eqn:source_GPE_from_GE}, and expressions for $\tilde{S}_E$ and $\tilde{S}_G$ are given in \cref{eqn:1Dsys_MKE_S,eqn:1Dsys_GPE_S}, thus
\begin{subequations}
\begin{multline}\label{eqn:production_TKE_from_MKE}
			h \mathcal{P} =
	- \frac{h}{\Upsilon} \ppar*{ \pbrk*{ \sigma_{uu} - 2 \sigma_{uu}^2 + \sigma_{uuu} } \dv{\Upsilon}{x} U + \pbrk*{ \sigma_{uu} - 2 \sigma_{uu} \sigma_{nu} + \sigma_{nuu} } U_n } \textfrac{1}{2} U^2
	\\
		- \frac{h}{\Upsilon} \ppar*{ \tilde{\sigma}_{uz\phi} \dv{\Upsilon}{x} U + \tilde{\sigma}_{nz\phi} U_n } \textfrac{1}{2} R g h \Phi \cos\theta
	\\
		+ \sigma_{uu} U u_\star^2
		+ \textfrac{1}{2} \sigma_{uu} U^2 w_e 
		+ \pbrk*{ \sigma_{u\phi} - \sigma_{uu} } U R g h \Phi \sin{\theta}
	\\
		+ \textfrac{1}{2} \pbrk*{ - \sigma_{uu} + 2 \sigma_{uu}^2 - \sigma_{uuu} } U^3 \pdv{h}{x}
		+ \textfrac{1}{2} \pbrk*{ - \sigma_{uu} + 4 \sigma_{uu}^2 - 3 \sigma_{uuu} } h U^2 \pdv{U}{x}
	\\
		+ \textfrac{1}{2} \pbrk*{ \sigma_{z\phi} \sigma_{uu} - \tilde{\sigma}_{uz\phi} } U R g h^2 \cos\theta \pdv{\Phi}{x}
		+ \pbrk*{ \sigma_{z\phi} \sigma_{uu} - \varsigma_{uz\phi} } U R g h \Phi \cos\theta \pdv{h}{x}
	\\
		+ \sigma_{v_y z\phi}' R g h^2 \Phi \cos{\theta} \frac{V}{\Upsilon}
		,
\end{multline}
\begin{multline}\label{eqn:production_GPE_from_TKE}
		h \mathcal{B}_K = \textfrac{1}{2} R g h \cos\theta \bigg\lgroup
		- \frac{h}{\Upsilon} \ppar*{ \pbrk*{ \sigma_{z\phi} + \sigma_{z\phi} \sigma_{u\phi} - \sigma_{uz\phi} } \dv{\Upsilon}{x} U + \pbrk*{ \sigma_{z\phi} + \sigma_{z\phi} \sigma_{n\phi}  - \sigma_{nz\phi} } U_n } 	\Phi
	\\
		+ \sigma_{z\phi} \Phi w_e 
		+ \pbrk*{ 2 - \sigma_{z\phi} \varsigma_{\phi} } w_s \Phi \cos\theta
		+ \sigma_{z\phi} E_s
	\\
		+ \pbrk*{ - \sigma_{z\phi} - \sigma_{z\phi} \sigma_{u\phi} + 2\varsigma_{uz\phi} } \Phi \pdv{}{x} \ppar*{h U}
		+ \pbrk*{ - \sigma_{z\phi} \sigma_{u\phi} + \sigma_{uz\phi} } U h \pdv{\Phi}{x}
	\\
		+ 2 \sigma_{v_y z\phi}' h \Phi \frac{V}{\Upsilon}
		\bigg\rgroup
		.
\end{multline}
\end{subequations}
To simplify, we consider flow along a deep channel of constant width ($U_n = V = \dv*{\Upsilon}{x} = 0$). Top-hat flows (\cref{tab:shape_factors}) then simplify to
\begin{subequations}\label{eqn:egytransfer_tophat}
\begin{align}
	\label{eqn:egytransfer_tophat_P}
	h \mathcal{P} &= 
	U u_\star^2 
	+ \textfrac{1}{2} U^2 w_e,
\\
	h \mathcal{B}_K &= 
	\textfrac{1}{2} R g h \cos\theta \pgrp*{ 
		\Phi w_e 
		+ \pbrk*{ 2 - \varsigma_\phi } w_s \Phi \cos\theta
		+ E_s
	},
\\
	h \mathcal{B}_E &= 
	- \textfrac{1}{2} R g h \Phi \cos{\theta} \pdv{U}{x},
\end{align}\end{subequations}
the first two matching the expressions from \cite{ar_Parker_1986}, and the third being consistent though never explicitly stated. However, using the values from \cite{ar_Islam_2010} (\cref{tab:shape_factors}) we find that
\begin{subequations}\label{eqn:egytransfer_shape}
\allowdisplaybreaks[1]
\begin{gather}
	\begin{multlined}[b][\multlinedwidth]
		h \mathcal{P} =
		1.50 U u_\star^2
		+ 1.50 \textfrac{1}{2} U^2 w_e 
		- 0.16 U R g h \Phi \sin{\theta}
		+ 0.24 U^3 \pdv{h}{x}
	\\
		- 0.03 h U^2 \pdv{U}{x}
		- 0.06 U R g h^2 \cos\theta \pdv{\Phi}{x}
		+ 0.05 U R g h \Phi \cos\theta \pdv{h}{x}
		,
	\end{multlined}
	\label{eqn:egytransfer_shape_P}
\\
		h \mathcal{B}_K = \textfrac{1}{2} R g h \cos\theta \bigg\lgroup
		  0.61 \Phi w_e
		+ 0.34 w_s \Phi \cos\theta
		+ 0.61 E_s
		+ 0.30 \Phi \pdv{}{x} \ppar*{hU}
		- 0.12 U h \pdv{\Phi}{x}
		\bigg\rgroup
		,
	\label{eqn:egytransfer_shape_BK}
\\
	h \mathcal{B}_E = - \textfrac{1}{2} R g \Phi \cos{\theta} \pgrp[\bigg]{ 1.03 h \pdv{U}{x} + 0.33 U \pdv{h}{x} }.
	\label{eqn:egytransfer_shape_BE}
\end{gather}
\end{subequations}
The difference made by shape factors in \cref{eqn:egytransfer_tophat} is not negligible, modifying the magnitude of terms already existing in \cref{eqn:egytransfer_shape} and introducing new terms. The shear production is boosted primarily by the transverse structure of the velocity field, existing terms increasing by ${\sim} 50\%$. The energy required for turbulent uplift, conversely, decreases, primarily due to the sediment being lower in the flow, existing terms reducing by ${\sim} 40\%$. The largest of the new terms is in \cref{eqn:egytransfer_shape_BK} and has a coefficient of $0.30$, and therefore is around half the size of the existing terms which now have coefficients of $0.61$.
The effect of the transverse structure of the flow cannot be neglected, especially if the evolution of the energy is to be modelled, because doing so removes from the governing equations leading order contributions to the transfer of energy between the stores within the system. 

The additional terms in the expressions can be understood as arising from imbalances in the evolution of the system caused by the imposed transverse structure. 
In \cref{eqn:egytransfer_shape_P}: the third and sixth terms is the difference between the work done directly on the MKE and the work done indirectly by exerting force on the momentum, the third representing the downslope component of gravity and the sixth the pressure force; the fourth and fifth are from the disparity of the downslope advection of MKE and momentum, the fifth also including the loss of MKE during transverse uplift of mass. 
In \cref{eqn:egytransfer_shape_BK}, the fourth, fifth, and sixth terms arise from an imbalance between four effects: the direct advection of GPE, its advection by the motion of particles and by the motion of volume, and the change in GPE due to the uplift of volume as velocity changes.

\Citet{ar_Skevington_F006_TCFlowPower} criticized the top-hat model, due to insufficient turbulent production $h \mathcal{P}$ to meet the energy requirements of suspending the sediment $h \mathcal{B}_K$. They found that $h \mathcal{P} < h \mathcal{B}_K + h \epsilon_T$ for a range of erosional ($E_s >  \varsigma_\phi w_s \Phi \cos\theta$) dilute ($\Phi<10^{-2}$) laboratory and environmental flows, violating the Knapp-Bagnold criterion \citep{ar_Knapp_1938,ar_Bagnold_1962}. \Citet{ar_Skevington_F006_TCFlowPower} argued that the transverse structure of real turbidity-currents resulted in a higher shear production and lower buoyancy production than the top-hat model, which is demonstrated in \cref{eqn:egytransfer_shape}. Further, by substituting the conditions of pseudo-equilibrium flow \cref{eqn:quasiequ_defn_simp,eqn:quasiequ_vol,eqn:quasiequ_part,eqn:quasiequ_mom}  into \cref{eqn:production_TKE_from_MKE,eqn:production_GPE_from_TKE} we obtain
\begin{subequations}
\begin{align}
	h \mathcal{P} &=
	\sigma_{u\phi} U u_\star^2
	+ \pbrk*{ 2 \sigma_{uu} \sigma_{u\phi} - \sigma_{uuu} } \textfrac{1}{2} U^2 w_e
	+ \pbrk*{ \sigma_{z\phi} \sigma_{u\phi} - \sigma_{uz\phi} } \textfrac{1}{2} R g h \Phi \cos\theta w_e
\notag\\ \label{eqn:TKE_production_quasieq}
	&= 
	1.34 U u_\star^2
	+ 1.50 \textfrac{1}{2} U^2 w_e
	+ 0.12 \textfrac{1}{2} R g h \Phi \cos\theta w_e
\displaybreak[1]\\
	h \mathcal{B}_K &= 
	\textfrac{1}{2} R g h \cos\theta \pgrp[\big]{
		\tilde{\sigma}_{uz\phi} \Phi w_e
		+ 2 w_s \Phi \cos\theta
	}
\notag\\
	&=
	\textfrac{1}{2} R g h \cos\theta \pgrp[\big]{
		1.03 \Phi w_e
		+ 2 w_s \Phi \cos\theta
	}.
\end{align}
\end{subequations}
Here, the numerical values for the shape factors are from \cite{ar_Islam_2010} (\cref{tab:shape_factors}). For this specific flow configuration, the turbulent production for the bed shear is boosted relative to the top-hat production \cref{eqn:egytransfer_tophat_P}. The buoyancy production is essentially the same as in the top-hat model, the slight increase more than made up for by the final term in \cref{eqn:TKE_production_quasieq}.

Returning to the full expressions \cref{eqn:production_GPE_from_MKE,eqn:production_TKE_from_MKE,eqn:production_GPE_from_TKE}, we see that for a current in a channel of varying width, or undergoing overspill, there is additional transfer of energy. These largely arise from imbalances as the previous extra terms did, however there are some terms which do not have vanishing coefficients even for top-hat flow. These terms are all negative, implying lower TKE production and buoyancy production than for flows not undergoing overspill. An alternative interpretation is that the values of $\mathcal{P}$ and $\mathcal{B}_K$ are the same in both confined and partially-confined currents, and other terms in the expressions change to compensate from the negative contribution from the levee effects, \eg the rate of entrainment increases. We do not pursue this alternative interpretation here.
\section{Hyperbolicity and characteristics} \label{sec:time}

The structure of the system \cref{eqn:1Dsys_vol,eqn:1Dsys_part,eqn:1Dsys_mom,eqn:1Dsys_egy} is that of a \emph{balance law} (distinct from a \emph{conservation law} because there are non-zero source terms). It may be manipulated and rewritten in the form (at least, for regions where the solution is continuous, away from hydraulic jumps)
\begin{align} \label{eqn:balancelaw_matrixform}
	\pdv{Q}{t} + A_F \pdv{Q}{x} &= A_S S,
&&\text{where}&
	Q &= \pgrp*{\begin{array}{c} h \\ \Phi \\ U \\ K \end{array}},
&
	S &= \pgrp*{\begin{array}{c} S_h \\ S_\Phi \\ S_U \\ S_T \end{array}},
\end{align}
and $A_F$ and $A_S$ are both $4 \times 4$ real valued matrices. For the case where the eigenvalues $\lambda$ of $A_F$ are real the system is hyperbolic, initial value problems are well posed, and information is transported along characteristic trajectories with speed $\dv*{x}{t} = \lambda$. (See \eg \cite{bk_Evans_PDE} for more details.)


For the non-linear system we consider, similarity flow (constant shape factors) results in a hyperbolic system. Indeed, the characteristic polynomial of $A_F$ is
\begin{subequations} \label{eqn:flux_char_poly}
\begin{gather}
	P_{F4}(\lambda) = ( \lambda - \sigma_{uk} U ) \cdot P_{F3}(\lambda)
\end{gather}
where
\begin{multline}
	P_{F3}(\lambda) =
	\lambda^3
	- \pbrk*{ 2\sigma_{uu} + \sigma_{u\phi} } U \lambda^2
	+ \pbrk*{ 2\sigma_{u\phi} + 1 } \sigma_{uu} U^2 \lambda
	- \textfrac{1}{2} \pbrk*{ \sigma_{u\phi} + 1 } G^2 \lambda
	- \sigma_{u\phi}\sigma_{uu} U^3
	+ \sigma_{u\phi} U G^2
\end{multline}
is the characteristic polynomial for the three equations system \cref{eqn:1Dsys_vol,eqn:1Dsys_part,eqn:1Dsys_mom}, and
\begin{equation}
	G = \sqrt{ \sigma_{z\phi} R g h \Phi \cos\theta } > 0.
\end{equation}
\end{subequations}
This structure arises due to the fluxes in the three equation system being independent of $K$, and we can immediately identify one eigenvalue $\lambda_K = \sigma_{uk} U$ associated with the transport of TKE, or energy more broadly. For the other eigenvalues, the exact expressions are unwieldy. Instead bounds can be established by looking for changes in sign of $P_{F3}$. Firstly,
\begin{subequations}
\begin{align}
	P_{F3}(U) &= \textfrac{1}{2} (\sigma_{u\phi} U - U) (2 [\sigma_{uu}-1] U^2 + G^2),
\\
	P_{F3}(\sigma_{u\phi} U) &= - \textfrac{1}{2}(\sigma_{u\phi} U - U) \sigma_{u\phi} G^2,
\end{align}
\end{subequations}
thus $\lambda_{\Phi}$ is between $U$ and $\sigma_{u\phi} U$ (using that $\sigma_{uu} \geq 1$). We identify this characteristic with the concentration field, $\Phi$, because for $\sigma_{u\phi}=1$ the corresponding characteristic equation is precisely the advection of concentration at speed $U$. For $\sigma_{u\phi} \neq 1$, the characteristic travels at a speed between that of the particles, $\sigma_{u\phi} U$, and the fluid, $U$. 

To identify the final two roots we use that:
\begin{itemize}
	\item If $\sigma_{u\phi} U < U$ then $P_{F3}(\sigma_{u\phi} U)>0$, and $P_{F3}(U)<0$. Thus by the asymptotic behaviour of $P_{F3}$ as $\lambda \to \pm \infty$ the roots satisfy $\lambda_- < \sigma_{u\phi} U$ and $\lambda_+ > U$.
	\item If $U < \sigma_{u\phi} U$ then $P_{F3}(U)>0$, and $P_{F3}(\sigma_{u\phi} U)<0$. Thus by the asymptotic behaviour of $P_{F3}$ as $\lambda \to \pm \infty$ the roots satisfy $\lambda_- < U$ and $\lambda_+ > \sigma_{u\phi} U$.
	\item If $U=0$ then $\lambda_{\pm} = \pm \sqrt{\textfrac{1}{2}(\sigma_{u\phi}+1)} G$, and $\lambda_-<0<\lambda_+$
	\item If $\sigma_{u\phi}=1$ then $\lambda_{\pm} = \sigma_{uu} U \pm \sqrt{ \sigma_{uu} (\sigma_{uu}-1) U^2 + G^2 }$, and $\lambda_-<U<\lambda_+$.
\end{itemize}

Further simplifying the last case by imposing $\sigma_{uu}=1$ we obtain $\lambda_{\pm} = U \pm G$, which are exactly the characteristic speeds of the classic shallow water equations \citep{bk_Stoker_WW}. Thus we identify $\lambda_{\pm}$ as waves of pressure and momentum.

An alternative definition of critical flow exists to the one used in \cref{sec:model_param}: that critical flow is when any of the characteristics is stationary. We investigate this by solving $P_{F4}(0)=0$, where
\begin{subequations}\begin{gather}
	P_{F4}(0) = \sigma_{uk} \sigma_{u\phi} U^2 \ppar*{ \sigma_{uu} U^2 - G^2 },
\\
	\begin{aligned}
	\text{thus}&&
		U&=0
	&\text{or}&&
		\Fro &\equiv \sqrt{\sigma_{uu}}\frac*{U}{G} = \pm 1.
	\end{aligned}
\end{gather}\end{subequations}
The condition $\Fro=\pm 1$ corresponds to $\lambda_\mp=0$, thus these are when waves of pressure and momentum are stationary, which corresponds to energy minimisation by the arguments in \cref{sec:model_param}.

Considering briefly hydraulic jumps, it is a well established result that the characteristics of one family must converge on a `shock' in a hyperbolic system \citep{ar_Lax_1957}, and consequently for a stationary jump at $x=x_J$ with $U>0$ we must have supercritical flow upstream ($\Fro>1$ on $x<x_J$) and subcritical flow downstream ($0<\Fro<1$ on $x>x_J$). By \cref{eqn:dimless_param_integral}, $\text{MKE}>\text{GPE}$ upstream and $\text{MKE}<\text{GPE}$ downstream, which is precisely the criterion established by \citet{ar_Thorpe_2010} for a jump to exist, though their reasoning is different. As discussed by \citet{ar_Thorpe_2010} and \citet{ar_Thorpe_2014}, the transverse structure changes across a jump in a physical system, and consequently the shape factors are different upstream and downstream. Additionally, even if the upstream flow is supercritical, internal waves may be able to travel upstream which could cause the shock to decay over time. Thus, the \emph{energetic Froude number} established here serves as a diagnostic tool for the existence of a shock, but not necessarily for its persistence long-term.

\section{Closure of of the turbulent and particulate processes} \label{sec:turbclose}


To build a full model from the general framework presented in \cref{sec:model}, closures are required for the turbulent processes, particulate processes, and transverse structure, which are the subject of the next three sections. This section tackles the turbulent processes of drag ($u_*^2$) and entrainment ($w_e$), and particulate processes of settling ($w_s$) and erosion ($E_s$). We also introduce the scaling required for turbulent dissipation ($\epsilon_T$).

\subsection{Drag} \label{sec:drag_closure}

We begin with a discussion of closures for drag; a variety of closures have been used in the context of both gravity currents and open channel hydraulics. These all fall into the general family
\begin{equation} \label{eqn:drag_family}
	u_*^2 = C_d U^{2p_d} K^{1-p_d}
\end{equation}
where $0 \leq p_d \leq 1$ and $\tilde{C}_d$ is a drag coefficient. Perhaps the most common is $p_d=1$ so that $u_*^2 \propto U^2$. This expression works well in flows where the TKE has reached a balance with the MKE so that $K \propto U^2$ (\ie $\gamma$ is a constant). However, for flows far from a statistical equilibrium, this closure is inappropriate because it does not account for the strength of turbulence in the turbulent drag.

An alternative closure was used by \cite{ar_Parker_1986}, who used $p_d = 0$ so that $u_*^2 \propto K$. However, in this closure, a stationary region of turbulent fluid ($U=0$, $K>0$) experiences a drag force. This indicates that this model, again, is only suitable for flows where the energy has reached a balance.


We desire a model that can be used for currents that have been driven away from proportionality, $K \not\propto U^2$. For this we use a mixing length model for the Reynolds stress, that is
\begin{align}
	\tau_{ij}^R 
	\eqdef - \reavg{\reflc{u_i}\reflc{u_j}} 
	= - \textfrac{2}{3} k \delta_{ij} + \mathscr{l} \sqrt{k} \ppar*{ \pdv{\reavg{u_i}}{x_j} + \pdv{\reavg{u_j}}{x_i} }
\end{align}
where $\mathscr{l}$ is the mixing length, and we write
\begin{align}
	\mathscr{l}(x,y,z,t) 		&= \xi_{\mathscr{l}}(x,\mathscr{y},\mathscr{z},t) \cdot h(x,t).
\end{align}
Note that $\xi_{\mathscr{l}}$ is not normalised as \cref{eqn:profile_normalisation}. Employing the boundary condition $\reavg{w} = 0$ at $z=0$, we may express the leading order basal drag as
\begin{subequations}\begin{gather} \label{eqn:turbclose_drag_mixlength}
	u_*^2 
	\eqdef \eval[\bigg]{ \tau_{31}^R }_{z=0} 
	= \eval[\bigg]{ \mathscr{l} \sqrt{k} \pdv{\reavg{u}}{z}  }_{z=0}
	= \eval[\bigg]{ \xi_{\mathscr{l}} \sqrt{\xi_k} \pdv{\xi_u}{\mathscr{z}} }_{\mathscr{z} = 0} \cdot \sqrt{K} U
	= C_d \sqrt{K} U,
\shortintertext{where}
	\begin{aligned}
		C_d &\eqdef C_\nu \varsigma_{u'} \eval[\bigg]{\xi_\nu}_{\mathscr{z} = 0},
	&
		\xi_\nu &\eqdef C_\nu^{-1} \xi_{\mathscr{l}} \sqrt{\xi_k},
	&\text{and}&&
		C_\nu &\eqdef \int_0^\infty \xi_{\mathscr{l}} \sqrt{\xi_k} \dd{\mathscr{z}}.
	\end{aligned}
\end{gather}\end{subequations}
Thus our model is of the class \cref{eqn:drag_family}, with $p_d=1/2$ and a non-constant drag coefficient.
The evaluation at `$z=0$' should be physically understood as being just above the viscous boundary layer, which is neglected from our model. The closure \cref{eqn:turbclose_drag_mixlength} satisfies our requirements: it responds appropriately to varying TKE and velocity, in that the drag is zero when either is zero, and is a monotonically increasing function of each. To determine the coefficient $C_\nu$ we will assume that $\xi_\nu$ takes on a top-hat profile \citep{ar_Claudin_2011,ar_Amy_2022}
\begin{align}
	\xi_\nu &=
	\begin{cases}
		1,								&	 -1 < \mathscr{y} < 1 \text{ and } 0 \leq \mathscr{z} \leq 1,		\\
		0, 								& 	\mathscr{z} > 1,
	\end{cases}
\end{align}
and use that that $K = \sigma_{uu} U^2 / 2 \gamma^2$. Using a model of equilibrium currents where $u_*^2 = \tilde{C}_d U^2$, and comparing the two models at equilibrium (subscript $\equl$), yields an expression for $C_\nu$. Explicitly
\begin{align}
	u_*^2 &
	= \frac{C_\nu U^2}{2^{1/2}} \cdot \ppar*{ \frac{\sigma_{uu}^{1/2} \varsigma_{u'}}{\gamma}  }_{\equl}
	= \tilde{C_d} U^2, 
&&\text{thus}&
	C_\nu &= \frac**{2^{1/2} \tilde{C_d} }{ \ppar*{ \frac{\sigma_{uu}^{1/2} \varsigma_{u'}}{\gamma}  }_{\equl} }.
	\label{eqn:eddy_viscosity_coefficient}
\end{align}

\subsection{Entrainment}


Entrainment is substantially harder to model than drag, due to it being a combination of several mechanisms which are spatially distributed. One mechanism is shear instability, \eg Kelvin-Helmholtz \citep[see][]{ar_Peltier_2003}, which indicates that the entrainment rate is proportional to the velocity of the flow, $w_e \propto \abs{U}$, with the coefficient a function of the gradient Richardson number (\ie dependent on $\Fro$ and the transverse structure of $\reavg{\phi}$ and $\reavg{u}$). Another mechanism is simply turbulent mixing: even for a flow that is stationary, $U=0$, if there is TKE available, $K>0$, then this will drive mixing with the ambient fluid above resulting in entrainment so that $w_e \propto \sqrt{K}$ (or $w_e \propto \abs{U}/\gamma$). Finally, for particle laden flows it is possible for settling to cause the ambient to detrain, provided the turbulence in the upper reaches of the current is weak (an effect governed by the strength of the settling velocity, \ie $\beta$, as used by \cite{ar_Pittaluga_2018} and \cite{ar_Traer_2018_p1}). In practice, settling is expected to suppress the other two mechanisms in some manner.

In this work, we capture only the first of these physical mechanisms using the model developed by \cite{ar_Fukushima_1985} and verified by \cite{ar_Parker_1987}, where
\begin{align}
	w_e &= C_e \abs{U},
&
	C_e &= \frac{0.00153}{0.0204 + \ElTu{\Fro}^{-2}},
&&\text{and}&
	\ElTu{\Fro} &= \frac{\sigma_{uu} \sigma_{z\phi}^{1/2}}{\sigma_{u\phi}^{1/2}} \Fro,
\end{align}
see \cref{app:ETshape}. However, we identify the modelling of entrainment as an area requiring further attention, in particular how $w_e$ depends on the transverse structure of the upper portion of the current, the strength of the TKE, and the settling velocity. As pointed out by \citet{ar_Caulfield_2021}, the mixing the occurs in stratified flows may depend in a non-monotonic fashion on a large number of parameters, which the simple closures developed thus far for gravity currents do not capture.

\subsection{Turbulent dissipation}


The turbulent energy that cascades down to the smaller scales is consumed by two processes: dissipation by viscosity, and turbulent uplift of particles. We reason that the energy lost to the turbulent uplift does not impact the rate of dissipation to viscosity, and that when both processes are present they combine to dampen the turbulence. Thus, using that for flows without a particle load the dissipation scales as $K^{3/2}$, have
\begin{align}	\label{eqn:turbclose_dissipation}
	h \epsilon_T = C_{\epsilon} K^{3/2}.
\end{align}
An expression for $C_\epsilon$ can be derived in reference to the pseudo-equilibrium conditions from \cref{sec:quasiequ}. We assume that, even for currents that are not in a pseudo-equilibrium, the small-scale turbulent structures are equivalent to those present in a pseudo-equilibrium balance. Specifically, one where the conservation of volume, momentum, and energy are in this balance, \ie $\gamma = \gamma_\equl$.
We do not expect the turbulent dissipation to depend directly on the downslope gravity (or the effect of the levees), and consequently we eliminate $\sin\theta$ between \cref{eqn:quasiequ_mom,eqn:quasiequ_egy} to obtain an expression for $C_\epsilon$. 
Moreover, following \citet{ar_Parker_1986}, we make the sweeping assumption that the turbulent dissipation within a compositional current ($w_s = E_s = 0$) is the same as in a particle driven current provided all other properties are the same. 
By this we obtain
\begin{multline}
	C_{\epsilon}
	=
	\ppar*{
		\pbrk*{ 2 \sigma_{u\phi} - \frac{\sigma_{uuu}}{\sigma_{uu}} } \gamma_{\equl}^2
		- \sigma_{uk}
		+ \pbrk*{\sigma_{u\phi} - \frac{2 \varsigma_{uz\phi} }{ \sigma_{z\phi} }} \frac{\gamma_{\equl}^2}{\Fro^2}
	} \sqrt{\frac{2}{\sigma_{uu}}} \gamma_{\equl}  C_e
	+ \sigma_{u\phi} \ppar*{\frac{2 }{\sigma_{uu}} \gamma_{\equl}^2}^{p_d+(1/2)} C_d.
\end{multline}
Employing the assumption of top-hat flow and setting $p_d=0$, we recover the dissipation closure used by \cite{ar_Parker_1986}. Thus the expression used here may be considered a generalisation. Herein $p_d=1/2$, and we must establish the shape factors and the equilibrium value of $\gamma_{\equl}$, which in general need not be a constant.

\subsection{Settling and erosion}


The settling velocity is calculated using \citep{bk_Soulsby_DMS}
\begin{gather}
	\begin{aligned}
		w_s &= \frac{\nu}{d} \ppar*{ \sqrt{10.36^2+1.049 \times d_s^3}-10.36 },
	&&\text{where}&
		d_s &= d \ppar*{ \frac{Rg}{\nu^2} }^{1/3}
	\end{aligned}
\intertext{is the the dimensionless grain size, $\nu = 10^{-6} \, \textrm{m}^2 \, \textrm{s}^{-1}$ is the viscosity of water, and $d$ is the particle diameter. For erosion we use \citep{ar_Dorrell_2018}}
	E_s = 4.90 \times \frac{ \max\pbrk{ u_{\star}^2 - u_{\star c}^2 , 0 }^{3/2} }{R g h}		\label{eqn:errosion}
\intertext{where the critical shear stress is given by \citep{bk_Soulsby_DMS}}
	u_{\star c}^2 = R g d \ppar*{ \frac{0.3}{1 + 1.2 \times d_s} + 0.055 \times \ppar*{ 1-\exp\ppar*{-0.02 \times d_s} } }.
\end{gather}
However, as is typical for erosion models, the theoretical justification of \cref{eqn:errosion} is only valid for fluvial systems, and its application to turbidity currents is questionable. In the long term, because erosion is a local process, it should be possible to quantify it purely in terms of the local dynamics independent of the type of flow considered.

\section{Transverse structure} \label{sec:vert}

The system of equations \cref{eqn:1Dsys_vol,eqn:1Dsys_part,eqn:1Dsys_mom,eqn:1Dsys_egy} has been derived carefully to account for the possibility that the shape factors (\cref{tab:shape_factors}) may be functions of $x$ and $t$. In establishing the evolving transverse structure, there is a similar situation to the derivation of the Reynolds Averaged Navier Stokes (RANS) equations. The RANS system includes not only the mean quantities, but also higher order moments containing the correlations of fluctuations, such as the TKE, $k = \textfrac{1}{2} \reavg{\reflc{u_i}\reflc{u_i}}$. An explicit evolution equation can be derived for the TKE, but this includes higher order correlations of three fluctuations $\reavg{\reflc{u_j}\reflc{u_i}\reflc{u_i}}$, and so on, the evolution equation for a $n$ component correlation containing an $n+1$ component correlation \citep[see, \eg,][]{bk_Davidson_TISE}. Thus the system of equations can never be closed by simply including more equations, and at some point a closure approximation must be made. Similarly, the evolution equation for the channel average of $\reavg{u}$ depends on the channel average of $\reavg{u}\vphantom{u}^2$, \cref{eqn:1Dsys_mom}, the evolution of which depends on the channel average of $\reavg{u}\vphantom{u}^3$, \cref{eqn:1Dsys_MKE}. In general, the evolution of the channel average of  $\reavg{u}\vphantom{u}^n$ will depend on the channel average of $\reavg{u}\vphantom{u}^{n+1}$, and thus at some order we must introduce a closure hypothesis. 

An alternative approach known as `slow manifold theory' has been developed \citep{bk_Roberts_MEDCS} and applied to open channel flow \citep{ar_Roberts_2002,ar_Cao_2016}, as well as boundary flows under an (effectively) infinite ambient \citep{ar_Suslov_1999}. These successes suggest that the technique may yield valuable insight into the transverse structure of turbidity current (indeed we will obtain a slow manifold in \cref{sec:phase}) though this would be a significant undertaking beyond the scope of the present work.


In the model presented here, we impose closure by having no further evolution equations, and instead approximating the shape factors as functions of the bulk averaged quantities already considered. Note, however, that this is still substantially more sophisticated that setting them all to unity, or to some constant value measured from experiment (see \cref{tab:shape_factors} for example values). To derive expressions for the structure we will assume that the current is in a local equilibrium, \ie the transverse structure evolves rapidly compared to the time scales of the overall flow. This may not be the case for flows where the particles settle out very slowly; the time scale of the settling may be on a similar timescale to that of the overall run-out of the current, but for the example of flows in the Congo system (\cref{sec:example}) it is certainly the case.

\subsection{The transverse structure of the bulk flow} \label{sec:vert_bulk}


Various models exist for the transverse structure of turbidity currents, e.g. the model by \cite{ar_Abad_2011}, employed by \cite{ar_Dorrell_2014}, was dependent on the Froude number as in the experiments of \cite{ar_Sequerios_2010}. However, it is clear that the turbulence (measured by TKE) is what uplifts the particles against selling, and thus the transverse structure should be a function of $K$ and $w_s$, \ie a function of $\beta$. This is important, because the fact that the transverse structure depends on $K$ presents mathematical problems for the time dependent system. Here we consider this dependence, to give insight both into the transverse structure of these currents, and the mathematical issues this presents.

To derive the transverse structure, we will neglect the lateral effects of varying channel width and levee overspill so that we effectively consider a 2-dimensional current. For the structure of the particles we follow \cite{ar_Rouse_1937} and balance the upward turbulent diffusion against settling
\begin{gather}
	\pdv{}{z} \ppar*{\reavg{\reflc{w}\reflc{\phi}}} = w_s \cos(\theta) \pdv{\reavg{\phi}}{z},
\intertext{where we use a mixing length model}
	\label{eqn:mixing_length_phi}
	- \reavg{\reflc{w}\reflc{\phi}} = D \xi_\nu \cdot h \sqrt{K} \pdv{\reavg{\phi}}{z}
\end{gather}
with $D \eqdef \Sch_\phi C_\nu$ and $\Sch_\phi$ the turbulent particle Schmidt number. From \cref{eqn:dimless_param,eqn:mixing_length_phi} we can immediately deduce
\begin{equation} \label{eqn:simplified_beta}
\begin{gathered}
	\mathcal{B}_K 
	= D \sqrt{K} \, Rg\Phi \cos\theta \int_0^{\mathrlap{\infty}} - \xi_\nu \pdv{\xi_\phi}{\mathscr{z}} \dd{\mathscr{z}},
\\
	\begin{aligned}
	&\text{thus}&
		\beta 
		&= \hat{\beta} \frac{ \varsigma_\phi \sigma_{z\phi} }{ 2 \int_0^{\mathrlap{\infty}} - \xi_\nu \pdv{\xi_\phi}{\mathscr{z}} \dd{\mathscr{z}} }
	&&\text{where}&
		\hat{\beta} &\eqdef \frac{w_s \cos\theta}{D \sqrt{K}}.
	\end{aligned}
\end{gathered}\end{equation}
Using that $\reavg{\phi} \to 0$ as $z \to \infty$,
\begin{align}
	\xi_\nu \dv{\xi_\phi}{\mathscr{z}} + \hat{\beta} \xi_\phi &= 0,
&&\text{thus}&
	\xi_\phi &= \xi_{\phi 0} \exp \ppar*{ - \int \frac{\hat{\beta}}{\xi_\nu} \dd{\mathscr{z}} }
\end{align}
where $\xi_{\phi 0}$ is set by normalisation \cref{eqn:profile_normalisation}.
For the simple case of $\xi_\nu = 1$ on $0 < \mathscr{z} < 1$ and $\xi_\nu \to 0^+$ on $\mathscr{z}>1$, as proposed by \citet{ar_Amy_2022} for turbidity currents, we get
\begin{align}
	\xi_\phi &= \xi_{\phi 0} \cdot
	\begin{cases}
		\exp \ppar*{ - \hat{\beta} \mathscr{z} }		&	\text{for } 0 < \mathscr{z} < 1,	\\
		0										&	\text{for } \mathscr{z}>1,
	\end{cases}
&&\text{and}&
	\beta &= \textfrac{1}{2} \sigma_{z\phi} \hat{\beta},
\end{align}
which has been evaluated under the assumption that, at the depth where $\xi_\nu$ vanishes, $\xi_\phi = 0$. Note that the expression for $\mathcal{B}_K$ in \cref{eqn:simplified_beta} is not consistent with the subdivided energy equations from \cref{sec:production}, and has been derived instead from simplified assumptions about the dynamics. Going forward, we will use $\hat{\beta}$ and avoid discussion of $\beta$ due to the difficulty in evaluating $\mathcal{B}_K$.

\begin{figure}
	\centering
	\includegraphics{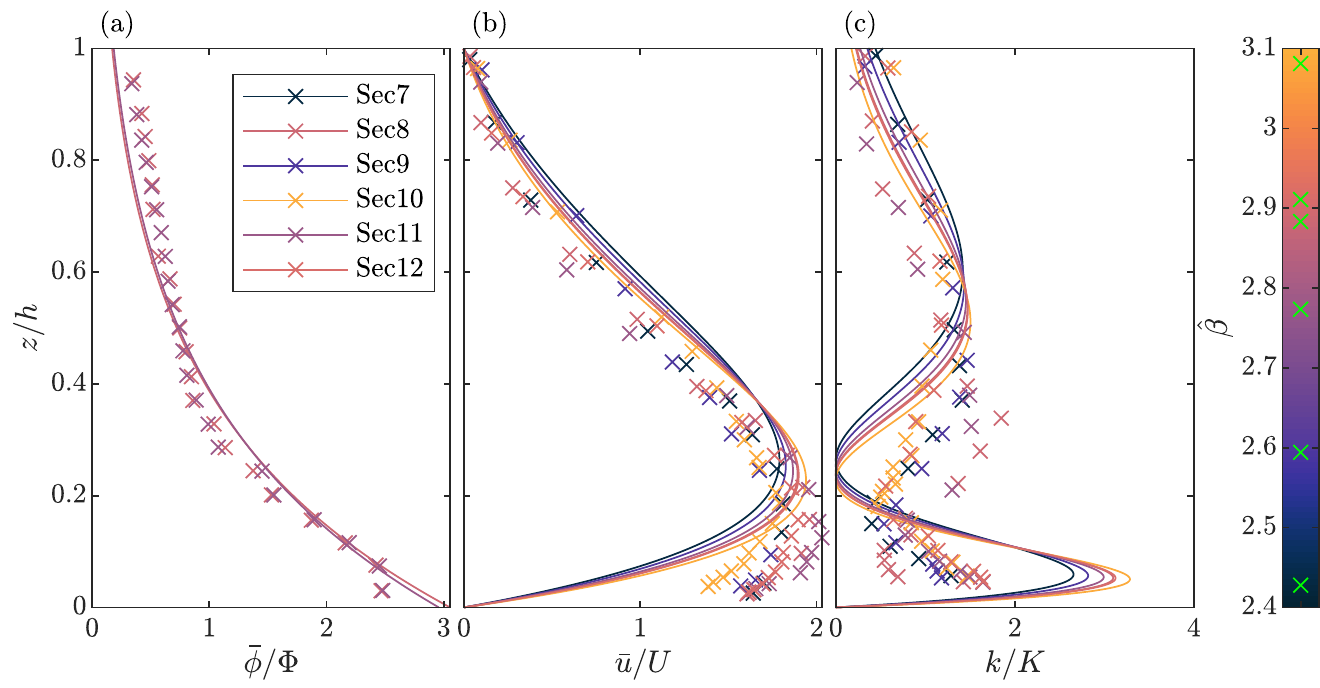}
	\caption{Comparison of the transverse structure in our model (curves) with the experimental data of \cite{ar_Islam_2010} (crosses), each experiment having its own modelled transverse structure due to the dependence on $\hat{\beta}$. The data used is that from their particle laden experiments on a flat bed, with the sections marked in the legend in table $1$ using the same notation as they do.}
	\label{fig:ComparisonExperiment}
\end{figure}

For velocity, we reason that the current is driven by the presence of particles, and so the velocity should be strongest in regions where there is a larger quantity of particles, that is $\xi_u \propto \xi_\phi$. However, we also know that the velocity goes to zero for $\mathscr{z} \in \pbrc{0,1}$. We capture these features by setting
\begin{align} \label{eqn:structure_velocity}
	\xi_u = \xi_{u0} \cdot \mathscr{z} (1-\mathscr{z}) \xi_\phi
\end{align}
where $\xi_{u 0}$ is set by normalisation \cref{eqn:profile_normalisation}. This structure is compared to experimental data in \cref{fig:ComparisonExperiment}, showing good agreement. The maxima of our structure is slightly higher (larger $z$) than measured, which could be improved by including a more sophisticated function that a parabola in \cref{eqn:structure_velocity}, however the function proposed here is sufficiently accurate to produce appropriate shape factors.


Finally, for TKE, the local production is
\begin{align}
	\tau_{31}^R \pdv{u}{z} = \frac{C_\nu \sqrt{K} U^2}{h} \xi_\nu \ppar*{ \pdv{\xi_u}{\mathscr{z}} }^2.
\end{align}
The transverse structure of the TKE is determined by the combined effects of production, diffusion, and dissipation. Here we assume that the structure of TKE is dominated by production, up to the fact that $k=0$ at $z=0$ which we impose similarly to \cref{eqn:structure_velocity}, that is 
\begin{align}
	\xi_k &= \xi_{k0} \cdot \mathscr{z} \ppar*{ \pdv{\xi_u}{\mathscr{z}} }^2 ,
\end{align}
where $\xi_{k 0}$ is set by normalisation \cref{eqn:profile_normalisation}. In this case the transverse structure approximates the experimental data less well, \cref{fig:ComparisonExperiment}(c), though by inspection it appears that the primary difference between our structure and that measured by experiment can be explained by the slight offset in velocity maximum, and a lack of diffusion which would smooth the profile out. We find that the structure presented here is good enough to produce physically interesting result, however we identify the transverse structure of TKE in gravity currents as an area in need of further research.

\begin{figure}
	\centering
	\includegraphics{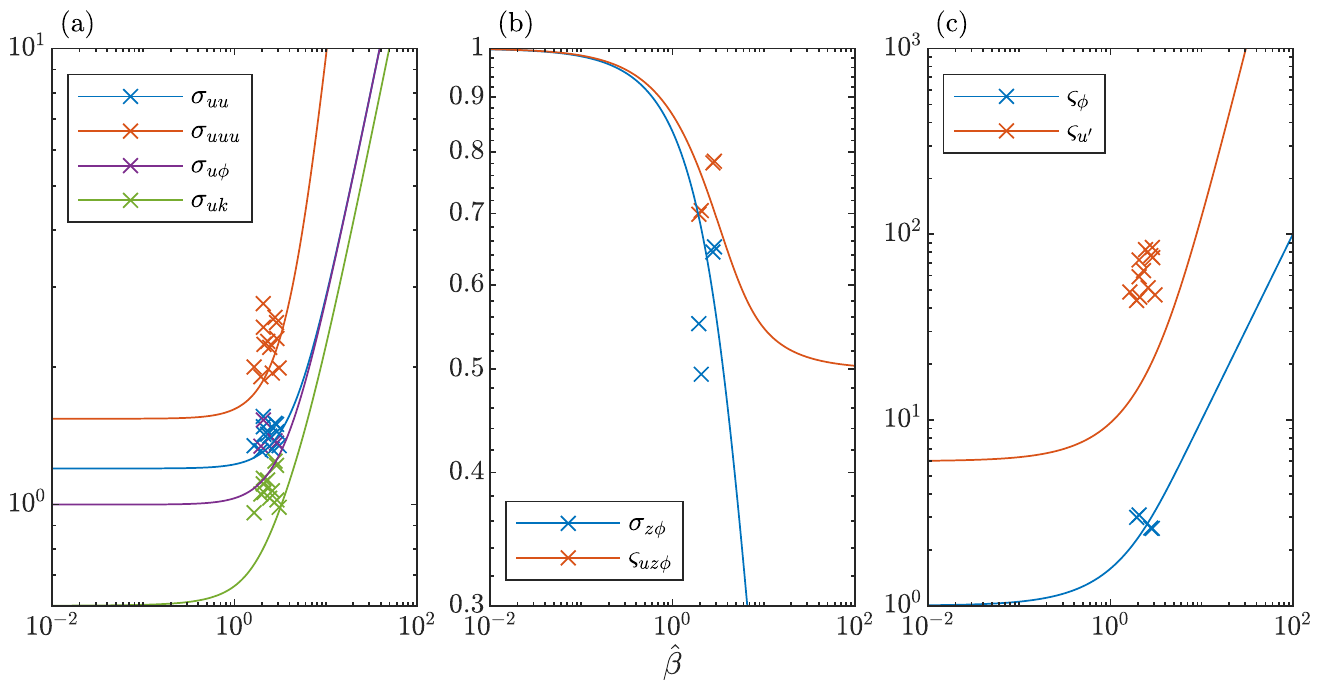}
	\caption{Comparison of the shape factors in our model (curves) with the experimental data of \cite{ar_Islam_2010} (crosses). The data used is that from their particle laden experiments, both for flat beds and sloped bed (Sec 1-12).}
	\label{fig:ShapeFactors}
\end{figure}

To calibrate our model, we compare to the experimental results of \cite{ar_Islam_2010}, who performed experiments with particles of diameter and density
\begin{align}
	d &= \qty{25e-6}{m}
&&\text{and}&
	\rho_p &= \qty{2.6e3}{kg.m^{-3}}
\end{align}
respectively. To scale their experimental data appropriately we first calculate the depth of the current as the location at which $\reavg{u} = 0$, which we extract from the data points by performing a linear fit through the top three velocity data-points. For $\reavg{u}$ we augment the data with $\reavg{u}=0$ at $z\in\pbrc*{0,h}$, while for $\reavg{\phi}$ and $k$ we recover values at these points (if the data does not extend to them) by again using a linear fit through the closest three points. The depth averages are then computed by integrating a piecewise cubic interpolation of the data, which allows us to appropriately scale the data for comparison, and also compute the shape factors (\cref{fig:ShapeFactors}). By setting 
\begin{align}
	\gamma_{\equl} &= 1,
&
	\Sch_\phi &= 0.7,
&&\text{and}&
	\tilde{C}_d &= 25^{-2},
\end{align}
and evaluating the equilibrium shape factors at $\hat{\beta}=0$, \cref{eqn:eddy_viscosity_coefficient}, we recover profiles similar to those measured.

In \cref{fig:ShapeFactors} we plot the shape factors as functions of $\hat{\beta}$, along with data for particle laden currents recomputed from the data from \cite{ar_Islam_2010}. The data shows a good correspondence to our model for the values of $\hat{\beta}$ considered, and it is physics based which also gives us confidence. The only notable problem is in $\varsigma_{u'} = \eval{\dv*{\xi_u}{\mathscr{z}}}_{\mathscr{z}=0}$, however due to the way this is used in the drag law it is only the ratio $\varsigma_{u'}/(\varsigma_{u'})_\equl$ that matters, and thus it may be out by a constant factor without consequence. Asymptotic expansions of the shape factors are given in \cref{tab:shape_asymptotic}, elucidating the behaviour in extreme regimes. 

\Cref{fig:ShapeFactors}(a) shows that the cross correlations $\sigma_{uu}$, $\sigma_{u\phi}$, and $\sigma_{uk}$ can be greater than $1$, and for large $\hat{\beta}$ are significantly greater. This means that, at high stratification, the momentum, particle volume, and turbulence are advected along at a much greater speed than the fluid volume, which is not captured by top-hat models. In a steady state system, the effect of source terms is weakened: for example \cref{eqn:1Dsys_part_DE} is equivalent to
\begin{align}
	\dv{}{x} \ppar*{h U \Phi} = \frac{1}{\sigma_{u\phi}} \ppar*{ S_{\Phi}  - h U \Phi \dv{\sigma_{u\phi}}{x} } .
\end{align}
Thus the effect of $S_\Phi$ is substantially weakened and the change in suspended load occurs over much greater length scales than would be predicted for a top-hat flow. This effect is more complicated for the energy equation at small to moderate $\hat{\beta}$, say $\hat{\beta} \approx 1$, because $\sigma_{uk}<1$ and $\varsigma_{uz\phi}<1$ while $\varsigma_{uuu}>1$, thus the effect of the source term $S_T$ may in some situations be amplified. More sophisticated effects can occur in the energy as the shape factors evolve. As $K$ decreases, $\hat{\beta}$ increases, and the current slumps, decreasing the shape factors associated with the gravitational potential energy and pressure $\sigma_{z\phi}$ and $\varsigma_{uz\phi}$. By \cref{eqn:1Dsys_egy_DE}, this reduces the contribution to the total energy (or total energy flux) of the gravitational terms, but crucially does not reduce the value of the total energy (or total energy flux) itself. Thus we conclude that as the current subsides in TKE and slumps, the GPE of the uplifted particles is released as MKE and TKE, accelerating the current and restoring the turbulence. In practice, this may manifest as a current that is simply more difficult to slow down. The converse effect may also be seen, where TKE is enhanced by levee overspill (see below), causing the particles to rise up, storing some of the increased energy density as GPE. This discussion is continued in \cref{sec:example}.

\begin{table}
	\centering
	$
	\renewcommand*{\arraystretch}{1.3}
	\begin{array}{ l | l | l }
								&	\hat{\beta} \to 0^+ 															&	\hat{\beta} \to \infty														\\
	\hline
		\sigma_{z\phi}			&	1 - \frac{1}{6} \hat{\beta} + \frac{1}{360} \hat{\beta}^3 						&	2 \hat{\beta}^{-1} - 2 e^{-\hat{\beta}} - 2 e^{-2 \hat{\beta}}				\\
		\sigma_{uu}				& 	\frac{6}{5}+\frac{9}{350} \hat{\beta}^{2}-\frac{1}{5250} \hat{\beta}^{4}		&	\frac{1}{4}\hat{\beta} + \frac{1}{4} + \frac{3}{4} \hat{\beta}^{-1}			\\
		\sigma_{uuu}			& 	\frac{54}{35}+\frac{27}{350} \hat{\beta}^{2}+\frac{81}{539000} \hat{\beta}^{4}	&	\frac{2}{27} \hat{\beta}^{2} + \frac{4}{27} \hat{\beta} + \frac{40}{81}		\\
		\sigma_{u\phi}			& 	1+\frac{1}{30} \hat{\beta}^{2}-\frac{11}{25200} \hat{\beta}^{4}					&	\frac{1}{4}\hat{\beta} + \frac{1}{4} + \frac{1}{2}\hat{\beta}^{-1} 			\\
		\sigma_{uk}				& 	\frac{3}{5}+\frac{9}{350} \hat{\beta} +\frac{9}{250} \hat{\beta}^{2}			&	\frac{16}{81}\hat{\beta}+\frac{16}{81}+\frac{112}{243}\hat{\beta}^{-1}		\\
		\sigma_{uz\phi}			& 	1-\frac{1}{5} \hat{\beta} + \frac{1}{30} \hat{\beta}^{2}						& 	\frac{1}{2}+\frac{1}{4}\hat{\beta}^{-1}+\frac{1}{2}\hat{\beta}^{-2}			\\
		\tilde{\sigma}_{uz\phi}	& 	1-\frac{1}{10} \hat{\beta} + \frac{1}{525} \hat{\beta}^{3}						&	\frac{1}{2}+\frac{1}{2}\hat{\beta}^{-1}+\hat{\beta}^{-2}					\\
		\varsigma_{uz\phi}		&	1-\frac{3}{20} \hat{\beta} +\frac{1}{60} \hat{\beta}^{2}						&	\frac{1}{2}+\frac{3}{8} \hat{\beta}^{-1}+\frac{3}{4} \hat{\beta}^{-2}		\\
	\hline
		\varsigma_{\phi}		&	1+\frac{1}{2} \hat{\beta} +\frac{1}{12} \hat{\beta}^{2}							& 	\hat{\beta} + \hat{\beta} e^{-\hat{\beta}} + \hat{\beta} e^{-2\hat{\beta}}	\\
		\varsigma_{u'}			&	6+3 \hat{\beta} +\frac{3}{5} \hat{\beta}^{2}									&	\hat{\beta}^{2}+2 \hat{\beta} +4
	\end{array}$
	\caption{Series expansions of the shape factors at small and large $\hat{\beta}$.}
	\label{tab:shape_asymptotic}
\end{table}

\subsection{Levee overspill rate and structure} \label{sec:vert_levee}


We assume that the fluid in the region $\mathscr{z}_B < \mathscr{z} < 1$ of the main body undergoes overspill, there being no lateral variation in the structure of concentration, longitudinal velocity, or TKE between the main body of the current and the levee crest. We justify this by noting that, due to the density stratification, there is an energy cost associated with raising fluid from below the levee crest, and thus the volume of fluid overspilling from this region should be small. This assumes that $h-B \ll \Upsilon$ to ensure that there is little variation in the depth of the upper, laterally flowing layer over the width. Consequently, the structure functions $\xi_u$, $\xi_\phi$, and $\xi_k$ may be evaluated at $\mathscr{y}=1$ (above the levee crest) using the expressions derived for the bulk. For the outflow velocity itself, we employ the same construction as for the longitudinal velocity in the body,
\begin{align}
	\xi_n (x,\mathscr{z},t) &= \xi_{n0} (\mathscr{z}-\mathscr{z}_B) (1-\mathscr{z}) \cdot \xi_{\phi}(x,1,\mathscr{z},t),
\end{align}
where $\xi_{n0}$ is set by normalisation.

The fact that the overspilling fluid is drawn from the upper reaches of the flow is significant for the effect of overspill on the current. Examining \cref{fig:ComparisonExperiment} we see that, for $B/h \geq 0.5$, the average concentration and longitudinal velocity of the overspilling fluid is below the average of the overall current, and for $B/h \geq 0.8$ this is true for TKE also. By removing fluid with below average values of these quantities, we increase the average values for the remaining fluid. Thus, overspill has the potential to rejuvenate a current, especially if the overspill is driven by primarily geometric effects, \ie decreasing channel cross section $\dv*{(B \Upsilon)}{x} < 0$. However, if the overspill balances entrainment of ambient, then the volume of fluid does not reduce and thus, in the absence of a sufficient source (erosion, downslope gravity, and drag/entrainment),  $\Phi$, $U$, and $K$ will decrease, see \cref{sec:equil}. The rejuvenating effect of a narrowing channel may be a common cause of long-run out turbidity currents, and we explore its effect in \cref{sec:example}.

\subsection{Symmetry group}

The general model (\cref{sec:model}) equipped with the closures (\cref{sec:turbclose,sec:quasiequ}) and the transverse structure (\cref{sec:vert}) constitutes a full model for the dynamics. This model may be analysed for symmetry under the rescaling of the variables, and it may be verified by substitution that the rescaling
\begin{subequations}\begin{gather}
	\label{eqn:symmetrygroup}
	\ppar*{\begin{array}{*{6}{c@{,\:\:}}c}
		h & B & \Upsilon & \Phi & Rg & x & t
	\end{array}}
	\mapsto
	\ppar*{\begin{array}{*{6}{c@{,\:\:}}c}
		a\,h & a\,B & a\,\Upsilon & \dfrac{1}{ab}\,\Phi & b\,Rg & a\,x & a\,t
	\end{array}}
\shortintertext{for any $a,b \neq 0$ implies}
	\ppar*{\begin{array}{*{3}{c@{,\:\:}}c}
		S_h & S_\Phi & S_U & S_T
	\end{array}}
	\mapsto
	\ppar*{\begin{array}{*{3}{c@{,\:\:}}c}
		S_h & \dfrac{1}{ab}\, S_\Phi & S_U & S_T
	\end{array}}
\end{gather}\end{subequations}
and consequently the system of equations \cref{eqn:1Dsys_vol,eqn:1Dsys_part,eqn:1Dsys_mom,eqn:1Dsys_egy} is unchanged. This symmetry will be used in the analysis of the system in \cref{sec:equil}.

\subsection{Hyperbolicity and degeneracy to mixed-type} \label{sec:vert_elliptic}

\begin{figure}
	\centering
	\includegraphics{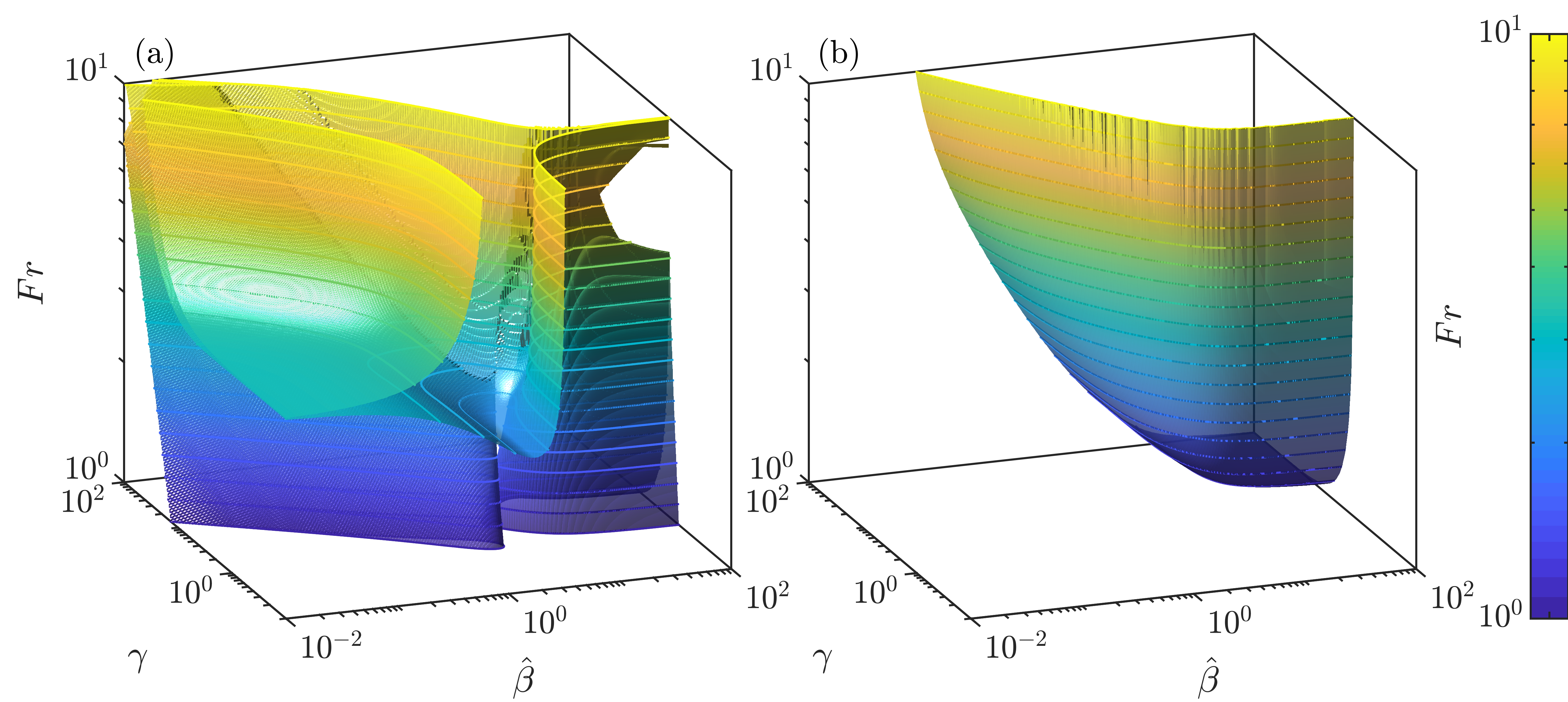}
	\caption{(a) The surface on which the determinant of the characteristic polynomial equals zero. (b) The surface on which the characteristic speed associated with the TKE field changes sign through infinity.}
	\label{fig:Temporal}
\end{figure}

With varying shape factors, the system becomes sufficiently complicated that an analytic investigation of the characteristics (as performed in \cref{sec:time}) is impractical, and we resort to numerical computations of the eigenvalues of $A_F$. The shape factors derived in \cref{sec:vert_bulk} depend on $K$, so that the energy equation is no longer decoupled from the other 3, and we have to investigate the full four equation system together. This presents a potential issue. The speeds $\lambda_K$ and $\lambda_\Phi$ from the constant shape factor case take very similar values, and this makes them vulnerable to becoming a repeated root, and then it only takes a very minor alteration to $P_{F4}$ for the real roots to become a pair of complex conjugate roots. Under these conditions, the system is of mixed elliptic-hyperbolic type, and initial value problems are not well posed. We find that there are regions of phase space where this happens. In \cref{fig:Temporal}(a) we plot the surface on which the determinant of the characteristic polynomial is zero. The system is hyperbolic across the majority of $0<\Fro<3$, $\gamma < 1$, and the surface divides the hyperbolic regime from the mixed regime. The structure of the mixed regime is complicated. There is a large portion in $\hat{\beta}<10$, $\Fro>2$, which is connected by two narrow strips down to $\Fro<1$. 

\Cref{fig:Temporal}(b) plots the surface 
\begin{align} \label{eqn:temporal_singular_TKE}
	\Fro^2 = \frac**{ \frac{\sigma_{z\phi}'}{\sigma_{z\phi}} }{\ppar*{ \frac{2}{\hat{\beta} \gamma^2} - \frac{\sigma_{uu}'}{\sigma_{uu}} }}
\end{align}
on which the evolution equation for TKE becomes singular. That is to say, all the entries in the bottom row of $A_F$ diverge, which causes $\lambda_K$ to flip between $-\infty$ and $\infty$. Local to this surface, and beyond the surface (that is, $\Fro$ larger than the right hand side of \cref{eqn:temporal_singular_TKE}), the system is hyperbolic, however it is not clear that the region where the characteristic velocity is divergent or reversed is physically meaningful.

The above discussion indicates that there are significant challenges with including varying shape factors in the model. These problems are important to address, because real currents do adjust their transverse structure. They arose due to the shape factors depending on $K$, but this \emph{must} be the case because it is the turbulent diffusion which supports the particles against settling (\cref{sec:vert}). In the majority of the region $\gamma \lesssim 1$ and $\Fro \lesssim 3$ the system is hyperbolic, which is where we focus our analysis, but because we only consider steady states the issue of hyperbolicity is not pressing.
\section{The phase-space of a steady confined current} \label{sec:phase}


\begin{figure}
	\centering
	\includegraphics{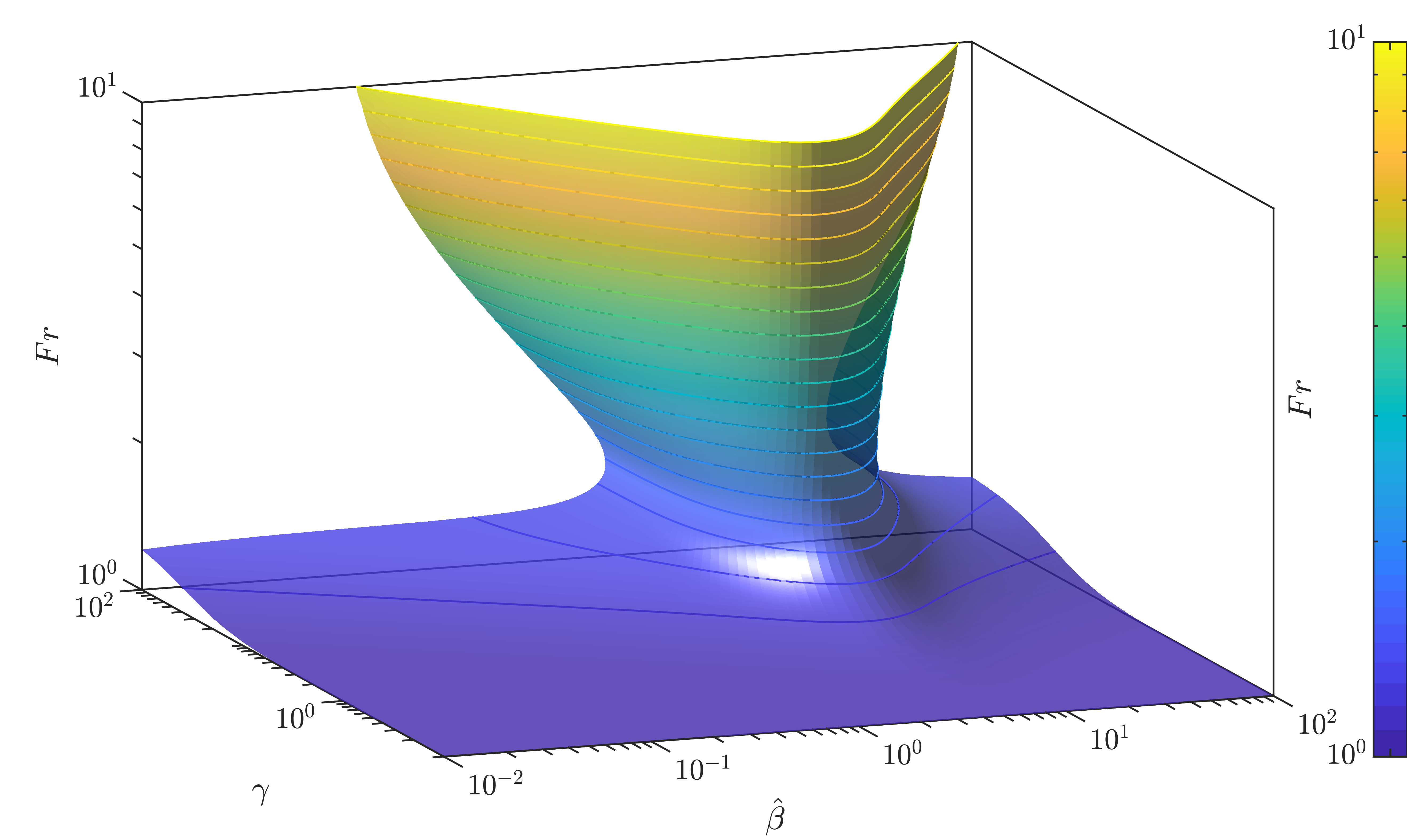}
	\caption{The dimensionless parameters at which the system becomes singular - critical flow \cref{eqn:steady_system_denominator}. The surface is colourised using $\Fro$. This surface is also included in \cref{fig:phase_traj}(a-d).}
	\label{fig:phase_singular}
\end{figure}

In this section we investigate the dynamics of a steady current ($\pdv*{}{t}=0$) confined in a deep channel of constant width and constant shallow slope ($B \to \infty$, $\dv*{\Upsilon}{x}=\dv*{\theta}{x}=0$); in particular the spatial evolution of $\Fro$, $\beta$, and $\gamma$. However, to start with we will examine when the system of equations becomes singular, which does not depend on the source terms of the system (and therefore is independent of $B$, $\Upsilon$ and $\theta$). For similarity flow, \cref{sec:model_param,sec:time} showed that criticality, and therefore a singularity in the steady state system, exists at $\Fro \in \pbrc*{-1,0,1}$. Including a dependence of the shape factors on the flow conditions significantly alters the structure of the system of equations, and thus the singularity and criticality conditions become substantially different. The steady state system becomes singular when
\begin{multline} \label{eqn:steady_system_denominator}
	\mathcal{D} 
	\eqdef
	\sigma_{uu} \sigma_{z\phi} \sigma_{uk} \ppar*{ 2 - \beta \frac{\sigma_{uk}'}{\sigma_{uk}}} \Fro^2 \ppar*{ \Fro^2 - 1 }
+ \beta \gamma^2 \Bigg[
		\sigma_{z\phi} \sigma_{uuu} \ppar*{2 \frac{\sigma_{uu}'}{\sigma_{uu}} - \frac{\sigma_{uuu}'}{\sigma_{uuu}}} \Fro^4 
	\\
		+ \sigma_{z\phi} \sigma_{uuu} \ppar*{- \frac{\sigma_{u\phi}'}{\sigma_{u\phi}} + \frac{\sigma_{uuu}'}{\sigma_{uuu}} + \frac{\sigma_{z\phi}'}{\sigma_{z\phi}}} \Fro^2
		+ 2 \sigma_{uu} \varsigma_{zu\phi} \ppar*{ \frac{ \sigma_{u\phi}' }{ \sigma_{u\phi} } - \frac{ \sigma_{uu}' }{ \sigma_{uu} } - \frac{ \varsigma_{zu\phi}' }{ \varsigma_{zu\phi} } } \Fro^2
	\\
		+ \sigma_{uu} \varsigma_{zu\phi} \ppar*{ - \frac{\sigma_{u\phi}'}{\sigma_{u\phi}} + 2 \frac{\varsigma_{zu\phi}'}{\varsigma_{zu\phi}} - \frac{\sigma_{z\phi}'}{\sigma_{z\phi}} }
	\Bigg]
\end{multline}
vanishes and we identify these Froude numbers with critical flow. The surface $\mathcal{D} = 0$ is plotted in \cref{fig:phase_singular} for the shape factors derived in \cref{sec:vert}. The surface tends towards $\Fro = 1$ at $\gamma \to 0$, $\hat{\beta} \to 0$, and $\hat{\beta} \to \infty$. However, in the region plotted the surface deforms away from $\Fro = 1$, and the region $\Fro < 1$ is connected all the way up to $\Fro \to \infty$ without a singularity. In this model, we consider the region connected, without singularity, to $\Fro < 1$ `subcritical', and the rest `supercritical'. However, the connection up to $\Fro \to \infty$ only appears in the regime where the speed of one characteristic has been reversed, \cref{fig:Temporal}(b), beyond the mixed regime, \cref{fig:Temporal}(a), and therefore is unlikely to have physical meaning.

\begin{figure}
	\centering
	\includegraphics{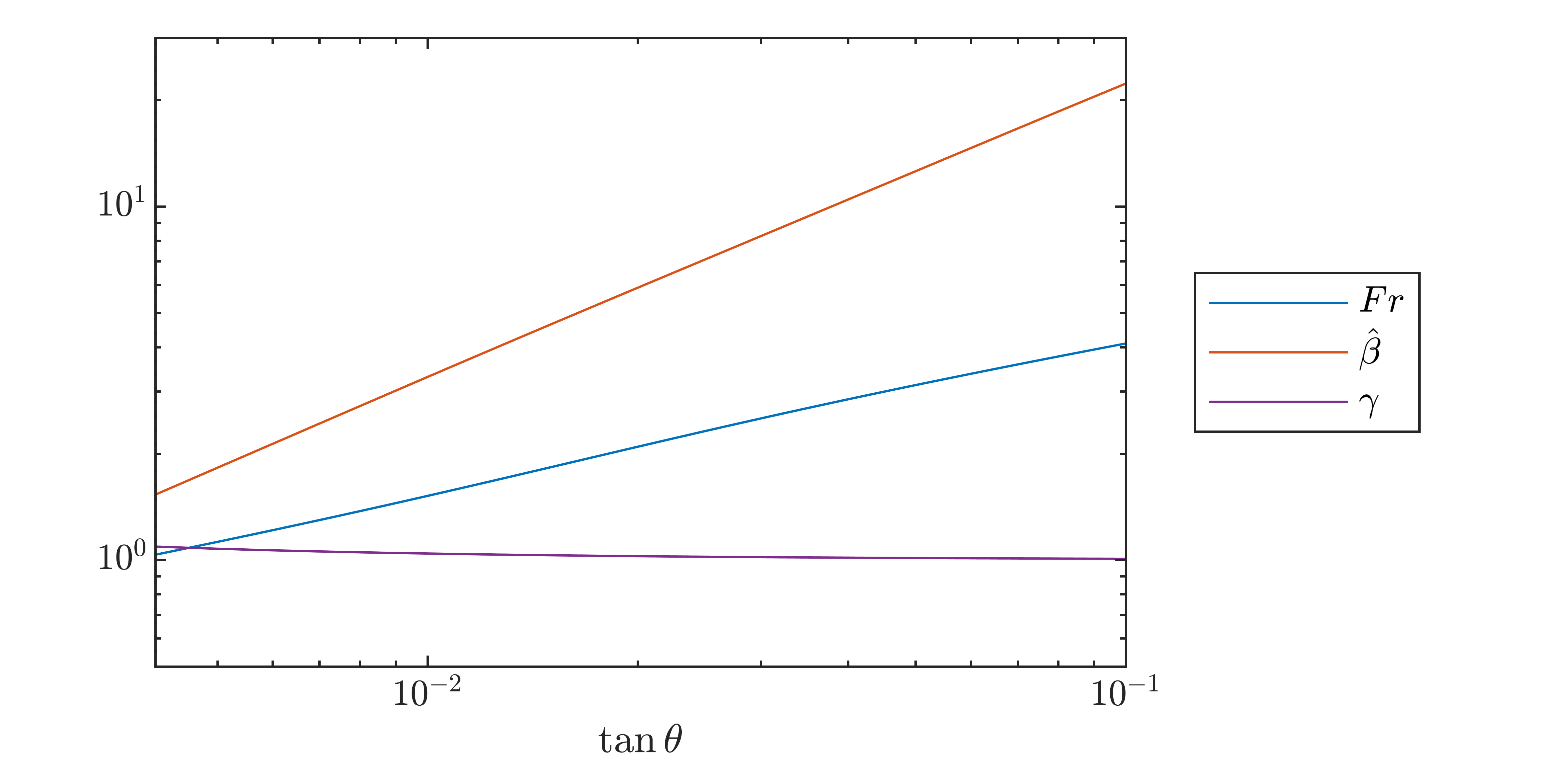}
	\caption{Pseudo-equilibrium conditions (\cref{sec:quasiequ}) for a confined flow, as a function of $\tan\theta$. The results are independent of $h$.}
	\label{fig:phase_quasiequl}
\end{figure}

We next consider the dynamics of a confined current. The special case of pseudo-equilibrium currents (\cref{sec:quasiequ}) is considered first, and the flow states for a range of slope angles are plotted in \cref{fig:phase_quasiequl}, note that the values of $\Fro$, $\hat{\beta}$, and $\gamma$ are only dependent on $\theta$ and importantly do not dependent on $h$. The values of $\gamma$ are marginally above $\gamma_{\equl}=1$ ($1<\gamma<1.27$), because the TKE experiences dissipation above that for a compositional current.

\begin{figure}[tp!]
	\centering
	\includegraphics{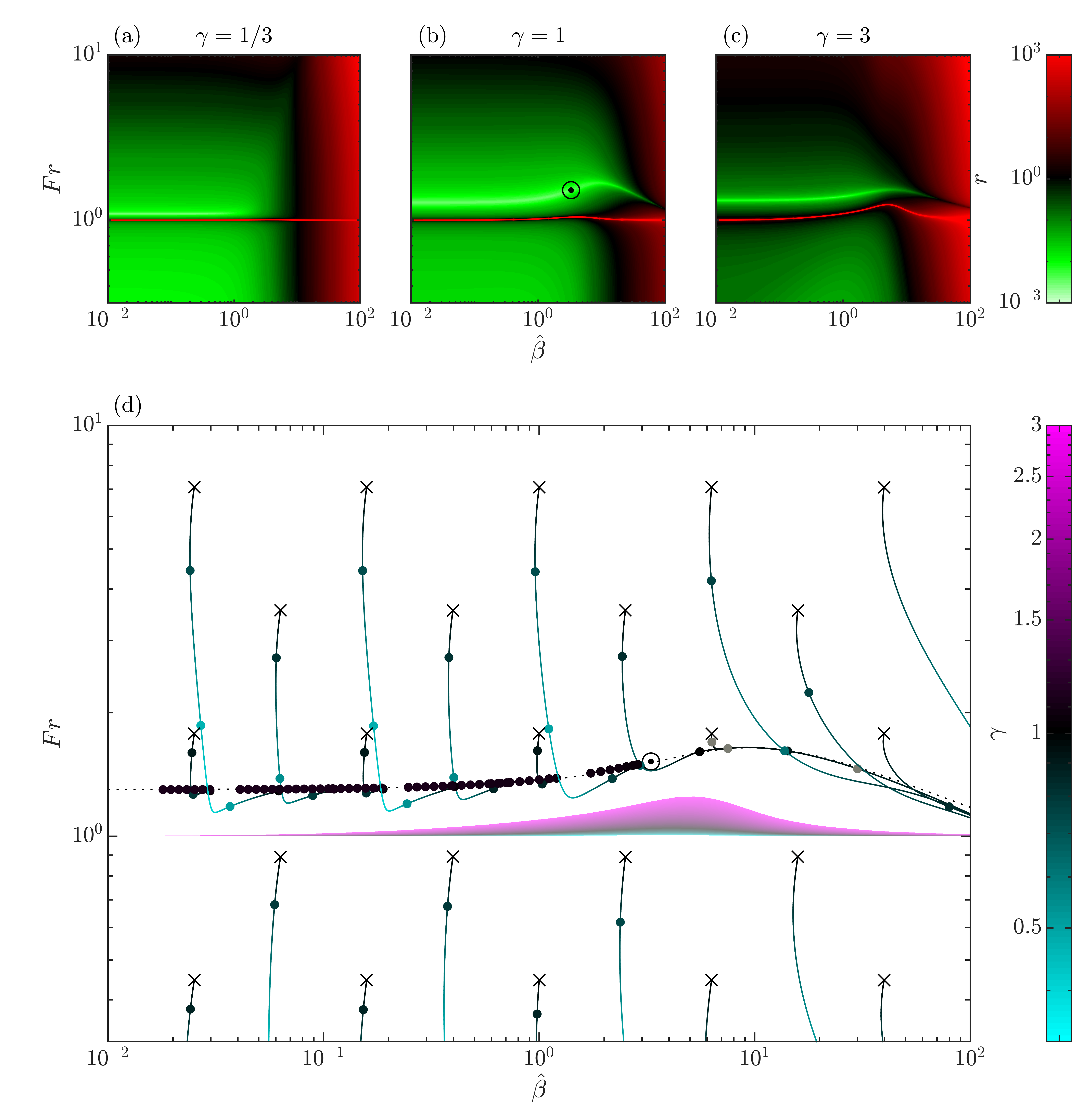}
	\caption{The phase space of the steady state confined system with $\tan(\theta)=0.01$. In (b), (c), and (d) planes of $\gamma = \gamma_{\equl}/3$, $\gamma_{\equl}$, and $3 \gamma_{\equl}$ are shown respectively, colourised using $r$ as defined in \cref{eqn:evolution_rate}. In each, the curve along which the system becomes singular is plotted in red (\cref{fig:phase_singular}), and the pseudo-equilibrium state is plotted in (b), (d) as a black circle with centre dot (\cref{fig:phase_quasiequl} and \cref{sec:quasiequ}). In (d) trajectories of the system are shown projected onto the $(\beta,\Fro)$-plane, all of which start from $\gamma = \gamma_{\equl}$, and $h=1 \mathrm{m}$. The value of $\gamma$ along the curve is shown using colour, a cross is plotted at $x=0$, and filled circles for $\log_{10} x \in \pbrc*{0,1,\ldots,10}$. Also shown, in faded colour, is the singular surface.}
	\label{fig:phase_traj}
\end{figure}

To examine the phase-space for flows that are not in a pseudo-equilibrium, we introduce the net rate of change $r$ of the dimensionless variables,
\begin{align} \label{eqn:evolution_rate}
	r 
	= \norm*{ \ppar[\bigg]{ \frac{h}{\Fro} \dv{\Fro}{x} , \frac{h}{\hat{\beta}} \dv{\hat{\beta}}{x} , \frac{h}{\gamma} \dv{\gamma}{x} } }_2
	\equiv \sqrt{ \ppar[\bigg]{ \frac{h}{\Fro} \dv{\Fro}{x} }^2 + \ppar[\bigg]{ \frac{h}{\hat{\beta}} \dv{\hat{\beta}}{x} }^2 + \ppar[\bigg]{ \frac{h}{\gamma} \dv{\gamma}{x} }^2 }.
\end{align}
which is a function of $(\Fro,\hat{\beta},\gamma)$, and independent of $h$.
This measure is shown as a contour plot in \cref{fig:phase_traj}(a-c). We see that slowly evolving flows exist in a narrow band of $\Fro$ values across a wide range of $\hat{\beta}$ by the sharp green band in \cref{fig:phase_traj}(b). Note that the slowly evolving flows are all in the region $\Fro>1$.

We close this section by examining the trajectories that result from various initial conditions. In \cref{fig:phase_traj}(d) we plot a selection of trajectories which all start from $\gamma=\gamma_{\equl}=1$. For subcritical currents, $\gamma$ and $\Fro$ decrease, stalling the flows. While supercritical currents may initially undergo the same behaviour, they converge on the region of slowly evolving flows and persist for a long time. We term this region of slowly evolving flows `the slow manifold', and includes the equilibrium point, though this is seen to be weakly unstable. The simulations that produced the curves in \cref{fig:phase_traj}(d) were stopped after a distance of $10^{10} \mathrm{m}$, which is much further than any real turbidity current. Moreover, the evolution of flows that reach the slow manifold may occur over length scales of $10^{6}h$ or greater. Thus, flows on the green curve in \cref{fig:phase_traj}(b) with $\hat{\beta} \lesssim 1$ may be considered to be practically in steady state, even if formally the $x$ derivative is non-zero in the model. This implies that, at least for confined flows, assuming flows are in equilibrium to simplify modelling may not be appropriate because it is very likely that most real world flows are on (or perhaps have not yet reached) a slow manifold \citep[\cf][]{bk_Roberts_MEDCS}.
 \section{Partially-confined equilibrium and pseudo-equilibrium flow states} \label{sec:equil}

\begin{figure}
	\centering
	\includegraphics{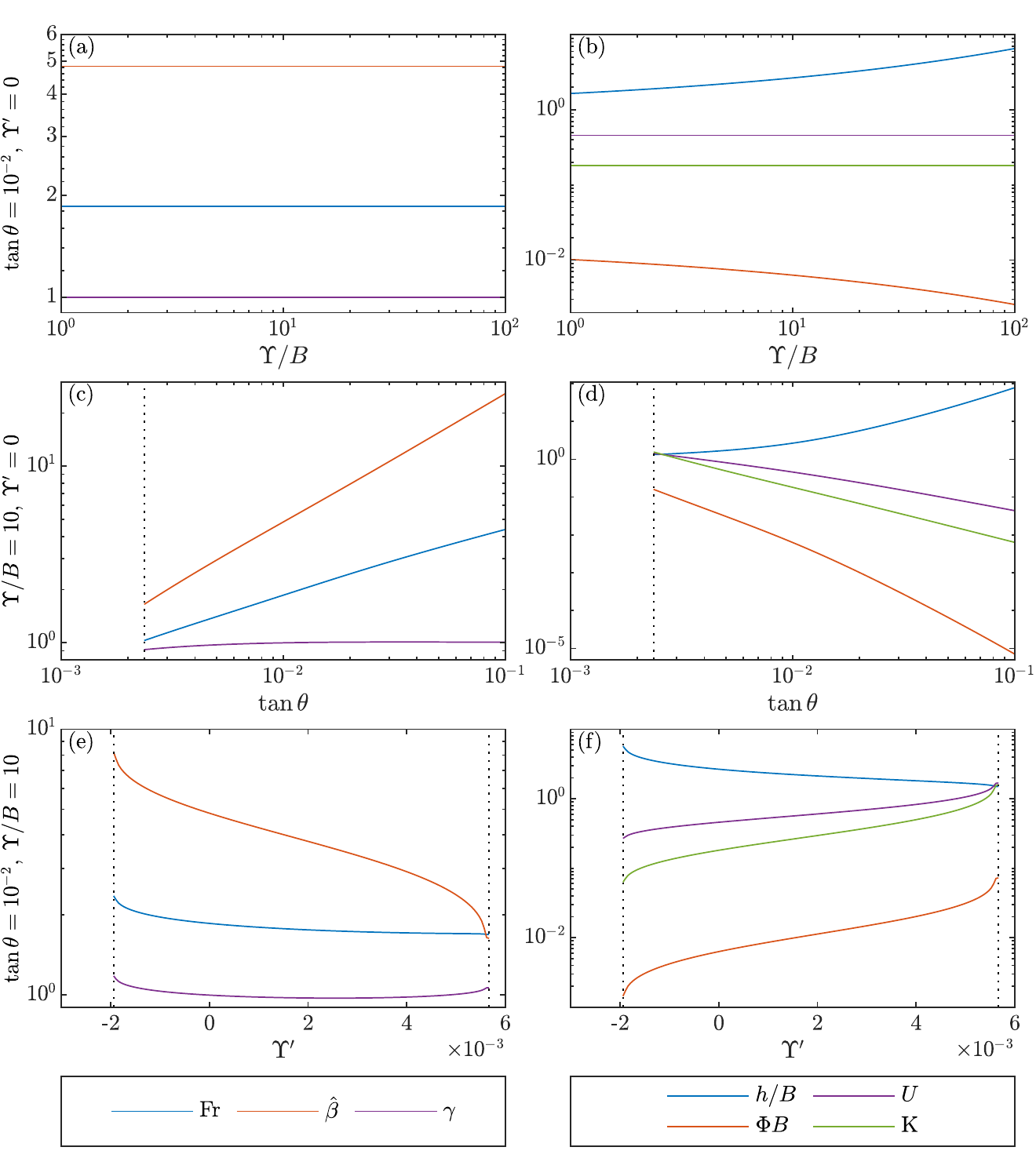}
	\caption{Equilibrium flow states. In (a,b), we set $\tan\theta = 10^{-2}$, and vary $\Upsilon/B$. In (c,d) we set $\Upsilon/B=10$ and vary $\tan\theta$. In (e,f) we plot the pseudo-equilibrium with $\tan\theta = 10^{-2}$, $\Upsilon/B = 10$ with varying $\Upsilon'$. In (a,c,e) we plot dimensionless parameters, and in (b,d,f) we plot scaled primitive variables.}
	\label{fig:equilibrium}
\end{figure}


Unlike the case of a confined channel, in a partially-confined flow ($h>B$) the levee overspill can balance with the entrainment to produce a current that does not vary in space at all, that is
\begin{gather}
	\dv{h}{x} = \dv{\Phi}{x} = \dv{U}{x} = \dv{K}{x} = 0,
\shortintertext{equivalently}
	S_h = S_\Phi = S_U = S_T = 0.
\end{gather}
Such a flow is in a fully equilibrium configuration, unlike the flows discussed in \cref{sec:quasiequ,sec:phase} where the depth was permitted to vary. Using the symmetry \cref{eqn:symmetrygroup} we can simplify our analysis, eliminating $B$ from the problem  by considering the variables $h/B$, $\Upsilon/B$ and $\Phi B$.


The equilibrium flow state depends on the channel aspect ratio $\Upsilon/B>1$, and the slope $\tan\theta$, and we plot this dependence in \cref{fig:equilibrium}(a-d). We note first of all that the equilibrium values of $\Fro$, $\hat{\beta}$, and $\gamma$ are independent of $\Upsilon/B$, and for the slope angle of $\tan\theta = 10^{-2}$ plotted we have
\begin{align}
	\Fro &= 1.88.
&
	\hat{\beta} &= 4.88,
&
	\gamma &= 1.03.
\end{align}
However, the depth does vary with both of these parameters. With varying $\theta$ all of the quantities vary, although for steeper slopes ($\tan\theta>10^{-2}$) the equilibrium concentration becomes very small. In all equilibrium flows the entrainment balances overspill in the volume equation, and there is a loss of particles, momentum, and TKE due to overspill, and consequently the flows must be erosional and on sufficiently steep slopes.


While these equilibrium states are interesting from a theoretical perspective, in practice we should expect to see channels that vary in cross section as the fluid flows along. In order to tackle these problems we generalise to situations where variation in $x$ constitutes a simple rescaling, which is formally a similarity solution \citep[\eg][]{bk_Barenblatt_SSSIA}, though we do not use that term due to the confusion with the similarity form of the transverse structure (\cref{sec:model}) and instead call them partially-confined pseudo-equilibria. To construct these we use \cref{eqn:symmetrygroup} to deduce that
\begin{gather}
	\label{eqn:symmetrygroup_steady_general}
	\ppar*{\begin{array}{*{5}{c@{,\;\;}}c}
		h & B & \Upsilon & \Phi & U & K
	\end{array}}
	=
	\ppar*{\begin{array}{*{5}{c@{,\;\;}}c}
		x\,\tilde{h}_0 & x\,\tilde{B}_0 & x\,\tilde{\Upsilon}_0 & x^{-1}\,\tilde{\Phi}_0 & \tilde{U}_0 & \tilde{K}_0
	\end{array}}
\end{gather}
is a solution, where $\tilde{\vphantom{h}\omitdummy}_0$ are constants. This is because $x$ only ever appears as a derivative so that $\pdv**{h}{x} = \frac**{h}{x}$, and similarly for $B$ and $\Upsilon$, while $\Phi$ only appears in a derivative when multiplied by $h$. In these solutions $\Fro$, $\hat{\beta}$, and $\gamma$ are constant in $x$, and so these flows are the partially-confined equivalent of \cref{sec:quasiequ}.

Due to the arbitrary nature of the constants, and the translational invariance of the system with respect to $x$, we rewrite \cref{eqn:symmetrygroup_steady_general} as
\begin{multline}
	\label{eqn:symmetrygroup_steady}
	\ppar*{\begin{array}{*{5}{c@{,\;\;}}c}
		h & B & \Upsilon & \Phi & U & K
	\end{array}}
	\\=
	\ppar*{\begin{array}{*{5}{c@{,\;\;}}c}
		\pbrk*{1-\frac{x}{x_v}} h_0 & \pbrk*{1-\frac{x}{x_v}} B_0 &  \pbrk*{1-\frac{x}{x_v}} \Upsilon_0 & \pbrk*{1-\frac{x}{x_v}}^{-1} \Phi_0 & U_0 & K_0
	\end{array}}
\end{multline}
where $\omitdummy_0$ are the values of the functions at $x=0$, and $x_v$ is the location at which the ideal linear channel vanishes. In this way, all of the length ratios $h/B$, $\Upsilon/B$, and $\Upsilon' \equiv \dv**{\Upsilon}{x}$ are constant, as is the total driving force $Rgh\Phi$, the velocity $U$, and the TKE $K$. We impose the values of $\Upsilon'$, $\Upsilon/B$ and $\tan\theta$, and from these solve for $h/B$, $\Phi B$, $U$, and $K$. We have already explored the variation of the solution with $\Upsilon/B$ and $\tan\theta$ for the case $\Upsilon'=0$, these are the equilibrium solutions in \cref{fig:equilibrium}(a-d). We now explore the generalisation to pseudo-equilibrium by letting $\Upsilon'$ vary.

We plot the values of the pseudo-equilibrium solution in \cref{fig:equilibrium}(e,f). Of particular interest is a narrowing channel, $\Upsilon'<0$, which are a common environmental feature. The fluid that is overspilling is from the upper portion of the current, which has a lower concentration of particles, momentum, and TKE than the overall current on average, and thus these values are enhanced by the overspill (\cref{sec:vert_levee}). For a current in a channel of constant width, the dilution by entrainment is stronger than this enhancement, and consequently other sources are required to balance the dilution. However, for a narrowing channel the volume is decreasing, and consequently the rate of overspill is higher than the rate of entrainment, so that the enhancing effect of overspill can balance or exceed the dilution by entrainment. This explains the higher $h$ and lower other values for more strongly narrowing channels, \cref{fig:equilibrium}(f): to achieve a balance the fluid that overspills must be from deeper in the current so that the overspilling fluid is closer to the average concentration \etc, and the magnitude of the enhancement from overspill is proportional to the depth averaged magnitude, so by having a lower magnitude of concentration \etc the enhancement is reduced proportionally.

The pseudo-equilibria are unstable in $x$, that is to say if we choose a boundary condition with $U$, $\Phi$, and $K$ slightly smaller than that of the pseudo-equilibrium value then the current collapses, these values go to zero and the depth diverges at large $x$. Conversely, if $U$, $\Phi$, and $K$ are slightly larger then they steadily increase downstream and $h$ decreases towards $B$. Note that this behaviours causes to the system of equations to become extremely stiff, particularly conservation of volume \cref{eqn:1Dsys_vol}, because $h-B \ll h$ in this regime. This behaviour occurs due to the enhancing effect of the levee overspill. We interpret the generalised equilibrium in \cref{fig:equilibrium}(e,f) as a threshold, with less energetic currents collapsing, and more energetic currents being enhanced. 

The effects discussed here may be important for the understanding of natural scale turbidity currents. The enhancing effect of levee overspill in a narrowing channel means that the particle concentration can increase downstream even without erosion. Consequently, it is possible for currents to be net depositional or bypass \citep{ar_Stevenson_2013} while not reducing in concentration, or maybe even increasing in concentration.
\section{Example solution for the Congo system} \label{sec:example}


As an example application of our system of equations, we simulate a current in the Congo system. This system has recently been the subject of several field studies, for example \citet{ar_AzpirozZabala_2017,ar_Simmons_2020}. Our goal here is to simulate a current which travels to the distal end of the \qty{1000}{km} channel; no current of this scale has been simulated previously. In this simulation, we use data for the geometry of the channel from \citet{ar_Savoye_2009}, approximating as a rectangular channel with the same levee elevation and cross sectional area. Experimenting with parameter values to obtain a current that travels a great distance, we use a particle diameter of $d = 10^{-4.5} \unit{m} \approx \qty{32}{\micro m}$, and an upstream boundary condition of $h=10^{1.5}\unit{m} \approx \qty{32}{m}$, $\Phi = 10^{-2.5} \approx 0.0032$. We enforce that the flow is initially on the slow manifold, that is we find values of $U$ and $K$ to minimise $r$ \cref{eqn:evolution_rate} in the supercritical regime.

\begin{figure}
    \centering
    \includegraphics{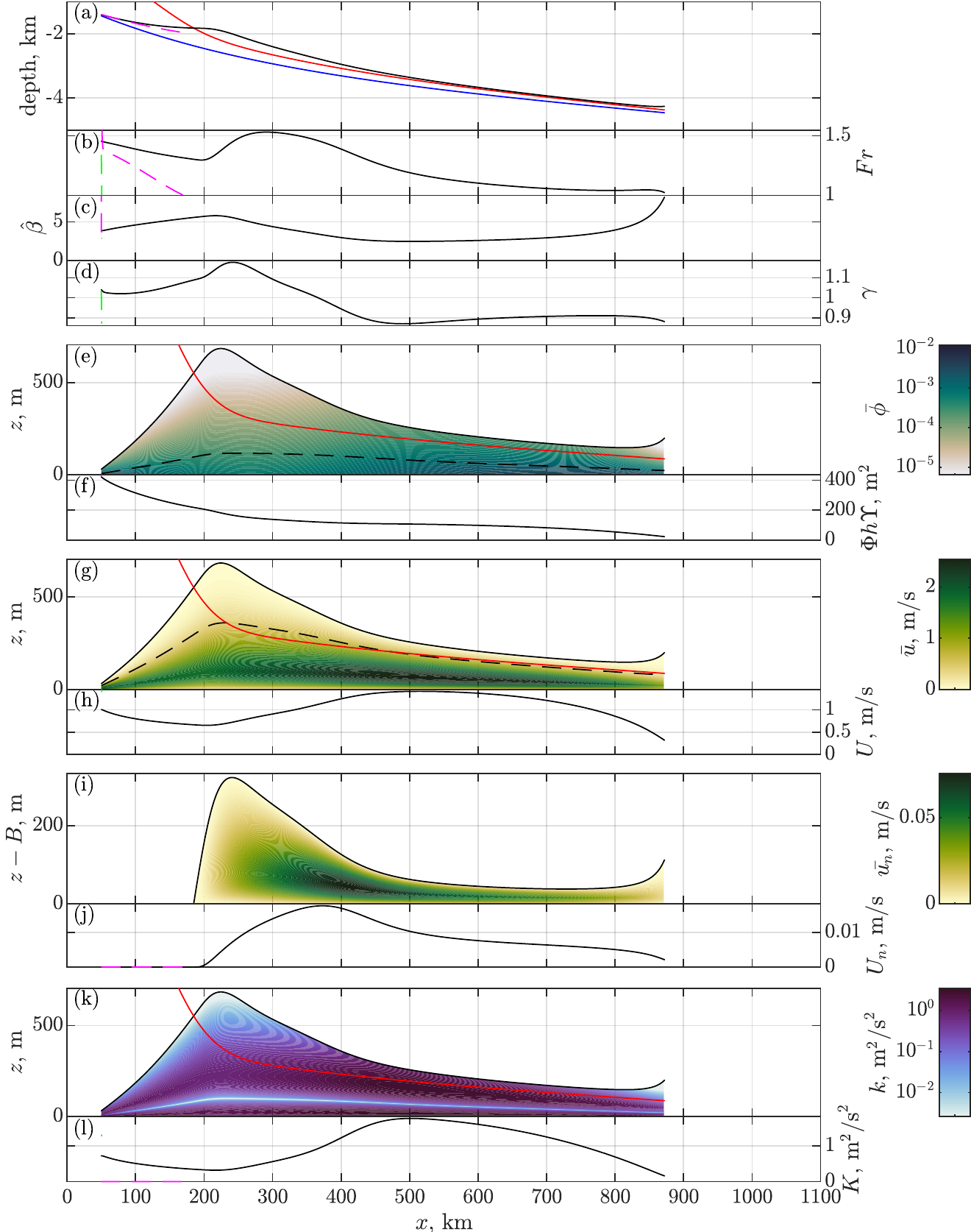}
    \caption{Example simulation of a current in the Congo system. In (a) the depth of the channel thalweg is plotted in blue, the levee elevation in red, and the current surface in black. Beneath this are the dimensionless measures $\Fro$ (b), $\hat{\beta}$ (c), and $\gamma$ (d). Then four pairs of figures show the dimensional values of concentration (e,f), longitudinal velocity (g,h), overspill velocity (i,j), and TKE (k,l). In each pair, the upper figure (e,g,i,k) plots the transverse structure as a contour plot including the levee elevation in red, while the lower (f,h,j,l) shows the depth average. In black dashed we plot the centre of mass $\sigma_{z\phi} h /2$ (e), and the depth computed from the square of velocity $\breve{h} \equiv h/\sigma_{uu}$ (g), see \cref{app:ETshape}. We also plot solutions using top-hat shape factors in green dashed, and the closures from \cite{ar_Parker_1986} in magenta dashed.}
    \label{fig:Example_Solution_1}
\end{figure}



The simulated current is plotted in \cref{fig:Example_Solution_1}. It initially deepens as it traverses the steeper ($\tan\theta \approx 10^{-2}$) narrowing  ($\Upsilon' \approx - 10^{-2}$) upstream portion of the channel. Along this portion, $\Fro$ is steadily decreasing, and heading towards subcriticality. If the flow becomes subcritical, it will collapse as seen in \cref{fig:phase_traj}, and we may anticipate that many currents in the Congo system do collapse in this manner. 
Indeed, \citet{ar_Pope_2022} found that (in the Bute Inlet, Canada) currents were progressively less common further down the channel, and all currents were supercritical suggesting that subcritical currents rapidly stall (though the calculation of bulk Richardson number was based on an assumption of local equilibrium, and is therefore prone to error).
However, by making a reasonable choice of initial conditions, it is possible to simulate a current that makes it to the point at which it begins overspilling at $x=\qty{184}{km}$. This immediately begins to enhance the concentration and velocity of the current, as discussed in \cref{sec:equil}, despite the total particle load reducing. Consequently, the Froude number reverses its decreasing trend and begins to increase. The TKE, conversely, is better mixed and so is enhanced less by the overspill, the larger effect of increased production being felt further down the channel once the velocity has substantially increased. This increase in TKE uplifts the particles, storing the energy of the current as GPE. The enhancing effect of overspill persists until $x \approx \qty{500}{km}$, driving $h$ towards $B$ and boosting the concentration, velocity, and TKE, allowing the current to run out on a very shallow slope ($\tan\theta \approx 2 \times 10^{-3}$). On $x > \qty{500}{km}$ the current undergoes a slow collapse of TKE, which is slowed by the liberation of GPE as the current slumps, maintaining the TKE and MKE of the flow as discussed in \cref{sec:vert_bulk}. The liberation of stored energy is a key feature of the new model, allowing the current to travel $~\qty{400}{km}$ even when the energy is being consumed. The simulation is halted at $x = \qty{872}{km}$ where the flow becomes critical; attempting to simulate beyond this point into the subcritical regime reveals an extremely rapid stalling of the current.

The model developed here is able to capture depositional supercritical currents that run out over great distances. We compare this to the same model, but with a top-hat transverse structure, shown as green dashed lines in \cref{fig:Example_Solution_1}. In order to compute $\hat{\beta}$, we set $\varsigma_{u'}=6$, though this does not alter the solution. In the top-hat case, it is challenging to find a plausible boundary condition that does not decrease to subcritical flow; the boundary condition for the simulation plotted was chosen using the same procedure as for the stratified case. We terminate the simulation when $\Fro=1$ is reached. To check that our choice of parametrisation of drag ($u_\star$), erosion ($E_s$), and equilibrium ($\gamma_\equl$) are not responsible for this failure, we also simulate with the parametrisation used by \cite{ar_Parker_1986} which is plotted in dashed magenta, and also experiences a collapse to subcritical flow. We expect that this is because the source terms have a much stronger effect on the current when the shape factors are smaller (\cref{sec:vert}), and thus the flow collapses much more rapidly in these models, and also because the condition of critical flow is different between the two models because of the shape factors in \cref{eqn:dimless_param}.

We conclude, then, that the transverse structure of the current is a vital component of its ability to flow over great distances: it permits the juvenile confined current to entrain and deepen without causing the current to become subcritical, enhances an established overspilling current in a narrowing channel, and allows for the store and release of available energy through redistribution of the particle load. To obtain a simulation that reaches the end of the channel several options are available. Firstly, it has been found through numerical experiment that a slightly smaller entrainment rate can produce an erosional current that travels to the end of the channel, and it may be possible that the effect of either transverse structure or levee overspill may be to reduce the entrainment rate in the distal reaches of the channel. Alternatively, including multiple particle sizes could allow for more complex dynamics, or else an improved model of the transverse structure may change the dynamics sufficiently to give a full run-out the the end of the system.
\section{Summary and conclusion} \label{sec:conclusion}

\begin{figure}
	\centering
	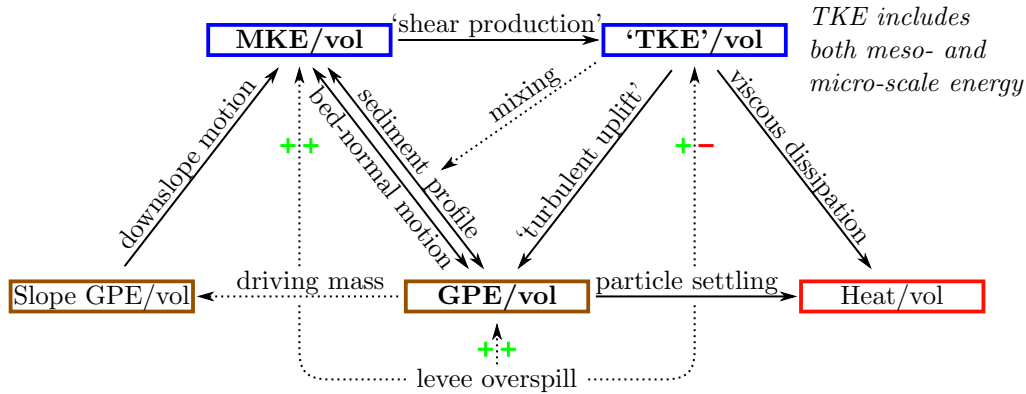
	\caption{Diagram of the energetic dynamics implied by \cref{eqn:1Dsys_MKE,eqn:1Dsys_TKE,eqn:1Dsys_GPE}, modifying and extending the insights of \cite{ar_Felix_2002}. Solid arrows represent the transfer of energy between the stores, the store at the tail loosing energy equal to that gained at the arrow. The dotted lines represent other interactions. The terms `TKE', `shear production', and `turbulent uplift' are clarified in the text, see also \citet{ar_Skevington_F006_TCFlowPower}.}
	\label{fig:energy_transfer}
\end{figure}

In this work we develop novel physics-based theory that, for the first time, explains system scale dynamics of environmentally crucial seafloor turbidity currents. This is achieved by developing a model (\cref{sec:model}) that captures the full effects of the evolving transverse flow structure of: velocity, specifying mean-flow kinetic energy (MKE); turbulent kinetic energy (TKE); and suspended particle concentration, specifying gravitational potential energy (GPE). Note that GPE here includes only the energy that would be released my moving the particles to the bed, the energy released by moving the particles downslope is accounted for separately. The model is sufficiently general to capture the effects of changing channel width and levee overspill.

New closures to specify transverse flow structure and turbulent dissipation are proposed and validated, along with parametrisations of drag, erosion, and entrainment (\cref{sec:turbclose,sec:vert}). With these a complete model of the temporal evolution of a turbidity current is obtained, which captures the flow along natural seafloor channel levee systems (\cref{sec:example}). This new modelling framework governs the evolution of the total energy stored in the flow (MKE+TKE+GPE) per unit volume. Energy may be dissipated to heat through two viscous effects: the dissipation of TKE at the Kolmogorov scale, and the viscosity acting against the particles as they settle against gravity. In partially-confined flow, levee overspill also reduces the \emph{depth total} energy present in all stores. However, the \emph{depth average} MKE and GPE per unit volume are increased by overspill because the fluid at the top of the current has a lower concentration of both (\cref{sec:vert_levee,sec:example}). These and other energetic dynamics are shown schematically in \cref{fig:energy_transfer}.

We stress that here we use `turbulence' to include the energy associated with all flow structures on length-scales smaller than the longitudinal length of the overall current. Under this definition, as highlighted by \cite{ar_Skevington_F006_TCFlowPower}, the `TKE' includes meso-scale structures with length-scale of the flow depth such as internal waves and large vortices, along with the micro-scale vortices much smaller than the depth. This is modelling simplification shared with all other four equation models \citep{ar_Parker_1986}, and simply means we have to be careful with how we interpret the predicted flow energetics.

The new model captures multiple mechanisms by which energy is transferred between the three stores (\cref{sec:production}): MKE, TKE, and GPE. The three principle means of transfer, shown in \cref{fig:energy_transfer}, are shear production ($\mathcal{P}$) converting MKE to TKE, the turbulent uplift of particles ($\mathcal{B}_K$) converting TKE to GPE, and overall bed-normal motion ($\mathcal{B}_E$) which transfers energy between GPE and MKE.  
Due to the inclusion of the meso-scale energy in the TKE, the `shear production' includes the energy transfer from the mean-flow to both the meso- and micro-scales, and the `turbulent uplift' includes the work done by the both the meso- and micro-scale flow on the particles.
Herein transverse flow structure is ascribed by a balance between turbulent uplift (TKE) and particle settling. The transverse structure of the velocity field enhances shear production; whilst the energy required for turbulent uplift decreases, primarily due to flow stratification. Thus, the Knapp-Bagnold auto-suspension criterion for near-equilibrium flows \citep{ar_Knapp_1938,ar_Bagnold_1962}, $\mathcal{P} > \mathcal{B}_K$, is satisfied more easily when transverse structure is included, as was conjectured by \cite{ar_Skevington_F006_TCFlowPower}. We conclude that the auto-suspension criterion is substantially weakened by the transverse flow structure; this should be included in all future models that simulate long run-out flows. 

Equally important is the non-equilibrium energy transfer mechanism that can only be captured with an evolving transverse structure of particle concentration (\cref{sec:vert_bulk}). As flows accelerate, TKE increases which distributes particles higher in the flow and increases GPE. Conversely as the current slows, e.g. on a transition onto a shallow slope, shear production decreases, TKE is reduced, and stored GPE is released as MKE maintaining the flow. Thus, variation in transverse flow structure highlights a new large scale energy storage mechanism. In particular, this mechanism enables depositional flows with waining turbulence to maintain their speed over great distances (\cref{sec:example}).

From here we focus on steady state systems (without time variation). For confined flows (\cref{sec:phase}) we identify new `pseudo-equilibrium' flow states where the Froude number, Rouse number, and ratio of MKE to TKE ($\gamma^2$) remain constant, but depth varies due to entrainment. However, these are not the only flow states that persist for a long time, in addition there is a slow manifold on which the current persists relatively unchanged over distances many orders of magnitude greater than its depth. For partially-confined flows (\cref{sec:equil}), we are able to establish full equilibrium solutions in which the concentration, momentum, and TKE lost to levee overspill are balanced by erosion, gravitational acceleration, and turbulent production respectively. Pseudo-equilibrium solutions also exist in narrowing channels where the width and height of the channel vary. Here levee overspill is of lower concentration and momentum than the channel average, enhancing the intensity of the flow that remains. However, solutions are unstable, and under the right circumstances the flow depth can be forced towards the levee elevation by continually enhancing concentration and momentum. This is seen in the distal end of our simulation of the Congo canyon-channel system (\cref{sec:example,fig:Example_Solution_1}). 

For the first time, the model proposed herein provides the ability to simulate sediment transport to the distal reaches of shallow sloped real-world systems 100s \unit{km} long (\cref{sec:example}). In contrast, extant models of such currents collapse almost immediately after the steep sloped sections. Thus, the framework provides a new toolkit to study auto-suspension and the role of turbidity currents at real world scale. To advance the accuracy of the new model to its fullest extent, and enable simulation to the end of the Congo system, future work should focus on developing more accurate parametrisations of the transverse structure and provide accurate drag, erosion, and entrainment closures calibrated for density driven flows.

\begin{myappendices}
	\section{Comparison to Ellison-Taylor variables and shape factors} \label{app:ETshape}


The choice of shape factors used in the main text is designed with consideration of the inclusion of levee overspill, and differs from the choice made by \citet{ar_Ellison_1959}, and developed by \citet{ar_Parker_1986,ar_Parker_1987}. There, the fluxes were simplified by defining
\begin{subequations}
\allowdisplaybreaks[1]
\begin{align}
	\reavg{u}(x,y,z,t) 		&= \ElTu{\xi_u}(x,\mathscr{y},\ElTu{\mathscr{z}},t) \cdot \ElTu{U}(x,t),
\\
	\reavg{v}(x,y,z,t) 		&= \ElTu{\xi_v}(x,\mathscr{y},\ElTu{\mathscr{z}},t) \cdot \ElTu{V}(x,t),
\\
	\reavg{\phi}(x,y,z,t) 	&= \ElTu{\xi_\phi}(x,\mathscr{y},\ElTu{\mathscr{z}},t) \cdot \ElTu{\Phi}(x,t),
\\
	k(x,y,z,t) 				&= \ElTu{\xi_k}(x,\mathscr{y},\ElTu{\mathscr{z}},t) \cdot \ElTu{K}(x,t),
\end{align}
where the quantities $\ElTu{\xi_u}$, $\ElTu{\xi_v}$, $\ElTu{\xi_\phi}$, $\ElTu{\xi_k}$ satisfy
\begin{align}
	\int_0^1 \int_0^\infty \ElTu{\xi_u} \dd{\ElTu{\mathscr{z}}} \dd{\mathscr{y}} &= 1,
&
	\int_0^1 \int_0^\infty \ElTu{\xi_u^2} \dd{\ElTu{\mathscr{z}}} \dd{\mathscr{y}} &= 1,
\\
	\int_0^1 \int_0^\infty \ElTu{\xi_u} \ElTu{\xi_\phi} \dd{\ElTu{\mathscr{z}}} \dd{\mathscr{y}} &= 1,
&
	\int_0^1 \int_0^\infty \ElTu{\xi_u} \ElTu{\xi_k} \dd{\ElTu{\mathscr{z}}} \dd{\mathscr{y}} &= 1.
\end{align}
\end{subequations}
In the above, we have identified variables specific to the Ellison-Taylor scaling by a breve accent. The definitions have been modified to be compatible with our consideration of a channel with finite width, and our lack of similarity assumption so that the variables $\ElTu{\xi_{\omitdummy}}$ depend on $x$, $\mathscr{y}$, and $t$ as well as $\ElTu{\mathscr{z}} \eqdef z/\ElTu{h}$. 

\begin{table}
	\resizebox{\textwidth}{!}{$
	\renewcommand*{\arraystretch}{1.3}
	\begin{array}{ l | r@{}l | c | r@{}l | r@{}l | r@{}l }
			& \multicolumn{2}{c|}{\text{Definition}}	& \text{Simplified}		& \multicolumn{2}{c|}{\text{Equiv.}}	& \multicolumn{2}{c|}{\text{PGF\&Y}} 	& \multicolumn{2}{c}{\text{I\&I}}	\\
	\hline
		a_0 &	&\frac**{h}{\ElTu{h}}	&&
				\sigma_{uu} &									& 1&.50 & 1&.50 \\
		a_1	&   \int_0^1 \int_0^\infty & \ElTu{\xi}_\phi \dd{\ElTu{\mathscr{z}}} \dd{\mathscr{y}}		&& 
				\sigma_{uu}&/\sigma_{u\phi}						& 0&.99		& 1&.12		\\
		a_2	& 2 \int_0^1 \int_0^\infty & \int_{\ElTu{\mathscr{z}}}^\infty \eval*{\ElTu{\xi}_\phi}_{\ElTu{\mathscr{z}}'} \dd{\ElTu{\mathscr{z}}'} \dd{\ElTu{\mathscr{z}}} \dd{\mathscr{y}}	
			& 2 \int_0^1 \int_0^\infty   \ElTu{\mathscr{z}} \ElTu{\xi}_\phi \dd{\ElTu{\mathscr{z}}} \dd{\mathscr{y}}	& 
				\sigma_{z\phi} \sigma_{uu}^2&/\sigma_{u\phi}	& 1&.00		& 1&.03 	\\
		a_3 &   \int_0^1 \int_0^\infty & \ElTu{\xi}_u^3 \dd{\ElTu{\mathscr{z}}} \dd{\mathscr{y}}		&& 
				\sigma_{uuu}&/\sigma_{uu}^2						& 1&.10		& 1&.12		\\
		a_4 & 2 \int_0^1 \int_0^\infty & \int_{\ElTu{\mathscr{z}}}^\infty \eval*{\ElTu{\xi}_u}_{\ElTu{\mathscr{z}}} \eval*{\ElTu{\xi}_\phi}_{\ElTu{\mathscr{z}}'} \dd{\ElTu{\mathscr{z}}'} \dd{\ElTu{\mathscr{z}}} \dd{\mathscr{y}}	& a_7 &
				&												& 1&.06		& 1&.00		\\
		a_5 &   \int_0^1 \int_0^\infty & \int_{\ElTu{\mathscr{z}}}^\infty \ElTu{\mathscr{z}} \eval*{\pdv{\ElTu{\xi}_u}{\ElTu{\mathscr{z}}}}_{\ElTu{\mathscr{z}}} \eval*{\ElTu{\xi}_\phi}_{\ElTu{\mathscr{z}}'} \dd{\ElTu{\mathscr{z}}'} \dd{\ElTu{\mathscr{z}}} \dd{\mathscr{y}}	& \textfrac{1}{2} (a_8-a_7)	&
			& 													& -0&.19	& -0&.20	\\
		a_6 & 2 \int_0^1 \int_0^\infty & \int_{\ElTu{\mathscr{z}}}^\infty \eval*{\ElTu{\xi}_u}_{\ElTu{\mathscr{z}}'} \eval*{\ElTu{\xi}_\phi}_{\ElTu{\mathscr{z}}'} \dd{\ElTu{\mathscr{z}}'} \dd{\ElTu{\mathscr{z}}} \dd{\mathscr{y}} & a_8 &
				&												& 0&.72		& 0&.55		\\
		a_7 & 2 \int_0^1 \int_0^\infty & \int_0^{\ElTu{\mathscr{z}}} \eval*{\ElTu{\xi}_u}_{\ElTu{\mathscr{z}}'} \eval*{\ElTu{\xi}_\phi}_{\ElTu{\mathscr{z}}} \dd{\ElTu{\mathscr{z}}'} \dd{\ElTu{\mathscr{z}}} \dd{\mathscr{y}}	&& 
				\tilde{\sigma}_{uz\phi} \sigma_{uu}&/\sigma_{u\phi}	& 1&.06		& 1&.15		\\
		a_8 & 2 \int_0^1 \int_0^\infty & \ElTu{\mathscr{z}} \ElTu{\xi}_\phi \ElTu{\xi}_u \dd{\ElTu{\mathscr{z}}} \dd{\mathscr{y}}	&& 
				\sigma_{uz\phi} \sigma_{uu}&/\sigma_{u\phi}  	& 0&.62		& 0&.78		\\
		a_9	&	\int_0^1 \int_0^\infty & \ElTu{\xi}_k \dd{\ElTu{\mathscr{z}}} \dd{\mathscr{y}}			&& 
				\sigma_{uu}&/\sigma_{uk} 						& ?&		& 1&.30		\\
		r_0 & 	\int_0^1 				& \xi_{\phi} \dd{\mathscr{y}} \big|_{\mathscr{z} = 0} 	&&
				\varsigma_\phi&/\sigma_{u\phi}					& 2&.00		& 2&.03
	\end{array}
	$}
	\caption{The shape factors used by \cite{ar_Parker_1987}, with $a_0$ being additional here. The first column is the symbols used for the shape factors. The second is their definition (modified here to account for possible lateral variation). The third is simplified expressions for the shape factors. In the fourth we express these shape factors in terms of the ones defined in \cref{tab:shape_factors}. The final two columns state the values for these shape factors reported by \cite{ar_Parker_1987} and \cite{ar_Islam_2010}.}\label{tab:shape_factors_old}
\end{table}

This transverse structure may be used to compute shape factors, and those from the measurements of \cite{ar_Parker_1987,ar_Islam_2010} are given in \cref{tab:shape_factors_old}. In this table, several of the shape factors have simplified expressions listed, and in each case this has been achieved by switching the order of integration of $\breve{z}$ and $\breve{z}'$. For $a_5$, we subsequently integrate the product $\breve{z} \dv*{\breve{\xi_u}}{\breve{z}}$ by parts. The simplifications for $a_2$, $a_4$, $a_5$ and $a_6$ do not rely on the presence of a width average, equivalent equalities may be derived for the laterally uniform case; nor on the idealisation of infinite transverse extent, in the upper bound of the integrals over $\ElTu{\mathscr{z}}$ the values of $\infty$ may be replace with an arbitrary constant (or function of $x$, $y$) without consequence. Thus, these equalities are satisfied by any transverse structure, including those measured in experiment. What is concerning, is that some experimental results do not satisfy the equalities proved above. In the measurements of \cite{ar_Parker_1987} we have 
\begin{align}
	\frac{a_4}{a_7} &= 1.00,
&
	\frac{a_5}{\textfrac{1}{2} (a_8-a_7)} &= 0.86,
&
	\frac{a_6}{a_8}	&= 1.16,
\intertext{while in those of \cite{ar_Islam_2010}}
	\frac{a_4}{a_7} &= 0.87,
&
	\frac{a_5}{\textfrac{1}{2} (a_8-a_7)} &= 1.08,
&
	\frac{a_6}{a_8}	&= 0.71.
\end{align}
If the integrals were evaluated exactly then all the ratios would be $1$. We expect that the discrepancy comes from under-resolved numerical integration, though differences of up to $30\%$ do suggest significant problems. In the computation of the values in \cref{tab:shape_factors}, $a_7$ and $a_8$ are used in preference to $a_4$, $a_5$, or $a_6$, because we suspect these are likely to have a smaller error in their evaluation.

To close this section, we demonstrate how to convert between the Ellison-Taylor variables (\cref{tab:shape_factors_old}) and those used here (\cref{tab:shape_factors}). Observe that
\begingroup\allowdisplaybreaks[1]
\begin{align*}
	\ElTu{U} \ElTu{h}
	= \ElTu{U} \ElTu{h} \! \int_0^{\mathrlap{1}} \int_0^{\mathrlap{\infty}} \ElTu{\xi}_u \dd{\ElTu{\mathscr{z}}} \dd*{\mathscr{y}}
	&= \! \int_0^{\mathrlap{\Upsilon}} \int_0^{\infty} \frac{\reavg{u} \dd{z} \dd*{y}}{\Upsilon}
	= U h \! \int_0^{\mathrlap{1}} \int_0^{\mathrlap{\infty}} \xi_u \dd{\mathscr{z}} \dd*{\mathscr{y}}
	= U h,
\shortintertext{similarly}
	\ElTu{U}^2 \ElTu{h}
	&= \! \int_0^{\mathrlap{\Upsilon}} \int_0^{\infty} \frac{\reavg{u}^2 \dd{z} \dd*{y}}{\Upsilon}
	= U^2 h \sigma_{uu},
\\
	\ElTu{U} \ElTu{\Phi} \ElTu{h}
	&= \! \int_0^{\mathrlap{\Upsilon}} \int_0^{\infty} \frac{\reavg{u} \reavg{\phi} \dd{z} \dd*{y}}{\Upsilon}
	= U \Phi h \sigma_{u\phi},
\\
	\ElTu{U} \ElTu{K} \ElTu{h}
	&= \! \int_0^{\mathrlap{\Upsilon}} \int_0^{\infty} \frac{\reavg{u} k \dd{z} \dd*{y}}{\Upsilon}
	= U K h \sigma_{uk}.
\end{align*}\endgroup
Consequently
\begin{equation}\label{eqn:ET_variables}
\begin{aligned}
	\ElTu{h} 			&= \frac{1}{\sigma_{uu}} h,
&
	\ElTu{U} 			&= \sigma_{uu} U,
&
	\ElTu{\Phi} 		&= \sigma_{u\phi} \Phi,
&
	\ElTu{K} 			&= \sigma_{uk} K,
\\
	\ElTu{\mathscr{z}}	&= \sigma_{uu} \mathscr{z},
&
	\ElTu{\xi}_u		&= \frac{1}{\sigma_{uu}} \xi_u,
&
	\ElTu{\xi}_\phi		&= \frac{1}{\sigma_{u\phi}} \xi_\phi,
&
	\ElTu{\xi}_k		&= \frac{1}{\sigma_{uk}} \xi_k.
\end{aligned}
\end{equation}
In this context, often a different definition of the Froude number is used
\begin{align} \label{eqn:ET_Froude}
	\ElTu{\Fro} 
	\eqdef \frac{\ElTu{U}}{(Rg\ElTu{\Phi}\ElTu{h} \cos\theta)^{1/2}}
	= \frac{\sigma_{uu}^{3/2}}{\sigma_{u\phi}^{1/2}} \frac{U}{(Rg \Phi h \cos\theta)^{1/2}}
	= \sigma_{uu} \sqrt{ \frac{\sigma_{z\phi}}{\sigma_{u\phi}} } \, \Fro.
\end{align}
Substitution of \cref{eqn:ET_variables} into the definitions in \cref{tab:shape_factors_old} yields the equivalent expressions listed, which can be inverted to obtain
\begin{equation}
\begin{aligned}
	\sigma_{z\phi} 		&= \frac{a_2}{a_0 a_1},
&
	\sigma_{uu}			&= a_0,
&
	\sigma_{uuu}		&= a_0^2 a_3,
&
	\sigma_{u\phi}		&= \frac{a_0}{a_1},
\\
	\sigma_{uk}			&= \frac{a_0}{a_9},
&
	\sigma_{uz\phi}		&= \frac{a_8}{a_1},
&
	\tilde{\sigma}_{uz\phi}	&= \frac{a_7}{a_1},
&
	\varsigma_\phi 		&= \frac{r_0 a_0}{a_1}.
\end{aligned}\end{equation}
These can in turn be substituted into \cref{eqn:1Dsys_vol_DE,eqn:1Dsys_part_DE,eqn:1Dsys_mom_DE,eqn:1Dsys_egy_DE} to yield (for the case where the $a_i$ are constants)
\begin{subequations}\begin{gather}
	a_0 \pdv{\ElTu{h}}{t} + \pdv{}{x}\ppar*{\ElTu{h} \ElTu{U}} = S_h,
\\
	a_1 \pdv{}{t}\ppar*{\ElTu{h} \ElTu{\Phi}} + \pdv{}{x} \ppar*{\ElTu{h} \ElTu{U} \ElTu{\Phi}} = S_{\Phi},
\\
	\pdv{}{t} \ppar*{\ElTu{h} \ElTu{U}}
	+ \pdv{}{x} \ppar*{\ElTu{h} \ElTu{U}^2 + \textfrac{1}{2} a_2 R g \ElTu{h}^2 \ElTu{\Phi} \cos\theta  }
	= S_U,
\\
	\begin{gathered}
		\pdv{}{t} \ppar*{ \ElTu{h} \pbrk*{ \textfrac{1}{2} \ElTu{U}^2 + a_9 K + \textfrac{1}{2} a_2 R g \ElTu{h} \ElTu{\Phi} \cos\theta}}  \qquad\qquad\qquad \\  \qquad\qquad\qquad
		+ \pdv{}{x} \ppar*{ \ElTu{h} \ElTu{U} \pbrk*{ \textfrac{1}{2} a_3 \ElTu{U}^2 + \ElTu{K} + \textfrac{1}{2} (a_7+a_8) R g \ElTu{h} \ElTu{\Phi} \cos\theta }} 
		= S_T,
	\end{gathered}
\end{gather}\end{subequations}
In steady state ($\pdv*{}{t}=0$), on shallow slopes ($\cos\theta \sim 1$), neglecting lateral effects in the source terms, and making use of the simplifications in \cref{tab:shape_factors_old}, this system of equations is equivalent to that presented in \cite{ar_Parker_1987}. In the time dependent case, the additional shape factor $a_0$ is required to properly convert between the depth variation and the transverse velocity in the far-field, see \cref{app:deriv_3D}.
	\section{Model derivation}  \label{app:deriv}

\subsection{The governing system for the full 3D flow}  \label{app:deriv_3D}

We derive the governing system following similar arguments to \cite{ar_Parker_1986}, but making fewer assumptions about the transverse structure of the current and including levees in our analysis. The current flows along a channel at an angle $\theta$ to the horizontal, and any variation in bed steepness (in the longitudinal direction) and channel direction is assumed to be over a sufficiently long distance that curvature terms may be neglected to leading order; we model a locally straight channel with varying cross section. We arrange the Cartesian coordinate system so that the $x$-axis lies longitudinally at the centre of the channel on the bed; $y$ is the lateral direction, the current is statistically symmetric about $y=0$; and $z$ is the transverse direction (\ie bed-normal, not necessarily vertical). We denote positions by $\vecb{x} = (x,y,z)$, times by $t$, and gravity by $\vecb{g}$.

The bed has elevation $z=\tilde{B}(x,y)$. We assume that $\tilde{B}(x,-y)=\tilde{B}(x,y)$ (the channel is symmetric), $\tilde{B}=0$ on $0 \leq y \leq \Upsilon_1(x)$ (the coordinate system is aligned with the thalweg), $\tilde{B}$ is increasing on $\Upsilon_1(x) \leq y \leq \Upsilon_2(x)$, and decreasing on $y \geq \Upsilon_2(x)$ (the up and down slopes of the levee are monotone). The channel is treated as approximately rectangular, so that $\Upsilon_2-\Upsilon_1 \ll \Upsilon_1$, and the channel half-width is
\begin{align}
    \Upsilon(x) \eqdef \frac{1}{B(x)} \int_{0}^{\Upsilon_2(x)} \tilde{B}(x,y) \dd{y}
\end{align}
where $B(x) = \tilde{B}(x,\Upsilon_2(x))$ is the channel height. How the width of the channel varies with elevation is captured by the non-decreasing function $\tilde{\Upsilon}(x,z)$, where $\tilde{\Upsilon}(x,\tilde{B}(x,y)) = y$ for $\Upsilon_1(x) \leq y \leq \Upsilon_2(x)$, and $\tilde{\Upsilon}(x,z) = \Upsilon_2(x)$ for $z > B(x)$.

The bed will be treated as locally separable, that is local to any point $x$ there is a function $\tilde{b}(\mathscr{y})$, $\mathscr{y} \eqdef y / \Upsilon$, so that
\begin{align} \label{eqn:seperable_bed}
    \tilde{B}(x,y) &= \tilde{b}(y/\Upsilon(x)) \cdot B(x),
&\text{and}&&
	\tilde{b}(1) &= 1.
\end{align}
Thus the $x$ derivative satisfies
\begin{align}
    \pdv{\tilde{B}}{x} 
    = - \frac{y}{\Upsilon^2} \dv{\Upsilon}{x} \cdot \dv{\tilde{b}}{\mathscr{y}} \cdot B + \tilde{b} \dv{B}{x}
    = - \frac{y}{\Upsilon} \dv{\Upsilon}{x} \pdv{\tilde{B}}{y} + \frac{\tilde{B}}{B} \dv{B}{x}.
\end{align}
The first term is much larger than the latter, and in the region where $\pdv*{\tilde{B}}{y}>0$, $y \simeq \Upsilon$. Thus to leading order
\begin{align} \label{eqn:dBdx_simplified}
    \pdv{\tilde{B}}{x}
    \simeq - \dv{\Upsilon}{x} \pdv{\tilde{B}}{y}.
\end{align}

The turbidity current is a suspension consisting of two parts, fluid and suspended sediment of volumetric concentration $\phi(\vecb{x},t)$. The fluid is the same as the ambient fluid, with density $\rho$ and viscosity $\nu$. The particles have density $\rho_p$ and settle out of the fluid at velocity $\tilde{\vecb{u}} = (\tilde{u},\tilde{v},\tilde{w})$. We define the relative density difference between the particles and the fluid to be $R = (\rho_p - \rho) / \rho$.

We begin the derivation of our model from the Boussinesq RANS equations, where we denote the Reynolds average of a variable $f(\vecb{x},t)$ by $\reavg{f}(\vecb{x},t)$, and the fluctuating component by $\reflc{f}(\vecb{x},t)$, so that $f = \reavg{f} + \reflc{f}$. The RANS system is
\begin{subequations}\label{eqn:3Dsys}
\allowdisplaybreaks[1]
\begin{align}
	\pdv{\reavg{u_j}}{x_j} &= 0,
	\label{eqn:3Dsys_vol}
	\\
	\pdv{\reavg{\phi}}{t} + \pdv{}{x_j} \ppar*{\pbrk*{\reavg{u_j} + \tilde{u}_j} \reavg{\phi} + \reavg{\reflc{u_j}\reflc{\phi}}} &= 0,
	\label{eqn:3Dsys_part}
	\\
	\pdv{\reavg{u_i}}{t} + \pdv{}{x_j}\ppar*{\reavg{u_j}\reavg{u_i} - \tau_{ji}^R} + \frac{1}{\rho} \pdv{\reavg{p}}{x_i} &= R g_i \reavg{\phi},
	\label{eqn:3Dsys_mom}
	\\
	\pdv{x_i \reavg{\phi}}{t} + \pdv{}{x_j} \ppar*{x_i\pbrk*{\reavg{u_j} + \tilde{u}_j} \reavg{\phi} + x_i \reavg{\reflc{u_j}\reflc{\phi}}} &=  \pbrk*{\reavg{u_i} + \tilde{u}_i} \reavg{\phi} +  \reavg{\reflc{u_i}\reflc{\phi}},
	\label{eqn:3Dsys_grav}
	\\
	\pdv{e}{t} + \pdv{}{x_j}\ppar*{\reavg{u_j}e + \textfrac{1}{\rho} \reavg{u_j}\reavg{p} - \tau_{ji}^R \reavg{u_i}} &= - P + R g_i \reavg{u_i} \reavg{\phi},
	\label{eqn:3Dsys_mech}
	\\
	\pdv{k}{t} + \pdv{}{x_j}\ppar*{\reavg{u_j}k + \textfrac{1}{\rho}\reavg{\reflc{u_j}\reflc{p}} + \textfrac{1}{2}\reavg{\reflc{u_j}\reflc{u_i}\reflc{u_i}}} &= P -\epsilon + R g_i \reavg{\reflc{u_i}\reflc{\phi}},
	\label{eqn:3Dsys_turb}
\end{align}\end{subequations}
which represent conservation of volume, particles, momentum, centre of mass (COM), mean-flow kinetic energy (MKE), and turbulent kinetic energy (TKE) respectively. We denote the velocity of the fluid by $\vecb{u}(\vecb{x},t) = (u,v,w)$, the pressure relative to the hydrostatic ambient by $p(\vecb{x},t)$ (so that the actual pressure is $p+\rho g_i x_i$), the MKE by $e \eqdef \textfrac{1}{2} \reavg{u_i} \reavg{u_i}$, and the TKE by $k \eqdef \textfrac{1}{2} \reavg{\reflc{u_i}\reflc{u_i}}$. The Reynolds stress, turbulent energy production, and turbulent energy dissipation are
\begin{align}
	\tau_{ij}^R &\eqdef - \reavg{\reflc{u_i}\reflc{u_j}},	&
	P &\eqdef \tau_{ji}^R\pdv{\reavg{u_i}}{x_j},	&
	\epsilon &\eqdef \nu \reavg{\pdv{\reflc{u_i}}{x_j}\pdv{\reflc{u_i}}{x_j}}
\end{align}
respectively. The boundary conditions at the bed are no-slip in the mean flow, \ie
\begin{subequations} \label{eqn:3Dsys_BC}
\begin{align} \label{eqn:3Dsys_BC_bed}
    \reavg{u} = \reavg{v} = \reavg{w} &= 0
    &&\text{for}&
    z = \tilde{B}.
\end{align}
We do not make similar conditions on the fluctuating components of velocity because we neglect the viscous boundary layer from our model (equivalently, $\tilde{B}$ is the elevation of the edge of the boundary layer). In the far field we require
\begin{align} \label{eqn:3Dsys_BC_infty}
	\reavg{u} = \reavg{v} = \reavg{\phi} = \reavg{p} = \reflc{u} = \reflc{v} = \reflc{w} = \reflc{\phi} = \reflc{p} &= 0
	&&\text{for}&
	z &\to \infty.
\end{align}
Note that $\reavg{w}$ is not set to zero, because the current entrains and displaces the ambient fluid, and this component of velocity is non-zero to account for the moving volume. To include this in or model, we carefully choose the location of the surface $z=h(x,t)$. For the idealised case of a current that is in approximate similarity form (\ie $\xi_u$, $\xi_\phi$, and $\xi_k$ are independent of $x$ and $t$), the location of the surface moves to track the evolution of the transverse structure. However, we must be careful as to where this surface is located in order to compare to the velocity $\eval{\reavg{w}}_{z\to\infty}$. If we consider a 2D current (neglect $y$ variation) and a surface $h(x,t)$ that moves with the fluid (neglect entrainment for now), then
\begin{align*}
	\pdv{h}{t} &
	= \eval*{\reavg{w}}_{z=h} - \pdv{h}{x} \eval*{\reavg{u}}_{z=h}
	= \eval*{\reavg{w}}_{z\to\infty} - \eval*{\reavg{w}}_{z=h}^{\infty} - \pdv{h}{x} \eval*{\reavg{u}}_{z=h}
\\&
	= \eval*{\reavg{w}}_{z\to\infty} + \int_h^\infty \pdv{\reavg{u}}{x} \dd{z} - \pdv{h}{x} \eval*{\reavg{u}}_{z=h}
	= \eval*{\reavg{w}}_{z\to\infty} + \pdv{}{z} \int_h^\infty \reavg{u} \dd{z}.
\end{align*}
We require the gradient of total longitudinal velocity above $z=h$ to be small, and for a 3D current a similar requirement is made about the gradient ($\pdv*{}{y}$) of lateral velocity. Thus, the interface should be above the bulk of the non-zero velocity field ($\reavg{u} \ll \reavg{u}_{\max}$ on $z>h$). However, to track the similarity form (if present) the interface cannot be too much higher than where the velocity field vanishes. Consider two elevations, $z=h_1$ and $z=h_2$, $h_1$ at the lowest elevation where $\reavg{u}$ becomes negligible, and initially $h_2(x,0) = c h_1(x,0)$, $c>1$. For both interfaces to track the fluid, they must both satisfy $\pdv*{h_i}{t} = \eval{\reavg{w}}_{z\to\infty}$, thus if after some time $t$ we have $h_1(x,t) = (1+\delta) h_1(x,0)$ then at the same time $h_2(x,t) = (c+\delta) h_1(x,0)$, and consequently the  elevation at which the velocity becomes negligible (the top of the similarity profile) has remained at $z=h_1$, but moved from $z = h_2/c$ to $z = h_2 \cdot \frac*{(1+\delta)}{(c+\delta)}$. This motion of the velocity field means that $h_2$ cannot capture a similarity profile (if present) while $h_1$ can. (For real flows where the similarity form is approximate at best, this argument needs to be extended to account for the properties of all fields over the entire depth, which will result in a small adjustment to the appropriate choice of $h$.)

Consequently, we conclude that the appropriate definition of the interface $z=h$ is the lowest elevation above which $\reavg{u}$ and $\reavg{v}$ become negligible. To account for some rate of entrainment of ambient fluid, $w_e$, we make the definition
\begin{align} \label{eqn:3Dsys_BC_entrain}
    w_e \eqdef \pdv{h}{t} - \eval*{\reavg{w}}_{z \to \infty}
\end{align}
\end{subequations}
where $h(x,t)$ is a measure of the depth of the fluid in the channel.

\subsection{The scales of the flow within the channel}  \label{app:deriv_channel_scales}

In an environmental setting, it is common for the longitudinal ($x$) length-scale to be greater then the lateral ($y$) which is greater than the transverse ($z$), a property we now exploit to simplify the system. In this section we study the scales of the flow \emph{within the channel}, the flow over the levee will be considered later. We employ a time-scale $\mathscr{T}$ , and length-scales $\mathscr{L}_i$ corresponding to the length ($i=1$), width ($i=2$), and depth ($i=3$) of the current ($\mathscr{L}_1 \gg \mathscr{L}_2 \gg \mathscr{L}_3$). We assume the Reynolds averaged velocities scale as $\mathscr{U}_i = \mathscr{L}_i/\mathscr{T}$. We denote the scale of the Reynolds averaged pressure by $\mathscr{P}$, the TKE scale by $\mathscr{K}$, the dissipation scale by $\mathscr{E}$, the scale of the Reynolds averaged concentration as $\varphi$, and define the gravitation scales $\mathscr{G}_i = R g_i \varphi$ so that $\mathscr{G}_1 = \mathscr{G}_3 \abs{\tan\theta}$.

All scales should be understood as the scale of the channel average \cref{eqn:CHavg_defn} of the given quantity, because this is the relevant scale for later analysis. It is possible to perform the channel average first and then the analysis of scales, which is preferable from the perspective of formal justification. However, this approach increases the complexity of the analysis so substantially that it becomes unintelligible, which is why the order of presentation here has been chosen.

We first consider the momentum equation \cref{eqn:3Dsys_mom}. To examine the scales of the system we require scales for the components of the Reynolds stress, for which we employ the eddy viscosity approximation, that is
\begin{align} \label{eqn:deviatoric_stress}
	\tau_{ij}^R &= - \textfrac{2}{3} k \delta_{ij} + \tau_{ij}^D,
	&\text{where}&&
	\tau_{ij}^D &= \nu_t \ppar*{ \pdv{\reavg{u_i}}{x_j} + \pdv{\reavg{u_j}}{x_i} }
\end{align}
is the deviatoric Reynolds stress, and the eddy viscosity $\nu_t(\vecb{x},t)$ has scale $\mathscr{N}$. We also employ the scales
\begin{align}
	\norm{\tilde{\vecb{u}}} &\lesssim \frac{\mathscr{L}_3}{\mathscr{T}},
	&
	\reavg{ \reflc{u_i} \phi } &\sim \frac{ \mathscr{N} \varphi }{ \mathscr{L}_i },
	&&\text{and}&
	\textfrac{1}{\rho}\reavg{\reflc{u_i}\reflc{p}} + \textfrac{1}{2}\reavg{\reflc{u_i}\reflc{u_j}\reflc{u_j}} &\sim \frac{ \mathscr{N} \mathscr{K} }{ \mathscr{L}_i }.
\end{align}
Now \cref{eqn:3Dsys_mom} becomes
\begin{align}
	\underbrace{ \pdv{\reavg{u_i}}{t} + \pdv{}{x_j}\ppar*{\reavg{u_j}\reavg{u_i}}   \vphantom{\Bigg)}}_{\textstyle \mathscr{L}_i/\mathscr{T}^2} &= 
	- \underbrace{ \frac{1}{\rho} \pdv{\reavg{p}}{x_i} 					            \vphantom{\Bigg)}}_{\textstyle \mathscr{P}/\rho\mathscr{L}_i}
	- \underbrace{ \frac{2}{3} \pdv{k}{x_i} 							            \vphantom{\Bigg)}}_{\textstyle \mathscr{K}/\mathscr{L}_i}
	+ \underbrace{ \pdv{}{x_j} \ppar*{ \nu_t \pdv{\reavg{u_i}}{x_j} }	            \vphantom{\Bigg)}}_{\textstyle \mathscr{N}\mathscr{L}_i/\mathscr{L}_j^2\mathscr{T}}
	+ \underbrace{ \pdv{\nu_t}{x_j} \pdv{\reavg{u_j}}{x_i} 				            \vphantom{\Bigg)}}_{\textstyle \mathscr{N}/\mathscr{L}_i\mathscr{T}}
	+ \underbrace{ R g_i \reavg{\phi}						                        \vphantom{\Bigg)}}_{\textstyle \mathscr{G}_i},
\end{align}
where the scales of the flow within the channel are given underneath. In the longitudinal direction ($i=1$) the driving force, scale $\mathscr{D}$, is provided by the larger of the pressure+TKE gradient and the longitudinal component of gravity, \ie
\begin{align}
	\mathscr{D} = \max \ppar*{ \frac{\textfrac{1}{\rho} \mathscr{P} + \mathscr{K}}{\mathscr{L}_1} , \mathscr{G}_1 }.
\end{align}
The driving force accelerates the flow until the turbulent viscous effects are sufficiently strong, causing all three effects to appear at leading order
\begin{align}
	\frac{\mathscr{L}_1}{\mathscr{T}^2} &= \mathscr{D} = \frac{\mathscr{N}\mathscr{L}_1}{\mathscr{L}_3^2\mathscr{T}},
	&\text{thus}&&
	\mathscr{D} &= \frac{\mathscr{L}_1}{\mathscr{T}^2},
	&
	\mathscr{N} &= \frac{\mathscr{L}_3^2}{\mathscr{T}}.
\end{align}
In the transverse ($i=3$) direction, the pressure+TKE gradient is generated by the effects of gravity,
\begin{align}
	\frac{\textfrac{1}{\rho}\mathscr{P} + \mathscr{K}}{\mathscr{L}_3} &= \mathscr{G}_3  \gg \frac{\mathscr{L}_3}{\mathscr{T}^2}.
\end{align}
How this balance interacts with the longitudinal balance depends on the slope. On a very shallow slope
\begin{equation}\begin{gathered}
	\begin{aligned}
		\abs{\tan\theta} &\leq \frac{\mathscr{L}_3}{\mathscr{L}_1}
	&\text{we have}&&
		\mathscr{G}_1 &\leq \frac{\textfrac{1}{\rho} \mathscr{P} + \mathscr{K}}{\mathscr{L}_1},
	&\text{thus}&&
		\mathscr{D} &= \frac{\textfrac{1}{\rho} \mathscr{P} + \mathscr{K}}{\mathscr{L}_1}
	\end{aligned}
\\
	\begin{aligned}
	&\text{so that}&&
		\mathscr{G}_3 \mathscr{L}_3 &= \textfrac{1}{\rho}\mathscr{P} + \mathscr{K} = \mathscr{D} \mathscr{L_1} = \frac{\mathscr{L}_1^2}{\mathscr{T}^2},
	\end{aligned}
\\
	\begin{aligned}	
	&\text{and}&
		\mathscr{G}_3 &\gg \frac{\mathscr{L}_3}{\mathscr{T}^2}
	&\text{implies}&&
		\mathscr{L}_1 &\gg \mathscr{L}_3;
	\end{aligned}
\end{gathered}\end{equation}
the current is driven by longitudinal pressure gradients. On a moderate to steep slope
\begin{equation}\begin{gathered}
	\begin{aligned}
		\abs{\tan\theta} &\geq \frac{\mathscr{L}_3}{\mathscr{L}_1}
	&\text{we have}&&
		\mathscr{G}_1 &\geq \frac{\textfrac{1}{\rho} \mathscr{P} + \mathscr{K}}{\mathscr{L}_1},
	&\text{thus}&&
		\mathscr{D} &= \mathscr{G}_1
	\end{aligned}
\\
	\begin{aligned}
	&\text{so that}&&
		\textfrac{1}{\rho}\mathscr{P} + \mathscr{K} &= \mathscr{G}_3 \mathscr{L}_3 = \frac{\mathscr{G}_1 \mathscr{L}_3}{\abs{\tan\theta}} = \frac{\mathscr{D} \mathscr{L}_3}{\abs{\tan\theta}} = \frac{\mathscr{L}_1 \mathscr{L}_3}{\mathscr{T}^2 \abs{\tan\theta}},
	\end{aligned}
\\
	\begin{aligned}	
	&\text{and}&
		\mathscr{G}_3 &\gg \frac{\mathscr{L}_3}{\mathscr{T}^2}
	&\text{implies}&&
		\mathscr{L}_1 &\gg \mathscr{L}_3 \abs{\tan\theta};
	\end{aligned}
\end{gathered}\end{equation}
the current is driven directly by the longitudinal component of gravity. Combining the two cases, the scales have bounds
\begin{align}
	\mathscr{G}_1 &\leq \frac{\mathscr{L}_1}{\mathscr{T}^2},
	&
	\mathscr{G}_3 &\leq \frac{\mathscr{L}_1^2}{\mathscr{T}^2 \mathscr{L}_3},
	&
	\textfrac{1}{\rho} \mathscr{P} + \mathscr{K} &\leq \frac{\mathscr{L}_1^2}{\mathscr{T}^2}.
\end{align}
where the first is equality on moderate to steep slopes, and the latter two are equality on shallow slopes. Analysing the turbulent production, the dominant contribution is
\begin{align}
	P &\simeq \nu_t \ppar*{ \pdv{\reavg{u}}{z} }^2 \sim \frac{\mathscr{N} \mathscr{U}_1^2}{\mathscr{L}_3^2} = \frac{\mathscr{L}_1^2}{\mathscr{T}^3},
	&\text{thus by \cref{eqn:3Dsys_turb}}&&
	\mathscr{K} &= \frac{\mathscr{L}_1^2}{\mathscr{T}^2},
	&
	\mathscr{E} &\leq \frac{\mathscr{L}_1^2}{\mathscr{T}^3}.
\end{align}
Using the developed scales, simplifications can be made to the system of equations \cref{eqn:3Dsys} by neglecting terms that are of order $\mathscr{L}_3/\mathscr{L}_1$ or $\mathscr{L}_2/\mathscr{L}_1$ smaller than the largest, or smaller, yielding
\begin{subequations}\label{eqn:3Dsys_scales}
\begingroup
\allowdisplaybreaks[1]
\begin{align}
	\pdv{\reavg{u_j}}{x_j} &= 0,
	\label{eqn:3Dsys_scales_vol}
	\\
	\pdv{\reavg{\phi}}{t} 
	+ \pdv{}{x_j} \ppar*{\reavg{u_j} \reavg{\phi}}
	+ \pdv{}{z} \ppar*{\tilde{w} \reavg{\phi} + \reavg{\reflc{w}\reflc{\phi}}}
	&\simeq 0,
	\label{eqn:3Dsys_scales_part}
	\\
	\pdv{\reavg{u}}{t} 
	+ \pdv{}{x_j}\ppar*{\reavg{u_j}\reavg{u}} 
	- \pdv{\tau_{31}^D}{z}
	+ \pdv{p^T}{x}
	&\simeq R g_1 \reavg{\phi},
	\label{eqn:3Dsys_scales_momx}
	\\
	\pdv{p^T}{y} &\simeq 0,
	\label{eqn:3Dsys_scales_momy}
	\\
	\pdv{p^T}{z} &\simeq R g_3 \reavg{\phi},
	\label{eqn:3Dsys_scales_hydro}
	\\
	\pdv{z \reavg{\phi}}{t} 
	+ \pdv{}{x_j} \ppar*{z \reavg{u_j} \reavg{\phi}}
	+ \pdv{}{z} \ppar*{z \tilde{w} \reavg{\phi} + z \reavg{\reflc{w}\reflc{\phi}}}
	&\simeq \pbrk*{\reavg{w} + \tilde{w}} \reavg{\phi} +  \reavg{\reflc{w}\reflc{\phi}},
	\label{eqn:3Dsys_scales_grav}
	\\
	\pdv{e}{t} 
	+ \pdv{}{x_j}\ppar*{\reavg{u_j}e + \reavg{u_j}p^T} 
	- \pdv{}{z}\ppar*{\tau_{31}^D \reavg{u}}
	&\simeq - P + R g_i \reavg{u_i} \reavg{\phi},
	\label{eqn:3Dsys_scales_mech}
	\\
	\pdv{k}{t} 
	+ \pdv{}{x_j}\ppar*{\reavg{u_j}k}
	+ \pdv{}{z}\ppar*{\textfrac{1}{\rho}\reavg{\reflc{w}\reflc{p}} + \textfrac{1}{2}\reavg{\reflc{w}\reflc{u_i}\reflc{u_i}}}
	&\simeq P -\epsilon + R g_3 \reavg{\reflc{w}\reflc{\phi}},
	\label{eqn:3Dsys_scales_turb}
\end{align}
\endgroup
Importantly, \cref{eqn:3Dsys_scales_momy,eqn:3Dsys_scales_hydro} describe a hydrostatic balance for the pressure field augmented by turbulence
\begin{align}
	p^T &\eqdef \textfrac{1}{\rho} \reavg{p} + \textfrac{2}{3} k.
\end{align}
If there is a lateral variation in particle concentration then this will generate a lateral flow field, which will be rapid in comparison to the long time scales it takes to flow over longitudinal distances. As a consequence we should expect that the particle field is uniformly distributed over $y$, generating a hydrostatic pressure field similarly distributed. The longitudinal velocity field is generated primarily by the action of pressure and gravity, thus this should be uniformly distributed also, except close to the boundaries.

We observe at this stage that the three contributions to the energy \cref{eqn:3Dsys_scales_grav,eqn:3Dsys_scales_mech,eqn:3Dsys_scales_turb} can be combined as $\text{\cref{eqn:3Dsys_scales_mech}}+\text{\cref{eqn:3Dsys_scales_turb}}-R g_3 \text{\cref{eqn:3Dsys_scales_grav}}$, which yields the equation for the total energy
\begin{multline}\label{eqn:3Dsys_scales_egy}
    \pdv{}{t} \ppar*{ e + k - R g_3 z \reavg{\phi} }
    + \pdv{}{x_j} \ppar*{ \reavg{u_j} \pbrk*{ e + k - R g_3 z \reavg{\phi} + p^T } }
    \\
    + \pdv{}{z} \ppar*{ - \tau_{31}^D \reavg{u} + \textfrac{1}{\rho}\reavg{\reflc{w}\reflc{p}} + \textfrac{1}{2}\reavg{\reflc{w}\reflc{u_i}\reflc{u_i}} - R g_3 z \pbrk*{\reavg{\reflc{w}\reflc{\phi}} + \tilde{w} \reavg{\phi}} }
    \\
    \simeq R g_i \reavg{u_i} \reavg{\phi}
    - \epsilon
    - R g_3 \pbrk*{\reavg{w} + \tilde{w}} \reavg{\phi}.
\end{multline}
\end{subequations}
This equation does not include turbulent production, $P$, or uplift of particles, $\reavg{\reflc{w}\reflc{\phi}}$.

\subsection{Averaging over the channel}

To average the system of equations we introduce the spatial averaging operator, $\spavg{\omitdummy}{}{}$, defined as
\begin{subequations}
\begin{align} \label{eqn:CHavg_defn}
	\spavg{f}{}{}(x,t) \eqdef \frac{1}{h(x,t) \Upsilon(x)} \int_{0}^{\Upsilon_2(x)} \ppar*{ \int_{\tilde{B}(x,y)}^{\infty} f(x,y,z,t)  \dd{z}}\dd{y}.
\end{align}
where $h(x,t)$ is a measure of the transverse extent of the current. We define the boundary average operators
\begin{align}
	\spavg{f}{0}{}(x,t) &\eqdef \frac{1}{h(x,t)}\int_{0}^{\infty} f(x,0,z,t) \dd{z},
	\\
	\spavg{f}{\Upsilon}{}(x,t) &\eqdef \frac{1}{h(x,t)} \int_{0}^{\infty} f(x,\tilde{\Upsilon}(x,z),z,t) \dd{z},
	\\
	\spavg{f}{B}{}(x,t) &\eqdef \frac{1}{\Upsilon(x)}\int_{0}^{\Upsilon_2(x)} f(x,y,\tilde{B}(x,y),t) \dd{y},
	\\
	\spavg{f}{\infty}{}(x,t) &\eqdef \lim_{z \to \infty} \frac{1}{\Upsilon(x)}\int_{0}^{\Upsilon_2(x)} f(x,y,z,t) \dd{y},
\intertext{and their differences}
	\spavg{f}{0}{\Upsilon}(x,t) &\eqdef \spavg{f}{\Upsilon}{}(x,t) - \spavg{f}{0}{}(x,t),
	\\
	\spavg{f}{B}{\infty}(x,t) &\eqdef \spavg{f}{\infty}{}(x,t) - \spavg{f}{B}{}(x,t).
\end{align}
\end{subequations}
The channel average of derivatives transforms as
\begin{multline*}
	h \Upsilon \spavg{\pdv{f_t}{t} + \pdv{f_x}{x} + \pdv{f_y}{y} + \pdv{f_z}{z}}{}{} 
	\simeq \pdv{}{t} \ppar*{ h \Upsilon \spavg{f_t}{}{} } + \pdv{}{x} \ppar*{ h \Upsilon \spavg{f_x}{}{} }- h \dv{\Upsilon}{x} \spavg{f_x}{\Upsilon}{} + h \spavg{f_y}{0}{\Upsilon} + \Upsilon \spavg{f_z}{B}{\infty},
\end{multline*}
where we have used \cref{eqn:dBdx_simplified} and $\dv*{\Upsilon_2}{x} \simeq \dv*{\Upsilon}{x}$ to leading order. Equivalently,
\begin{multline} \label{eqn:CHavg_deriv}
	h \spavg{\pdv{f_t}{t} + \pdv{f_x}{x} + \pdv{f_y}{y} + \pdv{f_z}{z}}{}{} 
	\simeq \pdv{}{t} \ppar*{ h \spavg{f_t}{}{} } + \pdv{}{x} \ppar*{ h \spavg{f_x}{}{} } + \frac{h}{\Upsilon} \dv{\Upsilon}{x} \ppar*{ \spavg{f_x}{}{} - \spavg{f_x}{\Upsilon}{} } + \frac{h}{\Upsilon} \spavg{f_y}{0}{\Upsilon} + \spavg{f_z}{B}{\infty}.
\end{multline}
An important fact, that will be used to simplify expressions, is that
\begin{align}
    \eval*{ - \dv{\Upsilon}{x} \reavg{u} + \reavg{v} }_{y=\tilde{\Upsilon}} \simeq \reavg{u_n} \eqdef \eval*{ \reavg{\vecb{u}} \dotp \vecb{n} }_{y=\tilde{\Upsilon}}
\end{align}
where $\reavg{u_n} $ is the component of fluid velocity in the direction normal to the levees
\begin{align}
	\vecb{n} \eqdef \ppar*{1 + \ppar*{\dv{\Upsilon}{x}}^2}^{-1/2} \, \ppar*{ -\dv{\Upsilon}{x} , 1 , 0 },
\end{align}
and we have used that $(1 + (\dv*{\Upsilon}{x})^2)^{-1/2} \simeq 1$ to leading order. Thus
\begin{align} \label{eqn:CHavg_normvel}
    - \dv{\Upsilon}{x} \spavg{\reavg{u} f}{\Upsilon}{} + \spavg{\reavg{v} f}{\Upsilon}{} \simeq \spavg{\reavg{u_n} f}{\Upsilon}{}.
\end{align}
to leading order.

To calculate the depth average of the pressure, which satisfies
\begin{align}
    p^T &\simeq - R \hat{g}_3 \int_{z}^{\infty} \eval*{\reavg{\phi}}_{z=z_2} \dd{z_2}
\end{align}
we employ the following transformation, exchanging the order of integration
\begin{align} \label{eqn:swap_order_int_z}
    \int_{\tilde{B}}^\infty \ppar*{ \int_{z_1}^\infty f(z_2) \dd{z_2} } \dd{z_1}
    = \int_{\tilde{B}}^\infty \ppar*{ \int_{\tilde{B}}^{z_2} f(z_2) \dd{z_1} } \dd{z_2}
    = \int_{\tilde{B}}^\infty  (z_2 - \tilde{B}) f(z_2) \dd{z_2},
\end{align}
consequently
\begin{align}
    \spavg{p^T}{}{} 
    \simeq - \frac{R g_3}{h \Upsilon} \int_{0}^{\mathrlap{\Upsilon_2}} \, \int_{\tilde{B}}^{\infty} (z-\tilde{B}) \reavg{\phi} \dd{z} \dd{y}
    \simeq - \textfrac{1}{2} \sigma_{z\phi} R g_3 h \sravg{\phi}{}{},
\end{align}
where
\begin{align}
	\sigma_{z\phi} \eqdef \frac{ \spavg{z\reavg{\phi}}{}{} }{ \textfrac{1}{2}h \sravg{\phi}{}{} }
	\simeq \frac{ \spavg{(z-{\tilde{B}})\reavg{\phi}}{}{} }{ \textfrac{1}{2}h \sravg{\phi}{}{} },
\end{align}
We now apply the operator $h \spavg{\omitdummy}{}{}$ using \cref{eqn:CHavg_deriv} to the system \cref{eqn:3Dsys_scales}, apply the boundary conditions \cref{eqn:3Dsys_BC} and statistical symmetry about $y=0$, and simplify using \cref{eqn:CHavg_normvel}.
\begin{subequations}	\label{eqn:1Dsys_noshape}
\begin{align}	\label{eqn:1Dsys_noshape_vol}
	0 &
	\simeq \pdv{h}{t} + \pdv{}{x}\ppar*{h \spavg{\reavg{u}}{}{}}
	+ \frac{h}{\Upsilon} \dv{\Upsilon}{x} \sravg{u}{}{}
	+ \frac{h}{\Upsilon} \sravg{u_n}{\Upsilon}{} - w_e,
\end{align}
\begin{multline}
	0
	\simeq \pdv{}{t}\ppar*{h \spavg{\reavg{\phi}}{}{}} + \pdv{}{x} \ppar*{h \spavg{\reavg{u}\reavg{\phi}}{}{} }
	+ \frac{h}{\Upsilon} \dv{\Upsilon}{x} \spavg{\reavg{u}\reavg{\phi}}{}{}	
	+ \frac{h}{\Upsilon} \spavg{\reavg{u_n}\reavg{\phi}}{\Upsilon}{}
	- \tilde{w}\spavg{\reavg{\phi}}{B}{}
	- \spavg{\reavg{\reflc{w}\reflc{\phi}}}{B}{},
\end{multline}
\begin{multline}
	0
	\simeq \pdv{}{t}\ppar*{h \spavg{\reavg{u}}{}{}}
	+ \pdv{}{x} \ppar*{h \pbrk*{ \spavg{\reavg{u}^2}{}{} + \spavg{p^T}{}{}  }}
	+ \frac{h}{\Upsilon} \dv{\Upsilon}{x} \pbrk*{ \spavg{\reavg{u}^2}{}{} + \spavg{p^T}{}{} - \spavg{p^T}{\Upsilon}{} }	\\
	+ \frac{h}{\Upsilon} \spavg{\reavg{u_n}\reavg{u}}{\Upsilon}{}
	+ \spavg{\tau_{31}^R}{B}{}
	- h g_1 R \spavg{\reavg{\phi}}{}{},
\end{multline}
\begin{multline}
	0
	\simeq \pdv{}{t}\ppar*{h \spavg{z\reavg{\phi}}{}{}}
	+ \pdv{}{x} \ppar*{h \spavg{\reavg{u}z\reavg{\phi}}{}{} }
	+ \frac{h}{\Upsilon} \dv{\Upsilon}{x} \spavg{\reavg{u}z\reavg{\phi}}{}{}	
	+ \frac{h}{\Upsilon} \spavg{\reavg{u_n}z\reavg{\phi}}{\Upsilon}{}
	- h \spavg{\reavg{w}\reavg{\phi}}{}{} 
	- h \tilde{w} \sravg{\phi}{}{} 
	- h \sravg{\reflc{w}\reflc{\phi}}{}{},
\end{multline}
\begin{multline}
	0 
	\simeq \pdv{}{t} \ppar*{ h \spavg{e}{}{} } 
	+ \pdv{}{x} \ppar*{ h \pbrk*{ \spavg{\reavg{u}e}{}{} + \spavg{\reavg{u}p^T}{}{} }}
	+ \frac{h}{\Upsilon} \dv{\Upsilon}{x} \pbrk*{ \spavg{\reavg{u}e}{}{} + \spavg{\reavg{u}p^T}{}{} } \\
	+ \frac{h}{\Upsilon} \pbrk*{ \spavg{\reavg{u_n}e}{\Upsilon}{} + \spavg{\reavg{u_n}p^T}{\Upsilon}{} }
	+ h \spavg{P}{}{}
	- R g_i h \spavg{\reavg{u_i}\reavg{\phi}}{}{},
\end{multline}
\begin{multline}
	0 
	\simeq \pdv{}{t} \ppar*{ h \spavg{k}{}{} }
	+ \pdv{}{x} \ppar*{ h \spavg{\reavg{u}k}{}{} }
	+ \frac{h}{\Upsilon} \dv{\Upsilon}{x} \spavg{\reavg{u}k}{}{} \\
	+ \frac{h}{\Upsilon} \spavg{\reavg{u_n}k}{\Upsilon}{}
	- \spavg{ \textfrac{1}{\rho}\reavg{\reflc{w}\reflc{p}} + \textfrac{1}{2}\reavg{\reflc{w}\reflc{u_i}\reflc{u_i}} }{B}{}
	- h \spavg{P}{}{}
	+ h \spavg{\epsilon}{}{}
	- R g_3 h \spavg{\reavg{\reflc{w}\reflc{\phi}}}{}{},
\end{multline}
\begin{multline}
	0 
	\simeq \pdv{}{t} \ppar*{ h \pbrk*{ \spavg{e}{}{} + \spavg{k}{}{} - R g_3 \spavg{z\reavg{\phi}}{}{} }} 
	+ \pdv{}{x} \ppar*{ h \pbrk*{ \spavg{\reavg{u}e}{}{} + \spavg{\reavg{u}k}{}{} - R g_3 \spavg{\reavg{u}z\reavg{\phi}}{}{} + \spavg{\reavg{u}p^T}{}{} }} \\
	+ \frac{h}{\Upsilon} \dv{\Upsilon}{x} \pbrk*{ \spavg{\reavg{u}e}{}{} + \spavg{\reavg{u}k}{}{} - R g_3 \spavg{\reavg{u}z\reavg{\phi}}{}{} + \spavg{\reavg{u}p^T}{}{} }  \\
	+ \frac{h}{\Upsilon} \pbrk*{ \spavg{\reavg{u_n}e}{\Upsilon}{} + \spavg{\reavg{u_n}k}{\Upsilon}{} - R g_3 \spavg{\reavg{u_n}z\reavg{\phi}}{\Upsilon}{} + \spavg{\reavg{u_n}p^T}{\Upsilon}{} } \\
	- \spavg{ \textfrac{1}{\rho}\reavg{\reflc{w}\reflc{p}} + \textfrac{1}{2}\reavg{\reflc{w}\reflc{u_i}\reflc{u_i}} }{B}{}
	+ h \spavg{\epsilon}{}{}
	- R g_1 h \spavg{\reavg{u}\reavg{\phi}}{}{} 
	+ R g_3 h \tilde{w} \sravg{\phi}{}{}.
\end{multline}
\end{subequations}
	\section{Derivation of the flow over the levees} \label{app:deriv_levee}

\subsection{The scales of flow over the levees}

\begin{figure}
	\centering
	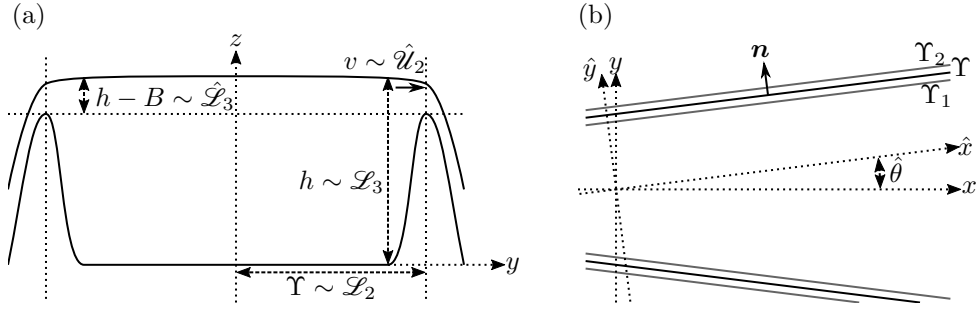
	\caption{(a) A cross section of the channel, showing the lengths and their associated scales. (b) A top-down view of the channel, showing the global coordinate system $(x,y)$ and the transformed local coordinate system $(\hat{x},\hat{y})$. Also depicted is the edge of the channel, $y=\Upsilon(x)$ in black and $y=\Upsilon_i(x)$ in grey ($i \in \{1,2\}$). In both figures, the axes have been non-linearly scaled to make visible the important features.}
	\label{fig:LeveeOverspillScales}
\end{figure}

To examine the flow over the levees, we transform into a coordinate system $(\hat{x},\hat{y},\hat{z})$ aligned with the levee on $y>0$, that is $\hat{y}$ is the direction $\vecb{n}$, and the transverse coordinate is unchanged, $\hat{z} = z$, see \cref{fig:LeveeOverspillScales}. The transformed velocity is $\hat{u}_i$ and gravity is $\hat{g}_i$. Denoting the angle of the levee walls by $\tan \hat{\theta} \eqdef \dv**{\Upsilon}{x} \sim \mathscr{L}_2/\mathscr{L}_1$, the velocities transform as
\begin{subequations}\begin{align}
	\hat{u} &\eqdef u \cos \hat{\theta} + v \sin \hat{\theta} \simeq u,
	&
	\hat{v} &\eqdef v \cos \hat{\theta} - u \sin \hat{\theta},
	&
	\hat{w} &\eqdef w,
\intertext{and $\eval{\hat{v}}_{y = \Upsilon_2} = u_n$; the gravity transforms as}
	\hat{g}_1 &\eqdef g_1 \cos \hat{\theta} \simeq g_1,
	&
	\hat{g}_2 &\eqdef - g_1 \sin \hat{\theta},
	&
	\hat{g}_3 &\eqdef g_3,
\end{align}\end{subequations}
Similar to the assumption \cref{eqn:seperable_bed}, the bed elevation is considered to be locally independent of $\hat{x}$ on the slope $\hat{\Upsilon}_1 < \hat{y} < \hat{\Upsilon}_2$, where $\hat{\Upsilon}_1$ and $\hat{\Upsilon}_2$ are also locally independent of $\hat{x}$. We introduce scales for the flow over the levee crest, in particular length scales $\hat{\mathscr{L}}_i$, where $\hat{\mathscr{L}}_1 = \mathscr{L}_1$, $\hat{\mathscr{L}}_2$ is the scale of the levee width $\hat{\Upsilon}_2 - \hat{\Upsilon}_1$, and $\hat{\mathscr{L}}_3$ the scale of the flow depth above the levee ($\mathscr{L}_1 = \hat{\mathscr{L}}_1 \gg \mathscr{L}_2 \gg \hat{\mathscr{L}}_2 \geq \mathscr{L}_3 \geq \hat{\mathscr{L}}_3$). We also introduce new velocity scales $\hat{\mathscr{U}}_i = \hat{\mathscr{L}}_i/\mathscr{T}$, where $\hat{\mathscr{U}}_1 = \mathscr{U}_1$.

For the effect of levee overspill to not dominate \cref{eqn:1Dsys_noshape_vol}, we require that $\mathscr{U}_2 \leq \mathscr{L}_2/\mathscr{T}$, and when the overspill effects the system at leading order (peak overflow) we will have $\mathscr{U}_2 = \mathscr{L}_2/\mathscr{T}$. From this we can establish the scales of velocity from the conservation of volume \cref{eqn:3Dsys_vol}. First, integrating over the region above the slope
\begin{align}
	0 
	&= \int_{\hat{\Upsilon}_1}^{\mathrlap{\hat{\Upsilon}_2}} \int_{\tilde{B}}^\infty \ppar*{ \pdv{\reavg{\hat{u}}}{\hat{x}} + \pdv{\reavg{\hat{v}}}{\hat{y}} + \pdv{\reavg{\hat{w}}}{\hat{z}} } \dd{\hat{z}} \dd{\hat{y}}
\\
	&= 
	\underbrace{ \int_{\hat{\Upsilon}_1}^{\mathrlap{\hat{\Upsilon}_2}} \int_{\tilde{B}}^\infty \pdv{\reavg{\hat{u}}}{\hat{x}} \dd{\hat{z}} \dd{\hat{y}}	}_{ \hat{\mathscr{L}}_2 \mathscr{L}_3/\mathscr{T} }
	+
	\underbrace{ \int_B^\infty \eval*{\reavg{\hat{v}}}_{\hat{y} = \hat{\Upsilon}_2} \dd{\hat{z}} }_{ \hat{\mathscr{U}}_2 \hat{\mathscr{L}}_3 }
	- 
	\underbrace{ \int_0^\infty \eval*{\reavg{\hat{v}}}_{\hat{y} = \hat{\Upsilon}_1} \dd{\hat{z}} }_{ \mathscr{U}_2 \mathscr{L}_3 = \mathscr{L}_2 \mathscr{L}_3/\mathscr{T} \vphantom{\hat{\mathscr{L}}} }
\end{align}
the fist term is substantially smaller than the last, thus $\hat{\mathscr{U}}_2 \hat{\mathscr{L}}_3 = \mathscr{U}_2 \mathscr{L}_3$. Examining the equation without the volume integral
\begin{align}
	0 =
	  \underbrace{\pdv{\reavg{\hat{u}}}{\hat{x}}	\vphantom{\bigg)}}_{1/\mathscr{T} \vphantom{\hat{\mathscr{L}}}}
	+ \underbrace{\pdv{\reavg{\hat{v}}}{\hat{y}}	\vphantom{\bigg)}}_{\hat{\mathscr{U}}_2/\hat{\mathscr{L}}_2}
	+ \underbrace{\pdv{\reavg{\hat{w}}}{\hat{z}}	\vphantom{\bigg)}}_{\hat{\mathscr{U}}_3/\hat{\mathscr{L}}_3},
\end{align}
the first term is substantially smaller than the second, which is balanced by the third, thus overall we have
\begin{align}
	\hat{\mathscr{U}}_1 &= \mathscr{U}_1 = \frac{\mathscr{L}_1}{\mathscr{T}},
&
	\hat{\mathscr{U}}_2 &= \frac{\mathscr{L}_3}{\hat{\mathscr{L}}_3} \mathscr{U}_2 = \frac{\mathscr{L}_3}{\hat{\mathscr{L}}_3} \frac{\mathscr{L}_2}{\mathscr{T}},
&
	\hat{\mathscr{U}}_3 &= \hat{\mathscr{L}}_3 \frac{\hat{\mathscr{U}}_2}{\hat{\mathscr{L}}_2} = \frac{\mathscr{L}_2}{\hat{\mathscr{L}}_2} \frac{\mathscr{L}_3}{\mathscr{T}}
\end{align}
at peak overflow. At the levee crest we expect the flow to be critical, that is 
\begin{align}\label{eqn:3Dsys_scale_levee_U2}
	\hat{\mathscr{U}}_2 &= \ppar*{\hat{\mathscr{G}}_3 \hat{\mathscr{L}}_3}^{\mathrlap{1/2}},
&\text{thus}&&
	\hat{\mathscr{L}}_3 &= \ppar*{ \frac{ \mathscr{L}_2^2 \mathscr{L}_3^2 }{ \hat{\mathscr{G}}_3 \mathscr{T}^2 } }^{\mathrlap{1/3}},
&\text{equiv.}&&
	\hat{\mathscr{G}}_3 &= \frac{ \mathscr{L}_2^2 \mathscr{L}_3^2 }{ \hat{\mathscr{L}}_3^3 \mathscr{T}^2 }.
\end{align}
Here $\hat{\mathscr{G}}_3$ is the scale of the gravitational driving force, similar to $\mathscr{G}_3$ but taking into account the reduced average particle concentration in $B<z<h$ so that $\hat{\mathscr{G}}_3 < \mathscr{G}_3$. Physically, the imposition of critical flow sets the transverse scale of the flow over the levee $\hat{\mathscr{L}}_3$, though we will instead use the condition to eliminate $\hat{\mathscr{G}}_3$ from expressions.

We examine the system \cref{eqn:3Dsys_part,eqn:3Dsys_mom,eqn:3Dsys_turb} for the flow of fluid down the levee slopes, and establish scales appropriate for peak overflow. The scale of the eddy viscosity we denote by $\hat{\mathscr{N}}$ on the downslope. We do not expect drag to be dominant in this region, and in the equation for the longitudinal component of momentum, for advection to not be dominated by the deviatoric Reynolds stress require
\begin{align}
	\pdv{}{\hat{y}} \ppar*{\reavg{\hat{v}} \reavg{\hat{u}}} &\gtrsim \pdv{\hat{\tau}_{31}^D}{\hat{z}} ,
&&\text{that is}&
	\hat{\mathscr{N}} &\lesssim \frac{ \mathscr{L}_2 \hat{\mathscr{L}}_3 }{ \hat{\mathscr{L}}_2 \mathscr{L}_3 } \mathscr{N}.
\end{align}
In the equation for the transverse component of momentum, we require that the effect of gravity dominates transverse acceleration so that the pressure is hydrostatic, 
\begin{align}
	R \hat{g}_3 \reavg{\phi} &\gg \pdv{}{\hat{y}} \ppar*{\reavg{\hat{v}} \reavg{\hat{w}}},
	&&\text{that is}&
	\hat{\mathscr{L}}_3 &\ll \hat{\mathscr{L}}_2.
\end{align}
(Note, this implies $\hat{\mathscr{U}}_2 \gg \hat{\mathscr{U}}_3$). Consequently, the scale of the total effective pressure is $\textfrac{1}{\rho}\hat{\mathscr{P}} + \hat{\mathscr{K}} = \hat{\mathscr{G}}_3 \hat{\mathscr{L}}_3$. Applying these scales we arrive at the system
\begin{subequations}\label{eqn:3Dsys_levee}
\allowdisplaybreaks[1]
\begin{align}
	\sum_{j=2}^3 \pdv{\reavg{\hat{u}_j}}{\hat{x}_j} &\simeq 0,
	\label{eqn:3Dsys_levee_vol}
\\
	\sum_{j=2}^3 \pdv{}{\hat{x}_j} \ppar*{\reavg{\hat{u}_j} \reavg{\phi}} + \tilde{w} \pdv{\reavg{\phi}}{\hat{z}} &\simeq 0,
	\label{eqn:3Dsys_levee_part}
\\
	\sum_{j=2}^3 \pdv{}{\hat{x}_j}\ppar*{\reavg{\hat{u}_j}\reavg{\hat{u}}} - \pdv{\hat{\tau}_{31}^D}{\hat{z}} &\simeq 0,
	\label{eqn:3Dsys_levee_mom_x}
\\
	\sum_{j=2}^3 \pdv{}{\hat{x}_j}\ppar*{\reavg{\hat{u}_j}\reavg{\hat{v}}} + \pdv{p^T}{\hat{y}} - \pdv{\hat{\tau}_{32}^D}{\hat{z}} &\simeq 0,
	\label{eqn:3Dsys_levee_mom_y}
\\
	\pdv{p^T}{\hat{z}} &\simeq R \hat{g}_3 \reavg{\phi}.
	\label{eqn:3Dsys_levee_hydro}
\\
	\sum_{j=2}^3 \pdv{}{\hat{x}_j}\ppar*{\reavg{\hat{u}_j}k} &\simeq 0.
	\label{eqn:3Dsys_levee_turb}
\end{align}
\end{subequations}

\subsection{Quantifying the flow over the levees} \label{app:deriv_levee_quant}

We depth average the system of equations over the interval ${\tilde{B}}<z<\infty$ using the operator
\begin{align} \label{eqn:DPavg_defn}
	\spavg{f}{D}{}(\hat{x},\hat{y},t) \eqdef \frac{1}{\hat{h}(\hat{x},\hat{y},t)} \int_{{\tilde{B}}(\hat{x},\hat{y})}^{\infty} f(\hat{x},\hat{y},\hat{z},t) \dd{\hat{z}}.
\end{align}
Here $\hat{h}$ is a local measure of depth that accounts for lateral variation, and will be equal the the bulk value at $\hat{y} = \hat{\Upsilon}_1$. The depth average of derivatives transform as
\begin{align}
	\hat{h} \spavg{\pdv{f_y}{\hat{y}} + \pdv{f_z}{\hat{z}}}{D}{} 
	&= \pdv{}{\hat{y}} \ppar*{\hat{h} \spavg{f_y}{D}{}} + \pdv{{\tilde{B}}}{\hat{y}} \eval*{f_y}_{\hat{z}=\tilde{B}} + \eval*{f_z}_{\hat{z}\to\infty} - \eval*{f_z}_{\hat{z}=\tilde{B}}.
\end{align}
For the pressure, which satisfies
\begin{align}
	p^T &= - R \hat{g}_3 \int_{\hat{z}}^{\infty} \reavg{\phi} \dd{\hat{z}}.
\end{align}
we can simplify the depth average employing \cref{eqn:swap_order_int_z} to obtain
\begin{align}
	\spavg{p^T}{D}{} 
	= - \frac{R \hat{g}_3}{\hat{h}} \int_{\tilde{B}}^{\infty} (\hat{z}-{\tilde{B}}) \reavg{\phi} \dd{\hat{z}} 
	= - \textfrac{1}{2} \hat{\sigma}_{z\phi} R \hat{g}_3 \hat{h} \sravg{\phi}{D}{}.
\end{align}
Here, and going forward, we employ the shape factors
\begin{subequations}
\allowdisplaybreaks[1]
\begin{gather}
	\begin{aligned}
		\hat{\sigma}_{z\phi} &\eqdef \frac{ \spavg{(\hat{z}-{\tilde{B}})\reavg{\phi}}{D}{} }{ \textfrac{1}{2}\hat{h} \sravg{\phi}{D}{} },
		&
		\hat{\sigma}_{vv} &\eqdef \frac{ \spavg{\reavg{\hat{v}^2}}{D}{} }{ \sravg{\hat{v}}{D}{2} },
	\end{aligned}
	\\
	\begin{aligned}
		\hat{\sigma}_{vu} &\eqdef \frac{ \spavg{\reavg{\hat{v}} \reavg{u}}{D}{} }{ \sravg{\hat{v}}{D}{} \sravg{u}{D}{} },
		&
		\hat{\sigma}_{vuu} &\eqdef \frac{ \spavg{\reavg{\hat{v}} \reavg{u}^2}{D}{} }{ \sravg{\hat{v}}{D}{} \sravg{u}{D}{2} },
		&
		\hat{\sigma}_{v\phi} &\eqdef \frac{ \spavg{\reavg{\hat{v}}\reavg{\phi}}{D}{} }{ \sravg{\hat{v}}{D}{}\sravg{\phi}{D}{} },
	\end{aligned}
\end{gather}\end{subequations}

Applying the operator $\hat{h} \spavg{\omitdummy}{D}{}$ to \cref{eqn:3Dsys_levee_vol,eqn:3Dsys_levee_part,eqn:3Dsys_levee_mom_x,eqn:3Dsys_levee_mom_y,eqn:3Dsys_levee_turb}, and using that $\pdv**{\hat{h}}{t}$ is negligible due to the scales of the flow so that $\eval{\hat{w}}_{\hat{z}\to\infty} = \hat{w}_e$, yields
\begin{subequations}\label{eqn:3Dsys_levee_shape}
\allowdisplaybreaks[1]
\begin{align}
	\pdv{}{\hat{y}} \ppar*{\hat{h} \sravg{\hat{v}}{D}{}} 
	&= \hat{w}_e,
	\label{eqn:3Dsys_levee_shape_vol}
\\
	\pdv{}{\hat{y}} \ppar*{\hat{\sigma}_{v\phi} \hat{h} \sravg{\hat{v}}{D}{} \sravg{\phi}{D}{}} 
	&= \tilde{w} \eval*{\reavg{\phi}}_{\hat{z} = \tilde{B}},
	\label{eqn:3Dsys_levee_shape_part}
\\
	\pdv{}{\hat{y}} \ppar*{\hat{\sigma}_{vu} \hat{h} \sravg{\hat{v}}{D}{} \sravg{u}{D}{}} 
	&= -\eval*{\hat{\tau}_{32}^D}_{\hat{z} = \tilde{B}},
	\label{eqn:3Dsys_levee_shape_momx}
\\
	\pdv{}{\hat{y}} \ppar*{\hat{\sigma}_{vv} \hat{h} \sravg{\hat{v}}{D}{2} - \textfrac{1}{2} \hat{\sigma}_{z\phi} R g_3 \hat{h}^2 \sravg{\phi}{D}{} } 
	- \pdv{{\tilde{B}}}{\hat{y}} R g_3 \hat{h} \sravg{\phi}{D}{} 
	&= -\eval*{\hat{\tau}_{31}^D}_{\hat{z} = \tilde{B}}.
	\label{eqn:3Dsys_levee_shape_momy}
\\
	\pdv{}{\hat{y}} \ppar*{\hat{\sigma}_{vk} \hat{h} \sravg{\hat{v}}{D}{} \spavg{k}{D}{}} 
	&= 0,
	\label{eqn:3Dsys_levee_shape_turb}
\end{align}\end{subequations}
Using \cref{eqn:3Dsys_levee_shape_vol,eqn:3Dsys_levee_shape_part} we can simplify \cref{eqn:3Dsys_levee_shape_part,eqn:3Dsys_levee_shape_momx,eqn:3Dsys_levee_shape_momy,eqn:3Dsys_levee_shape_turb} so that the system becomes (provided $\hat{h} \sravg{\hat{v}}{D}{} \neq 0$)
\begin{subequations}
\allowdisplaybreaks[1]
\begin{align}
	\pdv{}{\hat{y}} \ppar*{\hat{h} \sravg{\hat{v}}{D}{}} 
	&= \hat{w}_e,
\\
	\pdv{}{\hat{y}} \ppar*{\hat{\sigma}_{v\phi} \sravg{\phi}{D}{}} 
	&= \frac{ \tilde{w} \eval*{\reavg{\phi}}_{\hat{z} = \tilde{B}} - \hat{\sigma}_{v\phi} \sravg{\phi}{D}{} \hat{w}_e }{ \hat{h} \sravg{\hat{v}}{D}{} } = \hat{S}_\phi,
\\
	\pdv{}{\hat{y}} \ppar*{\hat{\sigma}_{vu} \sravg{u}{D}{}} 
	&= \frac{ -\eval*{\hat{\tau}_{32}^D}_{\hat{z} = \tilde{B}} - \hat{\sigma}_{vu} \sravg{u}{D}{} \hat{w}_e }{ \hat{h} \sravg{\hat{v}}{D}{} },
\\
	\pdv{}{\hat{y}} \ppar*{\textfrac{1}{2} \hat{\sigma}_{vv} \sravg{\hat{v}}{D}{2} - R g_3 \sravg{\phi}{D}{} \pbrk*{ \hat{\sigma}_{z\phi}\hat{h} + {\tilde{B}} } }
	&= - \textfrac{1}{2} \sravg{\hat{v}}{D}{2} \pdv{\hat{\sigma}_{vv}}{\hat{y}}
	- \textfrac{1}{2} R g_3 \hat{h} \sravg{\phi}{D}{} \pdv{\hat{\sigma}_{z\phi}}{\hat{y}}
	\notag \\*
	- \frac{R g_3}{\hat{\sigma}_{v\phi}} &\ppar*{ \textfrac{1}{2} \hat{\sigma}_{z\phi}\hat{h} + {\tilde{B}} } \ppar*{ \sravg{\phi}{D}{} \pdv{\hat{\sigma}_{v\phi}}{\hat{y}} - \hat{S}_\phi }
	\notag \\* &\qquad
	- \frac{\hat{\sigma}_{vv} \sravg{\hat{v}}{D}{} \hat{w_e} + \hat{\tau}_{31}^D }{h}
\\
	\pdv{}{\hat{y}} \ppar*{\hat{\sigma}_{vk} \spavg{k}{D}{}} 
	&= 0.
\end{align}\end{subequations}


We assume that, on the down-slope of the levee ($\hat{y}>\hat{\Upsilon}_2$), the flow is fast and shallow. Moreover we will assume that in some small region around the levee crest ($1<\hat{y}/\hat{\Upsilon}_2<1+\epsilon$) the shape factors do not vary in space. Consequently, in this region, the values of $\hat{q}$, $\sravg{\phi}{D}{}$, $\sravg{\hat{u}}{D}{}$, and $\hat{F}$ are constant to leading order, where
\begin{subequations}
\allowdisplaybreaks[1]
\begin{align}
	\hat{q} &\eqdef \hat{h} \sravg{\hat{v}}{D}{},
	\label{eqn:3Dsys_levee_downslope_vol}
	\\
	\hat{F} &\eqdef \textfrac{1}{2} \hat{\sigma}_{vv} \sravg{\hat{v}}{D}{2} - R g_3 \sravg{\phi}{D}{} \pbrk*{ \hat{\sigma}_{z\phi}\hat{h} + \tilde{B} }.
	\label{eqn:3Dsys_levee_downslope_momy}
\end{align}\end{subequations}
The value $\hat{q}$ is the volume flux, while $\hat{F}$ is related in a non-trivial way to the sum of the momentum flux and the basal pressure force, and may be interpreted as an energy.

Similar to the manipulations made in the literature on open-channel hydraulics \citep[\eg][though there it is typical to consider only the case of unitary shape factors]{bk_Chow_OCH}, we may ask the question: given a flux $\hat{q}$, energy $\hat{F}$, and concentration $\sravg{\phi}{D}{}$, what flow states are permitted? We rewrite \cref{eqn:3Dsys_levee_downslope_momy} as 
\begin{align} \label{eqn:levee_downslope_momy_condn}
	\hat{F} + R g_3 {\tilde{B}} \sravg{\phi}{D}{}  = \hat{\sigma}_{vv} \frac{\hat{q}^2}{2 \hat{h}^2} -  \hat{\sigma}_{z\phi} R g_3 \hat{h} \sravg{\phi}{D}{}.
\end{align}
For $\hat{h}>0$, the right hand side of \cref{eqn:levee_downslope_momy_condn} takes its minimal value at the critical depth $\hat{h} = \hat{h}_{\mathrm{crit}}$ where
\begin{align}\label{eqn:levee_downslope_momy_crit_h}
	\hat{h}_{\mathrm{crit}} 
	&\eqdef \ppar*{ \frac{ \hat{\sigma}_{vv} \hat{q}^2 }{ -  \hat{\sigma}_{z\phi} R g_3 \sravg{\phi}{D}{} } }^{1/3},
\end{align}
and thus to have any solutions to \cref{eqn:levee_downslope_momy_condn} we require that
\begin{align}\label{eqn:levee_downslope_momy_crit_E}
	\hat{F} + R g_3 B \sravg{\phi}{D}{} 
	&\geq \hat{F}_{\mathrm{crit}} 
	\eqdef \textfrac{3}{2} \ppar*{\hat{\sigma}_{vv} \hat{q}^2}^{1/3} \ppar*{-  \hat{\sigma}_{z\phi} R g_3 \sravg{\phi}{D}{}}^{2/3},
\end{align}
and we refer to $\hat{F}_{\mathrm{crit}}$ as the critical energy. When the left hand side of \cref{eqn:levee_downslope_momy_condn} exceeds the lower bound \cref{eqn:levee_downslope_momy_crit_E} then there are two solutions, a subcritical (slow and deep) solution with $\hat{h} > \hat{h}_{\mathrm{crit}}$ and a supercritical (rapid and shallow) solution with $\hat{h} < \hat{h}_{\mathrm{crit}}$. For a flow with constant shape factors and no drag, entrainment, or settling, the solution for flow over varying bottom topography has constant $\hat{q}$, $\hat{F}$, and $\sravg{\phi}{D}{}$, and thus tracks the varying solution to \cref{eqn:levee_downslope_momy_condn} as the left hand side varies due to varying ${\tilde{B}}$. For the flow over a levee where the shape factors are locally constant, and the flow is subcritical on the up-slope and supercritical on the down-slope, we conclude that the flow must become critical at the crest to transition between the two branches of the solution to \cref{eqn:levee_downslope_momy_condn}. For our situation, where the shape factors are only constant just downslope of the crest, we assume that the same criticality condition still applies at the crest. Manipulating \cref{eqn:levee_downslope_momy_crit_h} results in the condition that defines depth average velocity at the crest, namely
\begin{align}\label{eqn:levee_downslope_momy_crit_v}
	\eval*{ \sravg{\hat{v}}{D}{}}_{\hat{y} = \hat{\Upsilon_2}}
	&= \eval*{ \ppar*{ - \frac{\hat{\sigma}_{z\phi}}{\hat{\sigma}_{vv}} R g_3 \hat{h} \sravg{\phi}{D}{} }^{1/2} }_{\hat{y} = \hat{\Upsilon}_2}.
\end{align}
\end{myappendices}

\bibliographystyle{jfm}
\bibliography{Bibliography/Books,Bibliography/Mine,Bibliography/ShallowWater,Bibliography/GravityCurrents,Bibliography/ChannelHydraulics,Bibliography/TurbidityCurrents,Bibliography/SalinityCurrents,Bibliography/StratifiedTurbulence,Bibliography/SlowManifoldTheory,Bibliography/Sedimentology,Bibliography/AnalysisHyperbolic}

\end{document}